\documentclass[11pt,preprint]{aastex}
\usepackage{graphicx}
\usepackage{epstopdf}

\newcommand{\Teff}{\mbox{$T_\mathrm{eff}$}}
\newcommand{\Mjup}{\mbox{$M_\mathrm{Jup}$}}
\newcommand{\Msun}{\mbox{$M_{\odot}$}}

\begin{document}
\shorttitle{shorttitle}
\title{Planets Around Low-Mass Stars (PALMS). V. \\  Age-Dating Low-Mass Companions to Members and Interlopers of Young Moving Groups$^{*,**,***,\dagger}$}
\author{Brendan P. Bowler,\altaffilmark{1, 2, 21, 22, 23}, 
Evgenya L. Shkolnik\altaffilmark{3, \ddagger},
Michael C. Liu\altaffilmark{4, 23},
Joshua E. Schlieder\altaffilmark{5},
Andrew W. Mann\altaffilmark{6},
Trent J. Dupuy\altaffilmark{6}, 
Sasha Hinkley\altaffilmark{7},
Justin R. Crepp\altaffilmark{8},
John Asher Johnson\altaffilmark{9},
Andrew W. Howard\altaffilmark{4},
Laura Flagg\altaffilmark{10,3}, 
Alycia J. Weinberger\altaffilmark{11},
Kimberly M. Aller\altaffilmark{4}, 
Katelyn N. Allers\altaffilmark{12},
William M. J. Best\altaffilmark{4}, 
Michael C. Kotson\altaffilmark{4}, 
Benjamin T. Montet\altaffilmark{1, 9, 13}, 
Gregory J. Herczeg\altaffilmark{14}
Christoph Baranec\altaffilmark{4},
Reed Riddle\altaffilmark{1}, 
Nicholas M. Law\altaffilmark{15}, 
Eric L. Nielsen\altaffilmark{16, 17}, 
Zahed Wahhaj\altaffilmark{18},
Beth A. Biller\altaffilmark{19},
Thomas L. Hayward\altaffilmark{20}
\\ }
\email{bpbowler@caltech.edu}

\altaffiltext{1}{California Institute of Technology, 1200 E. California Blvd., Pasadena, CA 91125, USA}
\altaffiltext{2}{Caltech Joint Center for Planetary Astronomy Fellow}
\altaffiltext{3}{Lowell Observatory, 1400 W. Mars Hill Road, Flagstaff, AZ 86001, USA}
\altaffiltext{4}{Institute for Astronomy, University of Hawai`i at M\={a}noa; 2680 Woodlawn Drive, Honolulu, HI 96822, USA}
\altaffiltext{5}{NASA Postdoctoral Program Fellow, NASA Ames Research Center, MS-245-3, Moffett Field, CA 94035, USA}
\altaffiltext{6}{Department of Astronomy, University of Texas at Austin, TX, USA}
\altaffiltext{7}{University of Exeter, Physics and Astronomy, EX4 4QL Exeter, UK}
\altaffiltext{8}{Department of Physics, University of Notre Dame, 225 Nieuwland Science Hall, Notre Dame, IN, 46556, USA}
\altaffiltext{9}{Harvard-Smithsonian Center for Astrophysics, 60 Garden Street, Cambridge, MA, 02138 USA}
\altaffiltext{10}{Department of Physics and Astronomy, Northern Arizona University, P.O. Box 6010, Flagstaff, AZ 86011, USA}
\altaffiltext{11}{Department of Terrestrial Magnetism, Carnegie Institution of Washington, 5241 Broad Branch Rd NW, Washington, DC 20015 USA}
\altaffiltext{12}{Department of Physics and Astronomy, Bucknell University, Lewisburg, PA 17837, USA}
\altaffiltext{13}{NSF Graduate Research Fellow}
\altaffiltext{14}{Kavli Institute for Astronomy and Astrophysics, Peking University; Yi He Yuan Lu 5, Hai Dian Qu; Beijing 100871, P. R. China}
\altaffiltext{15}{Department of Physics and Astronomy, University of North Carolina at Chapel Hill, Chapel Hill, NC 27599-3255, USA}
\altaffiltext{16}{SETI Institute, Carl Sagan Center, 189 Bernardo Avenue, Mountain View, CA 94043, USA}
\altaffiltext{17}{Kavli Institute for Particle Astrophysics and Cosmology, Stanford University, Stanford, CA 94305, USA}
\altaffiltext{18}{European Southern Observatory, Alonso de Cordova 3107, Vitacura, Santiago, Chile}
\altaffiltext{19}{Institute for Astronomy, University of Edinburgh, Blackford Hill View, Edinburgh EH9 3HJ, UK}
\altaffiltext{20}{Gemini Observatory, Southern Operations Center c/o AURA, Casilla 603, La Serena, Chile}
\altaffiltext{21}{Visiting astronomer, Cerro Tololo Inter-American Observatory, National Optical Astronomy Observatory, which are operated by the Association of Universities for Research in Astronomy, under contract with the National Science Foundation.}
\altaffiltext{22}{Visiting Astronomer, Kitt Peak National Observatory, National Optical Astronomy Observatory, which is operated by the Association of Universities for Research in Astronomy (AURA) under cooperative agreement with the National Science Foundation.}
\altaffiltext{23}{Visiting Astronomer at the Infrared Telescope Facility, which is operated by the University of Hawaii under Cooperative Agreement no. NNX-08AE38A with the National Aeronautics and Space Administration, Science Mission Directorate, Planetary Astronomy Program.}
\altaffiltext{*}{Some of the data presented herein were obtained at the W.M. Keck Observatory, which is operated as a scientific partnership 
among the California Institute of Technology, the University of California and the National Aeronautics and Space Administration. 
The Observatory was made possible by the generous financial support of the W.M. Keck Foundation.}
\altaffiltext{**}{Based on observations obtained at the Southern Astrophysical Research (SOAR) telescope, which is a joint project of the Minist\'{e}rio da Ci\^{e}ncia, Tecnologia, e Inova\c{c}\~{a}o (MCTI) da Rep\'{u}blica Federativa do Brasil, the U.S. National Optical Astronomy Observatory (NOAO), the University of North Carolina at Chapel Hill (UNC), and Michigan State University (MSU).}
\altaffiltext{***}{Based in part on data collected at Subaru Telescope, which is operated by the National Astronomical Observatory of Japan.}
\altaffiltext{$\dagger$}{Based on observations collected at the European Organization for Astronomical Research in the Southern Hemisphere, Chile (ESO Program 090.A-9010(A)).}
\altaffiltext{$\ddagger$}{Based on observations obtained at the Canada-France-Hawaii Telescope (CFHT) which is operated by the National Research Council of Canada, the Institut National des Sciences de l'Univers of the Centre National de la Recherche Scientifique of France, and the University of Hawaii.}

\begin{abstract}

We present  optical and near-infrared adaptive optics (AO) imaging and spectroscopy 
of 13 ultracool ($>$M6) companions to late-type stars (K7--M4.5),
most of which have recently been identified as candidate members of nearby 
young moving groups (YMGs; 8-120~Myr) in the literature.
Three of these are new companions identified in our AO imaging survey and two
others are confirmed to be comoving with their host stars for the first time.
The inferred masses of the companions ($\sim$10--100~\Mjup)
are highly sensitive to the ages of the primary stars so we critically examine the kinematic and spectroscopic
properties of each system to distinguish bona fide YMG members from old field interlopers.
2MASS~J02155892--0929121~C is a new M7 substellar companion (40--60~\Mjup) 
with clear spectroscopic signs of low gravity and hence youth.
The primary, possibly a member of the $\sim$40~Myr Tuc-Hor moving group, is visually resolved into three components, making
it a young low-mass quadruple system in a compact ($\lesssim$100~AU) configuration.
In addition, \ion{Li}{1} $\lambda$6708 absorption in the intermediate-gravity M7.5 
companion 2MASS~J15594729+4403595~B provides unambiguous
evidence that it is young ($\lesssim$200~Myr) and resides below the hydrogen burning limit.
Three new close-separation ($<$1$''$) companions 
(2MASS~J06475229--2523304~B, PYC J11519+0731~B, and GJ~4378~Ab)
orbit stars previously reported as candidate YMG members, but instead
are likely old ($\gtrsim$1~Gyr) tidally-locked spectroscopic binaries without
convincing kinematic associations with any known moving group.
The high rate of false positives in the form of old active stars with YMG-like kinematics 
underscores the importance of radial velocity and parallax measurements 
to validate candidate young stars identified via proper motion and activity selection alone.
Finally, we spectroscopically confirm the cool temperature and
substellar nature of HD~23514~B, a recently discovered M8 benchmark brown dwarf orbiting 
the dustiest-known member of the Pleiades.

\end{abstract}

\section{Introduction}{\label{sec:intro}}

With masses between $\approx$13--75~$M_\mathrm{Jup}$, brown dwarfs directly link the lowest-mass stars 
and young giant planets in mass, radius, and temperature.   Nearly two thousand isolated brown dwarfs have been 
identified over the past 20 years, but their individual ages, metallicities, and masses are largely unknown.
Age benchmarks --- brown dwarfs orbiting well-characterized stars or members of coeval clusters --- are 
rare but valuable tools to test low-temperature atmospheric and evolutionary models for mutual consistency
and empirically calibrate isolated field objects because their ages, distances, and
compositions can be leveraged from the host star or cluster (e.g., \citealt{Naud:2014jx}; \citealt{Crepp:2014ce}; \citealt{Deacon:2014ey}).  

Over the past decade, dozens of nearby brown dwarfs with optical and near-infrared (NIR) spectroscopic 
signs of low surface gravity (youth) have been identified (e.g., \citealt{Rebolo:1995tr}; \citealt{Kirkpatrick:2008ec}; \citealt{Allers:2013hk}).  
Many of these objects have proper motions consistent with nearby ($<$80~pc) young moving groups (YMGs)
with ages of $\approx$10--120~Myr (e.g., \citealt{Scholz:2005fj}; \citealt{Looper:2007gc}; 
\citealt{Delorme:2012gv}; \citealt{Faherty:2012bc}; \citealt{Gagne:2014gp}) but because they are intrinsically faint, most of these
isolated brown dwarfs lack trigonometric distances and radial velocities (RVs) needed to unambiguously 
define their space motion.
As a result, only a handful of young field brown dwarfs have been kinematically linked 
with high confidence to YMGs like TWA (\citealt{Gizis:2007do}; \citealt{Weinberger:2013gk}), 
$\beta$~Pic (\citealt{Liu:2013gya}), and AB Dor (\citealt{Faherty:2012bc}; \citealt{Liu:2013ej}).
These low-gravity benchmarks are especially useful for empirical spectral classification 
sequences to map the effects of age on gravity-sensitive features like alkali absorption lines, $H$-band 
morphology, and H$_2$O steam bands at $\approx$1.4 and $\approx$1.9~$\mu$m spanning a large range
of temperatures (500~K--2500~K) and corresponding masses ($\approx$5--75~\Mjup).  
These empirically-based spectral sequences at fixed ages also enable detailed analyses of secondary effects like 
photospheric clouds and non-equilibrium chemistry, which become increasingly important near the L/T transition for young
giant planets (\citealt{Bowler:2010ft}; \citealt{Barman:2011fe}; \citealt{Marley:2012fo}; \citealt{Skemer:2014hy}).

Another approach to find these elusive young substellar benchmarks is to search for faint companions to known 
nearby young stars.
One advantage of this method is that precise parallaxes and RVs are more readily obtained for the 
primary stars, making kinematic association with YMGs relatively straightforward compared to faint isolated
brown dwarfs in the field.
Indeed, many substellar companions to YMG members have been discovered 
this way, primarily through space-based imaging (\citealt{Lowrance:1999ck}; \citealt{Webb:1999kf}; \citealt{Lowrance:2000ic})
and adaptive optics (AO) imaging from the ground (\citealt{Chauvin:2005dh}; 
\citealt{Biller:2010ku}; \citealt{Mugrauer:2010cp}; \citealt{Wahhaj:2011by}; \citealt{Bowler:2012cs}; 
\citealt{Delorme:2013bo}; \citealt{Bowler:2013ek}).
However, old interloping stars with $UVW$ space motions similar to YMGs can 
complicate efforts to identify young substellar companions in moving groups.
Distinguishing bona fide members from kinematic interlopers can be particularly challenging for M-type members of 
YMGs since both single mid- to late-M dwarfs and 
tidally locked M+M binaries can have short rotation periods (less than a few days) and strong magnetic dynamos 
even at old ($\gg$1~Gyr) ages.

Here we present a suite of spectroscopic and adaptive optics imaging observations of ultracool companions to 
M dwarfs.  The majority of these companions have been identified as part of the 
Planets Around Low-Mass Stars (PALMS) high-contrast imaging search for giant planets and brown dwarfs,
three of which are new here (2MASS~J06475229--2523304~B, PYC J11519+0731~B, and GJ~4378~Ab).  
Several additional systems were identified by the Astralux Lucky imaging survey (\citealt{Janson:2012dc}),
which obtained shallow diffraction-limited imaging of nearby, mostly active M dwarfs in the optical. 
The sky positions, proper motions, spectral types, distances, and ages of our targets are listed in Table~\ref{tab:targets}.
The physical properties of the companions can be found in Table~\ref{tab:comp}.
Most of the primary stars have been previously reported in
the literature as candidate or confirmed members of YMGs, making the companions excellent 
candidates for young benchmark brown dwarfs.  

We critically examine these moving group membership claims on a case-by-case basis using new 
RVs of the primary stars together with recently published parallaxes, when available.  
We also use our new spectroscopy of the companions to
independently constrain the ages of the systems via low-gravity spectroscopic features, which can be
more prominent and last longer ($\lesssim$200~Myr) at late-M and L spectral types than at earlier types.
Finally, we search for signs of orbital motion with multi-epoch high-resolution 
AO imaging to identify systems that will yield dynamical masses in the near future.

\section{Observations}{\label{sec:obs}}

\subsection{Adaptive Optics Imaging}

\subsubsection{Keck/NIRC2}

Our near-infrared imaging with NIRC2 at the 10-meter Keck~II telescope was carried out between 
2012 and 2014 using Natural Guide Star Adaptive Optics (NGSAO; \citealt{Wizinowich:2000hl}).
We utilized the narrow NIRC2 camera mode, resulting in a plate scale of 9.952~$\pm$~0.002~mas~pix$^{-1}$ 
(\citealt{Yelda:2010ig}) and a field of view of 10$\farcs$2~$\times$~10$\farcs$2.
The NIRC2 $J$, $H$, $K$, $K_S$, and $L'$ filters are on the Maunakea Observatory (MKO) filter system 
(\citealt{Simons:2002hh}; \citealt{Tokunaga:2005ch}).
The NIRC2 $Y$-band filter was installed in 2011 and has a central wavelength of 1.0180~$\mu$m with 
a bandwidth of 0.0996~$\mu$m, differing slightly from other $Y$-band filters (\citealt{Hillenbrand:2002ba};
see \citealt{Liu:2012cy} for additional details).
In addition, we also include a single epoch of $K_S$-band NGSAO~imaging 
of 1RXS~J034231.8+121622 from 2007 in our analysis of the system (Section~\ref{sec:rxs0342}).

Raw images were cleaned of cosmic rays and bad pixels, dark-subtracted, flat-fielded, and north-aligned
using the orientation of +0$\fdg$252~$\pm$~0$\fdg$009 from \citet{Yelda:2010ig}.
For our $L'$ data, the science frames were combined after masking the target to create a flat field.  
For close companions ($\lesssim$2$''$), astrometry and relative photometry are measured by fitting
three bivariate Gaussians to each binary component as described in \citet{Liu:2008ib}.  Aperture photometry
is used for companions widely-separated from their host stars.
The standard deviations from multiple measurements are adopted for our separation and position angle (P.A.)
uncertainties.

For PYC~J11519+0731 we also obtained deep imaging with the 600-mas diameter
coronagraph in angular differential imaging mode 
(\citealt{Liu:2004kk}; \citealt{Marois:2006df}) in $H$ band on 2012 May 22 UT.  We acquired 20 frames
each with an integration time of 30~sec spanning a total sky rotation of 11$^{\circ}$.  
Following image registration based on the star's position behind the partly opaque mask
and corrections for optical distortions (B. Cameron, 2007, priv. comm.), 
we performed PSF subtraction with the Locally-Optimized Combination of Images algorithm (LOCI; \citealt{Lafreniere:2007bg}) 
as described in \citet{Bowler:2015ja}.  However, because of the small sky rotation,
we also used a reference PSF library containing over 2000 NIRC2 coronagraphic images 
from our PALMS program.  At each geometric subsection the LOCI coefficients were computed from 100~library images,
which were chosen by minimizing the rms after scaling and subtracting every PSF reference image 
within that particular section.  For our final reduction we masked the region surrounding 
PYC~J11519+0731~B  to avoid self-subtraction when computing image coefficients.  
Coronagraph throughput measurements from \citet{Bowler:2015ja} were used to produce a
7~$\sigma$ contrast curve.  Results are described in Section~\ref{sec:pyc11519}.

\subsubsection{Subaru/IRCS}

We imaged 2MASS~J02155892--0929121, 2MASS~J06475229--2523304, and GJ 4378 A with the Infrared Camera and Spectrograph
(IRCS; \citealt{Tokunaga:1998wh}; \citealt{Kobayashi:2000uh}) coupled with the AO188 NGSAO system (\citealt{Hayano:2010kj}) 
at the 8.2-meter Subaru Telescope on UT 2012 October 2012 12 and 13.
The weather was clear throughout both nights but the seeing was highly variable (1--2$''$).
The A-B-C components of 2MASS~J02155892--0929121 were easily identified, but the (then unknown) fourth member,
Ab, was not resolved in these data.  We obtained MKO $K$-band images of 2MASS~J02155892--0929121 ABC and 
MKO $J$, $H$, and $K$ images of 2MASS~J06475229--2523304~AB and GJ 4378 Aab.  

The raw frames were cleaned of bad pixels, flat-fielded, and dark-subtracted.  Astrometry and
relative photometry were measured for each image individually by fitting three bivariate Gaussians to
each component (\citealt{Liu:2008ib}).
We adopt a plate scale and North orientation of 20.41~$\pm$~0.05 mas~pix$^{-1}$ and 
+89$\fdg$03~$\pm$~0$\fdg$09  measured in \citet{Bowler:2013ek} using the 2$\farcs$4 binary 
1RXS~J235133.3+312720~AB (\citealt{Bowler:2012cs}), which we observed with IRCS on UT~2012~October~12.

\subsubsection{Gemini/NICI}

On UT 2010 August 29 we targeted 1RXS~J034231.8+121622 with the Near-Infrared Coronagraphic Imager 
(NICI; \citealt{Chun:2008et}) on the 8.1-meter Gemini South telescope in queue mode 
as part of the NICI Planet-Finding Campaign (\citealt{Liu:2010hn}; Program GS-2010B-Q-500, PI: M. Liu).  
NICI is a simultaneous dual-channel imager with a tapered, partly opaque Lyot coronagraph and two 
1024$\times$1024 (18$''$$\times$18$''$ field of view) infrared arrays.  
A short imaging sequence of eight 60 s frames was acquired for 1RXS~J034231.8+121622 in $H$ 
and $CH_4 long$ (4\% throughput centered at 1.652~$\mu$m) bandpasses.
Ultimately this target it was not included in the
final NICI Planet Finding Campaign because the ADI sequence was aborted after a few frames.
The 0$\farcs$8 companion 1RXS~J034231.8+121622~B originally found by \citet{Bowler:2015ja} 
is easily detected in individual images without PSF subtraction.  

PYC~J11519+0731 was targeted in a similar fashion with NICI on 2013 April 19 UT as part
of a separate survey to search for giant planets around low-mass stars (Program GS-2013A-Q-54, PI: E. Nielsen).
A total of 10 short (1.14 s per coadd $\times$ 10 coadds) frames were obtained simultaneously in $H$ and $K_S$ bands, 
resulting in a total on-source integration time of 114 s per filter.  PYC~J11519+0731~B was easily resolved and
remained unsaturated in these short images.

We reduced the NICI images following the description in \citet{Wahhaj:2011by}, which
includes basic image reduction (bad pixel removal and flat fielding) and distortion correction.
Plate scale and sky orientation values were used from calibration measurements by the Gemini staff 
closest in time before the observations were taken (\citealt{Hayward:2014dk}).  
Astrometry and relative photometry incorporating the mask transmission measurement and uncertainty from
\citet{Wahhaj:2011by} is listed in Table~\ref{tab:astrometry}.

\subsubsection{Palomar 60-Inch/Robo-AO}

2MASS~J15594729+4403595, 2MASS~J11240434+3808108, and G~180-11 were imaged with the 
visible-light Robo-AO robotic laser guide star adaptive optics system (\citealt{Baranec:2013ey}; \citealt{Baranec:2014jc}) 
mounted on the automated Palomar 60-inch (1.5-meter) telescope (\citealt{Cenko:2006im}) on the night of 2014 June 13 UT.
Robo-AO utilizes a pulsed ultraviolet laser to generate a Rayleigh-scattered guide star for wavefront sensing, enabling
efficient observations with AO correction in the optical down to $V$$\approx$16~mag.
The electron-multiplying CCD camera is continually read out during the observations and 
a shift-and-add pipeline creates a final science image (\citealt{Law:2014is}).  
The nominal detector field of view is 44$''$ with a plate scale 0$\farcs$0431 pixel$^{-1}$ (\citealt{Baranec:2014jc}).
The pipeline processing produces a final image that is resampled at half the original pixel scale (0$\farcs$0216 pixel$^{-1}$).

All three targets were imaged in Sloan $r'$, $i'$, and $z'$ filters (\citealt{York:2000gn}) with exposure times of 60~sec each.
The late-M companions to 2MASS~J15594729+4403595 and 2MASS~J11240434+3808108 are clearly visible 
in the $i'$ and $z'$ images.  G~180-11~B is detected in all three filters.  The North orientation and 
plate scale are derived using the 
$z'$ image of 2MASS~J15594729+4403595~AB and astrometry of the pair from \citet{Bowler:2015ja}.
Based on its photometric distance of 27~pc, this system has a projected separation of $\approx$150~AU. 
No orbital motion is expected as the observations used for astrometric calibration were taken in 2012.
We measure a plate scale of 0$\farcs$0219~$\pm$~0$\farcs$0002 pixel$^{-1}$ and sky orientation
on the detector of 335$\fdg$4~$\pm$~0$\fdg$4.  
These are within 2~$\sigma$ of the values derived using a distortion solution based on $HST$-calibrated
images of M15 (A. Tokovinin, 2015, priv. communication).
The FWHM of the primary stars in these observations
are $\sim$0$\farcs$28, $\sim$0$\farcs$20, and $\sim$0$\farcs$16 in $r'$, $i'$, and $z'$, respectively, 
in 1$\farcs$3--1$\farcs$8 seeing conditions.
Flux ratios for the components are measured with aperture photometry and listed in Table~\ref{tab:astrometry}.

\subsection{Near-Infrared Spectroscopy}

\subsubsection{Keck/OSIRIS-NGSAO}

Near-infrared spectroscopy of low-mass companions 
allows for spectral classification and provides an independent diagnostic of youth 
in the form of gravity-sensitive atomic and molecular features.
We obtained moderate-resolution ($R$$\equiv$$\delta \lambda$/$\lambda$=3800) 
near-infrared spectra of the close-in companions ($<$3$''$) PYC~J11519+0731~B, G~180-11~B, HD~23514~B, GJ~4378~Ab,
2MASS~J06475229--2523304~B, and 2MASS J08540240--3051366~B
with the OH-Suppressing Infrared Imaging Spectrograph (OSIRIS; \citealt{Larkin:2006jd}) at the 10-meter Keck~I telescope
between 2012 June and 2014 Dec (Table~\ref{tab:specobs}).
Our 2013 and 2014 observations utilized the newly installed, more efficient grating described in \citet{Mieda:2014dt}.   
We used the 20~mas~pixel$^{-1}$ plate scale (0$\farcs$32$\times$1$\farcs$28 field of view) 
for PYC~J11519+0731~B, 2MASS~J06475229--2523304~B, and 2MASS J08540240--3051366~B.
The 50~mas~pixel$^{-1}$ scale (0$\farcs$8$\times$3$\farcs$2 field of view)  was used for the other 
companions (see Table~\ref{tab:specobs}).
All observations were taken in ABBA nodded pairs for sky subtraction.
Immediately before or after our science observations we targeted single A0V standards
a similar airmass and sky position for telluric absorption measurements.

Basic 2D image reduction, pairwise sky subtraction, and rectification into 3D cubes was carried out with version 3.2 of the 
OSIRIS Data Reduction Pipeline maintained by Keck Observatory.  Rectification matrices derived closest
in time prior to the observations were used for the pipeline processing.  
The science and standard spectra were extracted using aperture photometry, normalized to a common level, then median combined.
Telluric correction was carried out with \texttt{xtellcor\_general}, part of the Spextool reduction
package for IRTF/SpeX (\citealt{Vacca:2003wi}; \citealt{Cushing:2004bq}).
Individual bandpasses were flux-calibrated with photometry from \citet{Rodriguez:2012ef}
for HD~23514~B and from Table~\ref{tab:astrometry} for the remaining targets.
The median S/N per pixel of our (unsmoothed) OSIRIS spectra range from 25--120.

\subsubsection{IRTF/SpeX}

We obtained moderate-resolution spectra of the wide-separation (3$''$--8$''$) companions 
2MASS J11240434+3808108~B, G~160-54~B, and 2MASS~J02155892--0929121~C
with the Infrared Telescope Facility's SpeX spectrograph in SXD mode on the nights
of UT 2011 January 22, UT 2012 September 24, and UT 2014 January 18, respectively.
The 0$\farcs$5 slit width was used for all three targets, yielding a resolving power
of $R$$\approx$1200.  We also observed the unresolved primary star PYC~J11519+0731~Aab
in SXD mode with a 0$\farcs$8 slit width ($R$$\approx$750) on UT 2012 June 1.  All science data were taken in 
a nodded ABBA pattern.  A0V standards, internal flats, and 
arc lamps, were obtained at a similar sky position to the science target.
Individual spectra were extracted, median-combined, and telluric-corrected with the Spextool 
reduction package (\citealt{Vacca:2003wi}; \citealt{Cushing:2004bq}).
The median S/N per pixel of our SpeX spectra range from 30--60.

\subsection{Optical Spectroscopy}

\subsubsection{Mayall/RC-Spectrograph}

We obtained low-resolution optical spectra of most of the primary stars in our sample to provide uniform spectral classification
and measure H$\alpha$ line strengths. 
We targeted PYC~J11519+0731~Aab with the Ritchey-Chretien Spectrograph (RC-Spec) 
using the T2KA CCD mounted on the NOAO Mayall 4-meter telescope on 2014 May 22 UT.  
The 1$\farcs$5$\times$98$''$ slit was fixed in the North-South direction throughout
the night so we observed PYC~J11519+0731 at a low airmass (sec~$z$ = 1.1) near transit to minimize slit losses from
differential atmospheric refraction (\citealt{Filippenko:1982tl}).  
We used the 600 l~mm$^{-1}$ BL420 grating in first order which provided a resolving power of $R$$\approx$2600
spanning 6300--9200~\AA.
The raw 240~s image was corrected for bad pixels and cosmic rays, bias-subtracted, and flat-fielded.
We then extracted the spectrum by summing in the spatial direction and corrected for throughput losses
using our observations of the spectrophotometric standard HZ~44.

\subsubsection{SOAR/Goodman Spectrograph}

On 2014 June 28 UT we obtained low resolution optical spectra of the (unresolved) primary stars
2MASS~J01225093--2439505, 2MASS~J02155892--0929121, GJ~4378~A, 
and GJ~4379~B with the Goodman High Throughput Spectrograph at the
4.1-meter Southern Astrophysical Research (SOAR) Telescope located at
Cerro Pach\'on, Chile.  All targets were observed at low airmass (sec~$z$ $<$ 1.2)
with clear sky conditions.
The 0$\farcs$46 long slit was used with a CCD read rate of 400~kHZ and the
GG455 order blocking filter.
We acquired low-resolution ($R$$\sim$1800) spectra of all four targets 
with the 400~l/mm M2 grating spanning $\sim$5000--9000~\AA.
We also observed 2MASS~J02155892--0929121, GJ~4378~A, and GJ~4379~B with
a moderate-resolution setup ($R$$\sim$5900) using the 
1200~l/mm M5 grating spanning $\sim$6300--7500~\AA.
Integration times were between 240--300~sec for all of our observations
and the slit was rotated to the parallactic angle to avoid slit losses from chromatic dispersion.
The Goodman spectrograph suffers from strong fringing redward  
of $\approx$7000~\AA, which can change from target to target from
mechanical flexure.  To calibrate and correct these systematics we obtained 
arc lamp spectra for wavelength calibration and quartz flats after each target.  
For the 400~l/mm grating we used HgAr lamps and for the 1200~l/mm grating
we used CuHeAr lamps.  

After basic image reduction entailing bad pixel correction, cosmic ray removal, bias subtraction, and flat fielding, 
we removed night sky lines in the 2D image using the chip regions adjacent to the science spectrum in the spatial direction.
The spectrum was then horizontally rectified and optimally extracted using the 
method from \citet{Horne:1986bg}.  For each target, fringing was corrected using the same chip region
occupied by the science spectrum on the flat field.  By measuring the effects of fringing in real time at each telescope position over the
same pixels covered by the science data, we effectively divided out most of these artifacts in our target spectra.
Finally, we corrected for throughput 
losses with the standard star LTT~6248 (\citealt{Hamuy:1992ij}; \citealt{Hamuy:1994bn}),
which was observed with both grating settings.

\subsubsection{UH 2.2-meter/SNIFS}

We obtained low-resolution optical spectra of the host stars G~160-54, 2MASS~J15594729+4403595, and 2MASS~J06475229--2523304
with the SuperNova Integral Field Spectrograph (SNIFS; \citealt{Lantz:2004dk}) at the University of Hawai'i 2.2-meter telescope 
on the nights of UT 2012 Jan 29, UT 2012 Aug 23, and UT 2014 Dec 3.  SNIFS is an optical integral field spectrograph 
providing low-resolution ($R$$\approx$1300) spectroscopy spanning 3200--11000~\AA.
Several B and A-type stars and white dwarfs were targeted throughout the nights for use as 
spectrophotometric standards (Table~\ref{tab:specobs}).
The short and long-wavelength spectra were reduced, rectified into spectral cubes, and extracted
in real time with the SNIFS data reduction pipeline (\citealt{Aldering:2006fy}; \citealt{Scalzo:2010ec}).
This entailed flat fielding, bias subtraction, wavelength calibration using arc lamps, and spectral
extraction using a PSF model.  Additional flux calibration and telluric correction is described in 
\citet{Mann:2013fv}.  To summarize, we used spectrophotometric standards in conjunction with a model of the 
atmospheric transparency above Maunakea from \citet{Buton:2012gv} to derive airmass and time dependent 
corrections to the spectrum. The resulting flux calibration has been shown to be accurate to 1-2\% when compared 
to space-based spectra (\citealt{Mann:2013fv}).

\subsubsection{Keck/HIRES}

We obtained high-resolution spectra for most of the primary stars in our sample
to measure RVs, search for \ion{Li}{1} $\lambda$6708 absorption, and check for spectroscopic binarity.
Fully convective low-mass stars and high-mass brown dwarfs burn lithium on characteristic timescales that 
depend on their mass, so the presence or absence of the lithium feature provides an independent age 
constraint for the system (e.g., \citealt{Chabrier:1996kc}).
We targeted 2MASS J02155892--0929121 A,  
2MASS J02594789--0913109 A, and PYC J11519+0731 Aab with the High Resolution 
\'Echelle Spectrometer (HIRES; \citealt{Vogt:1994tb}) on the Keck I 10-m telescope. We used the 
0.861$\arcsec$ slit with HIRES to give a spectral resolution of $\lambda$/$\Delta\lambda$$\approx$58000. 
The detector consists of a mosaic of three 2048 $\times$ 4096 15-$\micron$ pixel CCDs, 
corresponding to a blue, green and red chip spanning 4900 -- 9300 \AA. We used the $GG475$ 
filter with the red cross-disperser to maximize the throughput near the peak of a M dwarf spectral 
energy distribution. 
	
Standard reduction was performed on each stellar exposure, i.e. bias-subtraction, flat-fielding, and the 
correction of bad pixels.  After optimal extraction, the 1-D spectra were wavelength calibrated with a 
Th/Ar arc taken within an hour of the stellar exposure. The resulting spectra reached S/N of  20 -- 50 
per pixel at 7000 \AA, depending on brightness. Each night, spectra were also taken of the RV standard GJ 908
and an A0V standard star for telluric line correction.

\subsubsection{CFHT/ESPaDOnS}

We acquired high-resolution \'echelle optical spectra of G160-54 A,  2MASS~J15594729+4403595 A, and PYC J11519+0731 Aab
using the \'Echelle SpectroPolarimetric Device for the Observation of Stars (ESPaDOnS; \citealt{Donati:2006vj}) 
on the Canada-France-Hawaii 3.6-m telescope.
ESPaDOnS is fiber fed from the Cassegrain focus to the Coud\'e focus. The fiber image is projected onto a Bowen-Walraven 
slicer at the  entrance to the spectrograph. We used the ``star+sky'' mode, which records the full spectrum over 
40 grating orders onto a 2048$\times$4608-pixel CCD detector covering 3700 to 10400 \AA\/ with a spectral resolution $\approx$68000. 
The data were reduced using {\it Libre Esprit}, a fully automated reduction package provided for the instrument and described in 
detail by \cite{Donati:1997wj,Donati:2006vj}.  RV standards used for these targets were GJ 273, GJ 628 and GJ625.

\subsubsection{Ir\'{e}nee du Pont/Echelle Spectrograph}

Four of our targets were observed with the optical echelle spectrograph mounted on the 2.5-m du Pont telescope at Las 
Campanas Observatory\footnote{The instrument manual can be found here: 
http://www.lco.cl/telescopes-information/irenee-du-pont/instruments/website/echelle-spectrograph-manuals/echelle-spectrograph-users-manual.}:  
2MASS~J06475229--2523304 A,  2MASS J08540240--3051366 A,  GJ 4378 A, and  GJ 4379 B. 
The TEK\#5 CCD with a 2k by 2k format and 24 micron pixels was employed for observations prior to December 2009.  
At this time, the SITe2K CCD was installed with the same format. In order to reach a spectral resolution of 45000, 
we used the 0.75\arcsec by 4\arcsec\ slit, and acquired simultaneous wavelength coverage from $\sim$3700 to 8500 \AA. 
We obtained a ThAr exposure at every pointing to avoid any flexure effects and day-time ``milky'' flats correct for pixel-pixel variations. 
In addition to a hot standard star for telluric line correction, RV standards GJ 273, GJ 433, GJ 908, and GJ 699 were also observed. 

Excluding our HIRES spectrum of PYC J11519+0731~A, which is detailed in Section~\ref{sec:pyc11519}, 
out RV measurements for the HIRES, ESPaDOnS, and du Pont telescopes are carried out in a similar fashion.
After correcting for the heliocentric velocity, we cross-correlated each spectral order between 7000 and 9000 \AA \ 
of each stellar spectrum with an RV standard of similar spectral type using IRAF's\footnote{IRAF (Image Reduction and 
Analysis Facility) is distributed by the National Optical Astronomy Observatories, which is operated by the Association 
of Universities for Research in Astronomy, Inc.~(AURA) under cooperative agreement with the National Science 
Foundation.} \texttt{fxcor} routine \citet{Fitzpatrick:1993um}. We excluded the Ca II infrared triplet (IRT) (the target stars 
exhibit Ca II emission that is not present in the RV standards) and regions of strong telluric absorption in the cross-correlation.  
The resulting cross-correlation function also allowed to quickly identify double- or triple-lined spectroscopic binaries: 
G160-54~A,  GJ 4378 A, and PYC J11519+0731 A.
In nearly all cases we are able to measure RVs to better than 1 km~s$^{-1}$.

\subsubsection{ESO-MPG 2.2-meter/FEROS}

As part of the ongoing CASTOFFS survey to identify young, low-mass stars in the solar neighborhood (Schlieder et al. 2015, in prep), 
PYC J11519+0731 Aab was observed with the Fiberfed Extended Range Optical Spectrograph (FEROS; \citealt{Kaufer:1999vj}) 
on the ESO-MPG 2.2-m telescope located at La Silla, Chile, on UT 22 February 2013.
FEROS is a fiber-fed optical \'{e}chelle spectrograph providing a resolution R$\approx$48000 from $\approx$3500--9200~\AA \ across 39 orders. 
We acquired three exposures each with integration times of 406~s at an airmass of 1.34 which resulted in a S/N per pixel of $>$50 in the red
orders of the median-combined spectrum.
FEROS contains two 2$\farcs$0 optical fibers separated by 2$\farcm$9 on sky. We observed in `object + sky' mode with one fiber on the science target 
and the other on sky. Data reduction was performed using the FEROS Data Reduction System (DRS) implemented within the ESO-MIDAS software package.  
The DRS performs flat-fielding, background subtraction, bad pixel removal, order extraction (using optimal extraction), and wavelength 
calibration from ThAr lamp lines.  The DRS then computes barycentric velocity corrections\footnote{\citet{Muller:2013kh} show that the barycentric correction 
computed by the FEROS-DRS is only accurate to $\sim$100 m/s because it does not take into account coordinate precession. This small systematic error 
is not significant for our analysis, which achieves an RV precision of $\sim$1~km/s.} and rebins the individual orders to the 
same wavelength scale to merge them and produce a continuous spectrum. The calibrations used by the DRS (bias, flats, and a ThAr lamp spectrum) 
are acquired in the afternoon prior to the observations. 
We also observed late-type RV templates drawn from \citet{Prato:2002hp} during observing runs in December 2011 and October 2012 
for use during cross-correlation analysis (Section~\ref{sec:pyc11519}). Each night of FEROS observations also includes observations of at least one RV standard 
drawn from the lists of \citet{Chubak:2012tv} or \citet{Nidever:2002vx} for internal calibration checks. The RV templates and standards were reduced in 
the same way as the science target.  

\subsubsection{CTIO-SMARTS 1.5-meter/CHIRON}

We obtained high-resolution ($R$$\approx$28000) optical spectra of 2MASS J02155892--0929121 A 
and 2MASS J06475229--2523304 A on the nights of UT 2014 December 08 and UT 2014 December 10,
respectively, with the CHIRON echelle spectrograph (\citealt{Tokovinin:2013hx}) at the CTIO 1.5-meter telescope operated by the SMARTS Consortium
at Cerro Tololo, Chile.  CHIRON is an extremely stable fiber-fed instrument built for high precision ($\approx$1 m/s) 
RV exoplanet searches.  Each spectrum covers the 4500--8900~\AA \ spanning 62 orders.
All observations are carried out in queue mode (\citealt{Brewer:2014jt}) and are reduced in an automated fashion
that includes bias subtraction, flat fielding, optimal spectral extraction, and wavelength
calibration of each order using a ThAr exposure acquired after each target.

We measure RVs for both targets by cross-correlating the science spectra with RV standards from \citet{Chubak:2012tv} 
that have similar spectral types and were also observed using the same setup with CHIRON.  Each order is first 
normalized and the science and standard spectra are then linearly interpolated onto a common wavelength grid.  
After applying a barycentric velocity correction to each spectrum, each order is cross correlated separately 
and the RV offset was determined by fitting a Gaussian to the cross correlation function.  The final RVs (Table~\ref{tab:rvs}) 
are the mean from 10 orders spanning 6120--6860~\AA, which were chosen to avoid telluric absorption features.  
Our adopted uncertainties incorporate both random errors (the standard deviation from the 10 orders) and an estimated 
systematic error of 0.5~km~s$^{-1}$ (measured by cross-correlating RV standards with each other) added in quadrature.

\subsubsection{Keck/ESI}

The wide separation of 2MASS~J15594729+4403595 B from its host star (5$\farcs$6)
makes it amenable to spectroscopy without the need for AO correction.
We observed 2MASS~J15594729+4403595 B using the Echellette Spectrograph and
Imager (ESI; \citealt{Sheinis:2002ft}) on Keck~II on UT 2012 September 9 under
near-photometric conditions. We used the 1$\farcs$0 slit,
which yielded a spectral resolution of $R\approx4000$ and wavelength coverage from
3000--10000 \AA. The slit was aligned perpendicular to the
star-companion position angle to avoid contamination from the primary.
We obtained three separate exposures, each 800 s, which yielded a S/N per pixel of
$\approx60$ redward of 7500~\AA. Observations of the G191B2B
and Feige~67 spectrophotometric standards were taken throughout the night.

We reduced the data using the ESIRedux package
(\citealt{Prochaska:2003wl}; \citealt{Bochanski:2009jt})\footnote{http://www2.keck.hawaii.edu/inst/esi/ESIRedux/index.html}.
ESIRedux performs bias, flat, and dark correction as well as
extraction of the two-dimensional data into a one-dimensional
spectrum. ESIRedux wavelength-calibrates the data using HgNe and CuAr
arcs (obtained at the start of the night) to construct a two-dimensional wavelength map. 
The orders are combined into a single spectrum and an approximate flux
calibration is applied (based on archived calibration data), which we improve upon
using the observed spectrophotometric standards and a model of the
atmosphere above Maunakea (\citealt{Buton:2012gv}; \citealt{Mann:2013fv}).

\section{Targets}

\subsection{2MASS J01225093--2439505 AB}

2MASS J01225093--2439505~B is a young mid-L dwarf companion identified by \citet{Bowler:2013ek} during the 
PALMS high-contrast imaging planet search.
Located at 1$\farcs$5 ($\approx$52~AU) from the young active mid-M dwarf 2MASS J01225093--2439505~A, 2MASS J01225093--2439505~B 
possesses red near-IR colors and an angular $H$-band spectral morphology, hallmarks of a dusty photosphere and
low surface gravity, respectively.  Our low-resolution optical spectrum of the primary from the SOAR/Goodman Spectrograph is shown in Figure~\ref{fig:pri_optspec}.
We find a spectral type of M4.0~$\pm$~0.5 from visual and index-based classification with the IDL-based classification package
 \texttt{Hammer} (Table~\ref{tab:optspec}; \citealt{Covey:2007bj}),
which is slightly later than the M3.5V classification from \citet{Riaz:2006du}.  
\citet{Malo:2013gn} found  the system probably belongs to the $\approx$120~Myr AB Dor moving group, which was bolstered by
Bowler et al. based on an RV measurement (9.6~$\pm$~0.7 km s$^{-1}$) and the 
photometric distance (36~$\pm$4~pc).  At this age, the companion 
has a luminosity that corresponds to both $\approx$13 and $\approx$25~\Mjup \ masses based on substellar evolutionary models,
which overlap in this region because of delayed onset of deuterium burning at lower masses 
(\citealt{Bowler:2013ek}).  More recently, \citet{Gagne:2014gp} suggested this system may belong to the 15--25~Myr $\beta$~Pic 
moving group using the RV of 11.4~$\pm$~0.2 km s$^{-1}$ from \citet{Malo:2014dk}, in which case the companion 
mass is as low as $\approx$6~\Mjup.  
A parallactic distance is needed to unambiguously distinguish these cases.
Note that the RVs from Bowler et al. and Malo et al. differ by 2.5~$\sigma$, indicating the primary could
be a single-line spectroscopic binary (SB1).  The Radial Velocity Experiment (RAVE; \citealt{Kordopatis:2013cd}) found a slightly larger 
higher RV of 12.3~$\pm$~2.5 km s$^{-1}$, which is formally consistent with the other two measurements. 

On UT 2013 August 17 we obtained the first $Y$-band images of 2MASS J01225093--2439505~B with NIRC2 (Figure~\ref{fig:twom0122y}).
Our astrometry is consistent with measurements between 2012--2013 from Bowler et al.  Based on the typical $Y$--$J$ color
of an M4.0V star from \citet{Rayner:2009ki}, our $Y$-band contrast of 7.6~$\pm$~0.3~mag implies an apparent magnitude
of $Y$~=~18.2~$\pm$~0.3 mag (Table~\ref{tab:astrometry}), an absolute magnitude of $M_Y$=15.4~$\pm$~0.3~mag, 
and a $Y$--$J$ color of 1.5~$\pm$~0.3~mag.  The relationship between $M_Y$ and spectral type has not yet been mapped for 
young brown dwarfs and giant planets, but compared to old field objects from \citet{Dupuy:2012bp}, our new measurement
best corresponds to L5--L7 objects.  This is in good agreement with our classification of L4--L6 based on $H$- and $K$-band spectroscopy
from \citet{Bowler:2013ek}.

\subsection{2MASS J02155892--0929121 AabBC}  

\citet{Riaz:2006du} classify 2MASS~J02155892--0929121 an M2.5 star displaying H$\alpha$ 
emission ($EW$=--6.9~\AA) in their catalog of $ROSAT$-detected M dwarfs.
We find a slightly later type of M3.5 with our Goodman spectrum (Table~\ref{tab:optspec}; Figure~\ref{fig:pri_optspec})
but a similar H$\alpha$ line strength of --5.6~\AA.  
\citet{Bergfors:2010hm} resolved this object into three components in a single epoch of imaging in 2008 
as part of the AstraLux Lucky Imaging survey: a close stellar companion at 0$\farcs$6 (``B'') and a 
fainter component at 3$\farcs$5 with red optical colors (``C''), which
\citet{Janson:2012dc} estimated has a spectral type of M8.  

Based on its activity and proper motion, \citet{Malo:2013gn} suggest this system is a possible member of 
the Columba, $\beta$~Pic, or Tuc-Hor moving groups.
\citet{Kraus:2014ur} also identify it as a candidate Tuc-Hor member, though their
measured RV (10.1~$\pm$~0.6~km~s$^{-1}$) disagrees with the expected value
assuming group membership by 5.2~km~s$^{-1}$.  Three additional epochs were presented
by \citet{Malo:2014dk}: 0.5~$\pm$~0.3~km~s$^{-1}$ and --0.6~$\pm$~0.3~km~s$^{-1}$ in 2010,
and 8.3~$\pm$~0.3~km~s$^{-1}$ in 2012.  A consistent but less precise RV of 0.6~$\pm$~5 km~s$^{-1}$
was measured by the RAVE survey in 2007 (\citealt{Kordopatis:2013cd}).  
These RV variations indicate the primary is an SB1 (Table~\ref{tab:rvs}).

We measure RVs of 6.2~$\pm$~0.4~km~s$^{-1}$ and 8.3~$\pm$~0.6~km~s$^{-1}$ for 2MASS~J02155892--0929121 
from spectra obtained in 2012 December and 2014 December with Keck/HIRES and SMARTS 1.5-m/CHIRON.   
Our Keck/HIRES measurement agrees well with the predicted RV of 4.5~$\pm$~1.4~km~s$^{-1}$ 
for Tuc-Hor membership from \citet{Malo:2013gn}.  Based on this RV, the web-based Bayesian tool BANYAN~II
to compute YMG membership probabilities\footnote{http://www.astro.umontreal.ca/$\sim$gagne/banyanII.php} 
by \citet{Gagne:2014gp} returns a 99.5\% likelihood that this system belongs to the Tuc-Hor moving group.  
The weighted mean RV (4.0~$\pm$~0.15 km s$^{-1}$) is also consistent with the predicted value for Tuc-Hor.
However, the changing RVs spanning 2010 to 2014 mean that
more spectroscopic monitoring will be needed to derive a systemic RV  and
unambiguously determine its group membership.

We imaged 2MASS J02155892--0929121 three times between 2012--2013 with Subaru/IRCS and Keck/NIRC2
(Figure~\ref{fig:twom0215abcnirc2}).  
We confirm the B and C components are comoving with the primary (Figure~\ref{fig:twom0215back})
and detect orbital motion between our epochs and those in 2008 from \citet{Bergfors:2010hm}.
For our August 2013 data with NIRC2, only the central 198$\times$248~pix$^2$ (1$\farcs$97$\times$2$\farcs$47) 
region of the detector encompassing the AB components was read out so we do not have astrometry 
of the C component at that epoch.

\subsubsection{Resolved AO Imaging of 2MASS J02155892--0929121 AabB}

From visual inspection of our NIRC2 images in August 2012 and August 2013 
we found that the A component appears slightly elongated compared to B at both
epochs, but in the opposite directions (SE in 2012 and NE in 2013).  
The separation between A and B is close enough (0$\farcs$6) that 
isoplanatic effects should be small, so any differences in the PSF shape are probably caused by
a marginally-resolved close-in companion.  

Using a 0$\farcs$4$\times$0$\farcs$4 region surrounding 2MASS J02155892--0929121~B as
a reference PSF, we jointly fit two identical PSF templates to the A component with 
the Nelder-Mead (``Amoeba'') downhill-simplex minimization technique (\citealt{Nelder:1965tk}; 
\citealt{Press:2007vx}).  Seven parameters were allowed to vary: an $x$-position, $y$-position,
and multiplicative scale factor for each PSF as well as an overall additive amplitude offset.  
Each image was fit separately (20 in 2012 and 10 in 2013) and the resulting coadded frames 
are shown in Figure~\ref{fig:twom0215nirc2}.  In each case, subtracting the primary reveals
a clear image of a companion and its first Airy ring.  The joint fits provide excellent matches to the data 
as demonstrated  by the low RMS residuals of $\approx$8~DN (70~DN) compared to a peak flux of 
$\approx$1600~DN (14000~DN) in our 2012 (2013) data.
The median and standard deviation of the separations, position angles, and $K_S$-band flux ratios 
are 42~$\pm$~7~mas, 112~$\pm$~2$^{\circ}$, and 1.2~$\pm$~0.3~mag for our 2012 data set.
For our 2013 data we measure 42~$\pm$~7~mas, 308~$\pm$~5$^{\circ}$, and 
1.17~$\pm$~0.05~mag.

Resolving the very close binary 2MASS J02155892--0929121~Aab at the same separation but with a nearly 
180$^{\circ}$ change in P.A. suggests an orbital period of $\sim$2~years (or shorter aliases).  
We perform a simple test to verify whether the amplitude of the RV variations are plausibly 
caused by this companion.  Based on the empirically-derived relations in \citet{Boyajian:2012eu},
the M3.5 spectral type of the Aa component corresponds to an effective temperature of $\approx$3300~K.
At an age of 40~Myr, this implies a mass of $\approx$0.2~\Msun \ for the primary (Aa) and 
$\approx$0.09~\Msun \ for the companion  (Ab) using on the measured $K$-band flux ratio and 
the evolutionary models of \citet{Baraffe:1998ux}.  Assuming a semi-major axis of $\approx$1~AU, 
which roughly corresponds to the observed separation of the pair, a circular orbit, and 
an inclination of 90$^{\circ}$, the induced velocity semi-amplitude is $\approx$5.0 km s$^{-1}$.  This is in good agreement
with the maximum total RV amplitude change of $\approx$11 km s$^{-1}$ from random RV sampling over the past decade.

There is no parallactic distance for this system, but spectrophotometric distance estimates 
to 2MASS J02155892--0929121~C (see below) and the unresolved primary 
(\citealt{Kraus:2014ur}; \citealt{Riaz:2006du}) imply distances of $\approx$20--30~pc.  
These correspond to physical separations of $\approx$1~AU for Aab.
Fortunately, the nearby components 2MASS J02155892--0929121~B and C (see below) offer excellent opportunities to
measure individual dynamical masses of pre-main sequence M dwarfs
using absolute astrometry of the system as opposed to a total mass with relative astrometry.

\subsubsection{The Ultracool Companion 2MASS J02155892--0929121 C}

Our 0.8--2.5~$\mu$m spectrum of 2MASS~J02155892--0929121~C is shown in Figure~\ref{fig:twom0215spec}.
Overall the NIR SED agrees fairly well with the M7 field template.  Individual bandpasses are most similar
to M7--M8 objects.  
The H$_2$O, H$_2$O-1, and H$_2$O-2 gravity-insensitive index-based classifications from \cite{Allers:2013hk},
which were originally defined by \citet{Allers:2007ja} and \citet{Slesnick:2004jy}, give spectral types of M6.6~$\pm$~1.7,
M7.0~$\pm$~2.3, and M6.7~$\pm$~2.2 with a weighted mean of M6.7~$\pm$~1.2.
Altogether we adopt a final NIR spectral type of M7~$\pm$~1.  Of note are the shallow 
1.244~$\mu$m and 1.252~$\mu$m \ion{K}{1} and 1.2~$\mu$m FeH features in $J$ band, hallmark signatures 
of low surface gravity and youth ($\lesssim$200~Myr) in brown dwarfs (e.g., \citealt{McGovern:2004cc}). 
We confirm this quantitatively using the indices from \citet{Allers:2013hk}, which yield an
intermediate-gravity (INT-G) classification for 2MASS~J02155892--0929121~C (Table~\ref{tab:grav}). 
This is strong evidence independent of kinematics that the system is quite young--- likely $\approx$30--150~Myr--- implying that
2MASS J02155892--0929121 C is indeed substellar.

The $K$-band photometric distance to 2MASS~J02155892--0929121~C is 29~$\pm$~5~pc using the 
field relations from \citet{Dupuy:2012bp}.  Since the system is young, this distance is probably underestimated.
Nevertheless, adopting the $K$-band bolometric correction from \citet{Liu:2010cw} gives a 
luminosity of log~($L$/$L_*$)=--3.23~$\pm$~0.16~dex, implying a mass of 42~$\pm$~15~\Mjup \ based on 
cooling models from \citet{Burrows:1997jq}.  A larger distance of 40~pc corresponds to a luminosity
of --2.95~$\pm$~0.13~dex and a mass of 57~$\pm$~19~\Mjup.

\subsection{2MASS J02594789--0913109 AB}  

This late-M companion to the M4.0 star 2MASS~J02594789--0913109~A (\citealt{Reid:2007gl}) was identified by \citet{Janson:2012dc} 
from a single epoch of imaging in 2008.  
Aside from its large proper motion ($\mu_\mathrm{tot}$=0$\farcs$6 yr$^{-1}$), the primary is otherwise unremarkable.
It was not detected by $ROSAT$ or $GALEX$ and there are no signs the system is young.
We do not detect lithium in our high resolution Keck/HIRES spectrum and H$\alpha$ is in absorption ($EW$=0.34~\AA), implying
an old age of several Gyr (\citealt{West:2008eq}).
Based on our RV measurement of 4.9~$\pm$~0.5 km~s$^{-1}$, 2MASS~J02594789--0913109 is not consistent with any known 
YMGs at any distance. 
Our shallow imaging of the system with Keck demonstrates 
the pair's shared proper motion.  At its estimated distance of 36~$\pm$~13~pc, an angular separation of 0$\farcs$6 
corresponds to $\sim$22~AU.  \citet{Janson:2012dc} estimate a spectral type of M9.5 and a companion mass of $\sim$0.08~\Msun.
We find a similar value of 87~$\pm$~13~\Mjup \ using a $K$-band bolometric correction from \citet{Liu:2010cw},
a uniform age distribution from 4--10~Gyr, and the \citet{Burrows:1997jq} evolutionary models.

\subsection{1RXS~J034231.8+121622 AB (2MASS~J03423180+1216225 AB)}{\label{sec:rxs0342}}

1RXS~J034231.8+121622 is a young (60--300~Myr; \citealt{Shkolnik:2009dx}; \citealt{Shkolnik:2012cs}) M4.0 star 
at 23.9~$\pm$~1.1~pc (\citealt{Dittmann:2014cr}) with
H$\alpha$ emission and saturated X-ray luminosity (\citealt{Riaz:2006du}).  
\citet{Janson:2012dc} found a nearby (0$\farcs$8) candidate companion with red optical colors, which was confirmed
to be a comoving brown dwarf and classified as L0~$\pm$~1 by \citet{Bowler:2015ja} 
with near-infrared AO imaging and spectroscopy.  Based on the age and luminosity of the system,
Bowler et al. inferred a mass of 35~$\pm$~8~\Mjup \ for 1RXS~J034231.8+121622~B.

Here we present additional astrometry of 1RXS~J034231.8+121622~AB between 2007 and 2013 
with Keck/NIRC2 and Gemini-S/NICI.  Combined with published astrometry (Table~\ref{tab:rxs0342_ast}) 
there is clear orbital motion in the system.  Although the coverage is quite limited, 
only a subset of Keplerian orbits are consistent with the observed astrometry.  
Our first attempt to constrain the orbital parameters and total mass of the system 
with orbit fitting is described below.

The orbital elements are calculated by fitting the resolved astrometric observations to a Keplerian orbit 
using \texttt{emcee} (\citealt{ForemanMackey:2013io}), an implementation of the affine-invariant MCMC 
ensemble sampler of \citet{Goodman:2010eu}. Our model has 8 free parameters: $\sqrt{e} \cos \omega, 
\sqrt{e} \sin \omega$, the time of periapsis, $\log$(period), $\log$(total mass), inclination, the position angle
of the ascending node, 
and distance. For the distance, we apply a prior following the parallax measured by \citet{Dittmann:2014cr}.  
At each step, we forward model the orbits of the two components, calculating their relative projected positions 
on the sky as viewed from the Earth at the time of each astrometric observation and determining the likelihood 
of these orbital elements (Figure~\ref{fig:rxs0342orbit}). We find that even though observations cover only 
1.5\% of the orbit, we are able to constrain the orbital period to 12\% and the total mass to 18\%.

The median and 68\% confidence intervals of the posteriors are 411~$\pm$~51~years for the orbital period,
0.16~$\pm$~0.07 for the eccentricity, and 0.152~$\pm$~0.027~\Msun \ for the total mass (Figure~\ref{fig:rxs0342posteriors}).  
This is somewhat lower than the total mass predicted by evolutionary models.  Using an $H$-band 
bolometric correction from \citet{Casagrande:2008dt} and the parallactic distance to the system, we find a 
luminosity of log~$L$/$L_{\odot}$~=~--2.17~$\pm$~0.05~dex for the primary star.  This corresponds to masses
of 0.15, 0.20, 0.23, and 0.25 for ages of 60, 100, 200, and 300~Myr using the evolutionary models of \citet{Baraffe:1998ux}.
The only way the total mass is consistent with these models is for young ages ($\lesssim$100~Myr), though it is also
possible that systematic errors exist in the evolutionary models (\citet{Dupuy:2014iz}).  
Unfortunately, the relatively uncertain age of this system makes
it difficult to use as a robust test of models.

\subsection{HD 23514 AB (2MASS~J03463839+2255112)} 

HD~23514 is an F6 member of the Pleiades with an unusually high fractional infrared luminosity
of $\approx$2\%, indicating the presence of a large amount of collisionally-ground dust within $\sim$1~AU 
(\citealt{Rhee:2008tp}).  \citet{Rodriguez:2012ef} discovered a substellar companion to HD~23514
located at 2$\farcs$6 ($\approx$360~AU) with AO imaging. HD~23514~B was also independently 
found by \citet{Yamamoto:2013uq} as part of the Subaru SEEDS exoplanet imaging program.
As a companion to a well-established member of the Pleiades cluster, HD~23514~B is an excellent
intermediate-age (120~$\pm$~10 Myr) benchmark for comparative analysis with field objects of unknown ages.

Figure~\ref{fig:hd23514b} shows our $H$- and $K$- band (1.5--2.4~$\mu$m) spectra of HD~23514~B
taken with Keck/OSIRIS.  We have not made any correction for reddening since extinction to the Pleiades
is small ($E(B-V)$ $<$ 0.06~mag; \citealt{Taylor:2008kx}).
HD~23514~B exhibits typical features of late-M dwarfs including FeH absorption,
strong CO bands, and \ion{Na}{1} absorption.  Interestingly, the $H$-band shape is only slightly angular, 
implying that late-M dwarfs showing even more pronounced angular morphology are younger than 120~Myr.
This agrees with the conclusions by \citet{Allers:2013hk}, who found that intermediate- and very low-gravity
brown dwarfs have ages $\lesssim$200~Myr.
Compared to field templates, HD~23514~B best resembles M9--L0 objects
in $H$ band and M7--M8 objects in $K$ band.  We therefore adopt a NIR spectral type of M8~$\pm$~1.
In the future, $J$-band spectroscopy of HD~23514~B will be useful to better calibrate the age-dependency
of $J$-band alkali absorption line strengths.

\subsection{G~160-54 AabBC (2MASS J04134585--0509049 AabBC)} 

G~160-54 is a little-studied system with an (unresolved) spectral type of M4.5~$\pm$~0.5 (Figure~\ref{fig:pri_optspec}; Table~\ref{tab:optspec})
and weak H$\alpha$ emission ($EW$=--0.8~\AA \  from UH~2.2-m/SNIFS and $EW$=--0.54~\AA \  from CFHT/ESPaDOnS).
Although not detected by $ROSAT$, a UV counterpart appears in $GALEX$ ($FUV$ = 21.7~$\pm$~0.4~mag, 
$NUV$~=~20.79~$\pm$~0.17~mag) and was resolved into a close visual triple system by \citet{Bowler:2015ja}
with NIR AO imaging at Keck and Subaru.
G~160-54~B is a stellar companion 0$\farcs$2 ($\approx$4~AU) from the primary G~160-54~A and G~160-54~C is a 
fainter ($\Delta$$K$=2.8~mag) comoving tertiary located at 3$\farcs$3 ($\approx$70~AU.
No parallax or RV has been measured for this system, but we note that 
at a distance of $\approx$18.5~pc and RV of $\approx$13~km~s$^{-1}$ its $UVW$ kinematics (\{--9.9, --20.7, --0.9\} km s$^{-1}$)
would precisely match known members of the Tuc-Hor moving group.  At that age ($\sim$30--40~Myr), the outer companion
G~160-54~C would fall below the hydrogen burning limit.

The very weak H$\alpha$ emission, however, disagrees with established Tuc-Hor members, which without exception have
line strengths larger than --6~\AA \ at a spectral type of M4 (\citealt{Kraus:2014ur}).
Cross-correlating our high resolution spectrum of G~160-54 AB with RV standards 
yields three strong peaks, implying there is a \emph{fourth} 
component in the system since G~160-54~C is too distant and too faint in the optical to produce such a large cross-correlation peak.  
Since we only have a single epoch we cannot identify whether this fourth member 
orbits G~160-54 A or B, but for this work we assume it orbits the primary ``A'' component in a similar configuration
as 2MASS~J02155892--0929121.

Our SpeX SXD spectrum of G~160-54~C is shown in Figure~\ref{fig:g160sxd}.  Compared to field templates,
it best resembles the M7 dwarf across the entire 0.8--2.4~$\mu$m spectrum and among individual bandpasses.
There are no obvious signs of low surface gravity.  This is corroborated with index-based spectral and
gravity classifications from \citet{Allers:2013hk}, which yield a NIR spectral type of M7~$\pm$~0.5 and a field
gravity (FLD-G).  Altogether, despite the potentially coincidental kinematic overlap with Tuc-Hor, this system is certainly
older than 150~Myr and most likely much older ($>$1~Gyr).  At these old ages, all four components in this system
are low-mass stars.  Based on its distance (21~$\pm$~9~pc), age (1--10~Gyr), and $K_S$-band magnitude 
(12.24~$\pm$~0.06; \citet{Bowler:2015ja}), G~160-54~C has a mass of 85~$\pm$~16 using a $K$-band bolometric 
correction from \citet{Liu:2010cw} and the \citet{Burrows:1997jq} evolutionary models.
The probability it is substellar ($<$75~\Mjup) is only 7\%.

\subsection{2MASS J05464932--0757427 AB} 

The M3.0 primary star 2MASS J05464932--0757427~A is not particularly noteworthy as it was not detected
by $ROSAT$ nor $GALEX$.
\citet{Janson:2012dc} identified a moderate-contrast companion ($\Delta$$z'$=5.6~mag) at 2$\farcs$8 in two epochs of imaging
in 2008 and 2009.  We imaged the system in $K_S$ band with NIRC2 in 2013.  The known secondary was clearly detected
and there are no signs of additional companions in the system.  Janson et al. estimate a spectral type of $>$L0 for the
companion based on $z'$- and $i'$-band contrasts.  Our $K_S$-band contrast of 3.9~mag corresponds to an apparent 
magnitude of 13.7~mag and an absolute $K_S$-band magnitude of $\approx$10.4~mag at the estimated distance of
$\approx$45~pc from Janson et al.
Assuming a 20\% uncertainty in distance,  the implied mass of 2MASS J05464932--0757427~B is 
87~$\pm$~3~\Mjup \ for an age range of 1--10~Gyr using a bolometric correction from \citet{Liu:2010cw}
and evolutionary models from \citet{Burrows:1997jq}.

\subsection{2MASS~J06475229--2523304 AabB}

2MASS~J06475229--2523304 is an active late-K-type star first recognized by \citet{Riaz:2006du}, 
who found weak H$\alpha$ emission ($EW$=--1.1~\AA) and a saturated X-ray fractional luminosity
(log~$L_X$/$L_\mathrm{bol}$=--3.13 dex).  
There are various conflicting classifications in the literature.  \citet{Riaz:2006du} find a spectral type of
K7V from low-resolution optical spectroscopy, \citet{Torres:2006bw} find K1IIIe from high-resolution optical spectroscopy,
and \citet{Pickles:2010ks} infer K3III from photometric fits to synthetic spectra spanning optical to near-infrared wavelengths.
To better characterize this star we obtained a UH~2.2-m/SNIFS low-resolution optical spectrum spanning $\approx$3500--8500~\AA.  
Figure~\ref{fig:twom0647_optspec} compares our UH~2.2-m/SNIFS spectrum to late-type templates 
from \citet{Bochanski:2007it} and \citet{Mann:2013fv}.
2MASS~J06475229--2523304 broadly resembles K7/M0 dwarfs, and we find a consistent index-based spectral type of
K7 using \texttt{Hammer} (\citealt{Covey:2007bj}).  We therefore adopt a K7.0~$\pm$~0.5 
classification for 2MASS~J06475229--2523304.

The band depths in our UH~2.2-m/SNIFS spectrum do not match dwarf templates in 
Figure~\ref{fig:twom0647_optspec}, however, which is confirmed via spectral indices from \citet{Reid:1995kw}.
We measure values of 0.890, 0.950, and 0.952 for the TiO5, CaH2, and CaH3 indices, which sits above the
dwarf locus in CaH2 + CaH3 versus TiO5 diagram in a region occupied by giant stars (e.g., \citealt{Mann:2012ei}).
This points to a low surface gravity, indicating 2MASS~J06475229--2523304 is most likely a very young or 
evolved (subgiant or giant) star.
The near-infrared colors of 2MASS~J06475229--2523304 ($J$--$H$ = 0.59 $\pm$ 0.04~mag, $H$--$K_S$ = 0.21 $\pm$ 0.04~mag)
 are similar to both late-K/early-M dwarfs and giants and cannot be used to distinguish between these evolutionary stages; only at later 
 spectral types do their branches diverge in near-infrared color-color
 diagrams (e.g., \citealt{Bessell:1988uv}).  The reduced proper motion of 
 2MASS~J06475229--2523304 ($H_V$ $\equiv$ $V$ + 5log($\mu$) + 5, where $\mu$ is the total proper motion in arcseconds per year)
 is 10.3~mag, which is more consistent with field dwarf-like kinematic and photometric properties instead of distant giant stars with 
 high luminosities and low tangential velocities (e.g., \citealt{Lepine:2011gl}).  
 
Based on its activity and proper motion, \citet{Malo:2013gn} suggest 2MASS~J06475229--2523304 is a 
possible member of Columba or AB~Dor, but they note that neither
group assignment is compatible with the measured RV of --56.5~km~s$^{-1}$ from \citet{Torres:2006bw}.
\citet{Malo:2014dk} find an RV of --24.4~$\pm$~0.8 km~s$^{-1}$ measured in 2011 and 
the RAVE survey (\citealt{Kordopatis:2013cd}) find an RV of --67 $\pm$ 6 km~s$^{-1}$ measured in 2009, 
suggesting the primary is an SB1.  Indeed, Torres et al. also listed this target as a possible binary (``SB1?'').  
We confirm this with our own RVs of --2.1~$\pm$~0.9~km~s$^{-1}$ and --5.5~$\pm$~0.8~km~s$^{-1}$ obtained in 2013 and 2014.
Our du Pont spectrum from 2013 shows slight H$\alpha$ emission ($EW$=--0.5~\AA) with an asymmetric line profile;  
however, there are no signs of a companion in the cross-correllation function.
Hereinafter this spectroscopic binary (SB) is referred to as 2MASS~J06475229--2523304~Aab.
 
 There is no parallax measurement for this system yet.  
\citet{Riaz:2006du} find a $J$-band spectrophotometric distance of 45~$\pm$~17~pc and
\citet{Ammons:2006uh} estimate a distance of 94$^{+88}_{-35}$~pc from SED fits.
Assuming a Pleiades-like age, we infer an absolute $V$-band magnitude of 8.11~$\pm$~0.09~mag and a 
photometric distance of 36~$\pm$~2 pc using the relation from \citet{Bowler:2013ek}.  
If 2MASS~J06475229--2523304~Aab is younger than $\sim$120~Myr or if it is an evolved star 
then its distance will be farther.
Although we do not have a systemic RV, we note that no combination of RV between 
--50 to --2 km~s$^{-1}$ and distance between 10--100~pc is consistent with the $XYZ$ 
spatial positions of known moving groups.

 \subsubsection{The Comoving Companion 2MASS~J06475229--2523304 B}

In 2012 we resolved a close (1$\farcs$1), modest-contrast ($\Delta$$K$=5.5~mag) point source 
near 2MASS~J06475229--2523304 with Subaru/IRCS AO imaging (Figure~\ref{fig:twom0647imgs}).
Follow-up astrometry with Keck/NIRC2 in 2013 and 2014 confirm the candidate is comoving with 
the primary star (Figure~\ref{fig:back_twom0647}).  No orbital motion is detected.

 \citet{Torres:2006bw} cite 60~m\AA \ \ion{Li}{1} absorption from their high resolution spectrum of the primary star. 
Regardless of its kinematics, this would imply an age range of $\sim$40--150~Myr assuming the system 
is indeed young rather than evolved,
which raises the possibility that 2MASS~J06475229--2523304 B is a new young brown dwarf companion.
However, we do not detect any  \ion{Li}{1} absorption with our own high resolution du Pont/Echelle and SMARTS~1.5-m/CHIRON spectra, 
pointing to an age $\gtrsim$100~Myr and a companion mass of about 40, 85, and 100~\Mjup \ for distances of 
40, 70, and 100~pc, respectively.

In December 2014 we obtained a $K$-band spectrum of 2MASS~J06475229--2523304~B with Keck/OSIRIS NGSAO.  
The spectrum does not resemble a late-M/early-L spectral type
as expected for a young ($\lesssim$200~Myr), nearby (within $\sim$100~pc) system.  Instead, the best match is an M3
template with no obvious signs of low surface gravity (Figure~\ref{fig:twom0647b_osiris}).  This has forced us
to re-evaluate the nature of this triple system since the component spectral types of K7 and M3 conflict with our measured 
NIR flux ratio of 5.5~mag.  The
corresponding effective temperatures of K7 and M3 dwarfs are 3960~K and 3412~K (\citealt{Boyajian:2012eu}), respectively,
which translates into a $K$-band contrast ratio of $\approx$1.7--1.9~mag regardless of the system age (\citealt{Baraffe:1998ux}).
In other words, 2MASS~J06475229--2523304~B is at least 3.5~mag fainter than it should be assuming all stars
in the system are on the main or pre-main sequence.

\subsubsection{A Subgiant and M Dwarf Binary?}

Since it appears unlikely that 2MASS~J06475229--2523304 AabB is composed of normal stars on the main or pre-main sequence, 
here we discuss a few alternative explanations.  One option is that the system is very young
and an edge-on protoplantary disk surrounding 2MASS~J06475229--2523304 B is producing
$\sim$3.5--4~mag of extinction in the NIR, similar to TWA~30~B (\citealt{Looper:2010dy}) 
and possibly FW~Tau (\citealt{Kraus:2014tl}; \citealt{Bowler:2014dk}).
However, this would require the system to be younger than the disk dissipation timescale of 
$\sim$10~Myr, in which case we would expect lithium absorption in the primary to be  
similar to TWA members (at least 300~m\AA; e.g., \citealt{Mentuch:2008gb}).  This age is definitively ruled out with the nondetection of \ion{Li}{1}
in our high resolution optical spectrum, so this edge-on disk scenario is unlikely.

Another option is that the primary is an evolved star while the companion is on the main sequence.
Using the $J$-band contrast, the typical $V$--$J$ color of M3 dwarfs from \citet{Pecaut:2013ej},
and the absolute $V$-band magnitude of M3 dwarfs from \citet{Drilling:2000vo}, we find a 
spectrophotometric distance of 240~$\pm$~40~pc to 2MASS~J06475229--2523304~B.
If we assume the primary is a red giant, its apparent $V$-band magnitude implies a distance of 
1700~$\pm$~100~pc.  Instead, to match the distance estimate of 240~pc for the companion,
2MASS~J06475229--2523304~A would need to have an absolute $V$-band magnitude of $\sim$4.0~mag.
This is above the main sequence even if the primary is an equal-flux binary and is most consistent with
evolved stars on the subgiant branch.  \citet{Pojmanski:1997vo} and \citet{Richards:2012ea} 
measure a rotation rate of 7.56 days for 2MASS~J06475229--2523304~A which is somewhat faster than
the typical rotation period for subgiants of $\sim$10--100~days
(\citealt{doNascimentoJr:2012bl}; \citealt{vanSaders:2013bs}).  However, the fast rotation period can be explained by a
tidally-locked short-period companion (e.g., \citealt{Ryan:1995be}), which can also account for its RV variations, H$\alpha$,
and X-ray activity.
We conclude that this system is most likely a distant ($\sim$240~pc) triple composed of an active K7 subgiant primary 
with a close, likely tidally-locked companion and a much wider ($\sim$260~AU) M3 dwarf.

\subsection{2MASS J08540240--3051366~AB} 

This nearby active M4.0 star was detected in $ROSAT$ and $GALEX$ and shows fairly strong H$\alpha$ emission
(\citealt[$EW$=--8.4~\AA]{Riaz:2006du}).
Its $GALEX$ photometry ($FUV$=21.1 $\pm$ 0.4 mag, $NUV$ = 19.53 $\pm$ 0.15) points to activity levels
comparable to known moving group members in UV/NIR color-color diagrams (e.g., \citealt{Bowler:2012dc}).
\citet{Chauvin:2010hm} estimate an age of 100 Myr based
on its saturated X-ray luminosity (log~$L_X$/$L_\mathrm{bol}$=--2.79~dex; \citealt{Riaz:2006du}).  
With no RV, the BANYAN tool for calculating moving group membership probabilities (\citealt{Malo:2013gn}; \citealt{Gagne:2014gp})
predicts a high \emph{a priori} value (87\%) that 2MASS~J08540240--3051366 is a member of $\beta$~Pic. 
We measure an RV of 44.5 $\pm$ 0.6 km~s$^{-1}$ from our high resolution optical spectrum 
obtained in 2009.  Based on this value, 2MASS J08540240--3051366 does not appear to be a member 
of a known moving group.  

\citet{Chauvin:2010hm} first imaged this star as part of their AO imaging search for
planets around young southern stars with VLT/NaCo.  They identified a modest-contrast ($\Delta$$K$=2.8~mag) companion
located at 1$\farcs$7.  Assuming an age less than 200~Myr and a distance of $\approx$10--15~pc as inferred by Riaz et al. and Chauvin et al., 
the mass of 2MASS J08540240--3051366~B is below $\approx$50~\Mjup.
\citet{Janson:2012dc} and \citet{Janson:2014gz} present additional Lucky imaging of the system between 2010 and 2012
and find hints of orbital motion, which we confirm with our own NIRC2 images in 2012.

In December 2014 we obtained Keck/OSIRIS $H$- and $K$-band spectroscopy of 2MASS J08540240--3051366~B.
Compared to field templates in Figure~\ref{fig:twom0854_hgcomp}, it best resembles M4--M6 spectral types with no
significant deviations from field objects that would point to an especially low surface gravity.  It is therefore inconsistent
with $\beta$~Pic membership.  We adopt a near-infrared
spectral type of M5~$\pm$~0.5, which is slightly earlier than the classification of $\sim$M7~$\pm$~2 suggested by
our photometry in Table~\ref{tab:astrometry}.

Assuming an age of 100--200~Myr and a distance of 13~$\pm$~5~pc, the implied mass of 2MASS J08540240--3051366~B
is 42~$\pm$~21~\Mjup \ based on the $K$-band bolometric correction from \citet{Liu:2010cw}
and evolutionary models from \citet{Burrows:1997jq}.  On the other hand, if the system is older 
then its mass is closer to the hydrogen burning limit (e.g., 78~$\pm$~19~\Mjup \ at 0.5--1~Gyr).

\subsection{2MASS~J11240434+3808108 AB} 

This M4.5 star is a likely member of the $\sim$500~$\pm$~100~Myr Ursa Major moving group (see \citealt{Bowler:2015ja} 
for a detailed summary).  \citet{Close:2003ie} and \citet{Cruz:2003fi} first noticed the late-M companion
2MASS~J11240434+3808108~B located at 8$\farcs$3 ($\approx$170~AU) from the primary star.  
No additional companions were identified in deep AO imaging (\citealt{Bowler:2015ja}).
We obtained a moderate-resolution spectrum of 2MASS~J11240434+3808108~B with IRTF/SpeX (Figure~\ref{fig:twom1124})
and optical AO imaging with Robo-AO (Figure~\ref{fig:roboao}).
We find a NIR spectral type of M9.5~$\pm$~0.5 using the spectral indices from \citet{Allers:2013hk} and 
visual comparison to field templates, which is somewhat later than 
the optical type of M8.5 found by \citet{Cruz:2003fi}.  
The gravity indices of \citealt{Allers:2013hk} return a field surface gravity (FLD-G) with no signs 
of youth.  This independently constrains the age of the system to $\gtrsim$100~Myr, consistent with the 
tentative association to the Ursa Major moving group.

\subsection{PYC J11519+0731 AabB (2MASS J11515681+0731262 AabB)}{\label{sec:pyc11519}}

\subsubsection{Spectral Classification and Binarity of PYC J11519+0731 Aab}{\label{sec:pycbin}}

PYC~J11519+0731 has received little previous attention in the literature aside from identification as a $ROSAT$ 
source counterpart (\citealt{Zickgraf:2003cr}; \citealt{Haakonsen:2009iq}) and measurement of a 2.391 day rotationally-modulated 
photometric period (\citealt{Kiraga:2012wj}). 
It was identified as a candidate member of the $\beta$ Pic moving group by \citet{Schlieder:2012gj} 
from its proper motion and strong activity traced by $ROSAT$ X-ray and $GALEX$ UV emission. 

\citet{Schlieder:2012iw} and \citet{Lepine:2013hc} classify PYC~J11519+0731 as M3 and M2.5, respectively, from moderate-resolution 
optical spectroscopy.  
Our Mayall/RC-Spectrograph optical spectrum and IRTF/SpeX near-infrared spectrum are shown 
in Figures~\ref{fig:pycoptcomp} and \ref{fig:pycnircomp}.  
Even at the relatively low resolving power of our Mayall spectrum ($R$$\sim$2500), the H$\alpha$ emission line ($EW$=--2.8~\AA)
is marginally resolved 
into two components (inset).  This is easily split into a double-lined spectroscopic binary at higher resolutions (see below).  
Visual comparison to M dwarf templates from \citet{Bochanski:2007it} indicates an M2 optical spectral type. 
Our spectrum deviates somewhat from the templates below $\approx$7100~\AA, which is probably caused by slit loss from 
differential chromatic refraction and/or the unresolved binarity.  Altogether we adopt a spectral type 
of M2.0~$\pm$~0.5.  Compared to the NIR spectrum of the M2.0V star Gl~806 from \citet{Rayner:2009ki} in Figure~\ref{fig:pycnircomp},
PYC~J11519+0731 shows some peculiarities in the form of deeper molecular and atomic absorption features 
primarily between 1.50--1.65~$\mu$m, which could reflect its double-line spectroscopic binary (SB2) nature or a super-solar metallicity. Indeed,
Schlieder et al. found rather large \ion{Na}{1}~$\lambda$8200 absorption ($EW$ = 4.3~\AA), which is above 
the median value for field-age dwarfs of this type (see Figure 3 in \citealt{Schlieder:2012iw}) and 
points to a high surface gravity and old age.

We obtained three epochs of high-resolution optical spectroscopy of PYC~J11519+0731 with 
Keck/HIRES, ESO-MPG 2.2-m/FEROS, and CFHT/ESPaDOnS between 2012 and 2014 (Table~\ref{tab:specobs}).
Visual inspection of each spectrum reveals prominent emission and absorption features split into doublets (Figure~\ref{fig:pychalpha}), 
clearly indicating the primary star (PYC~J11519+0731~A) is itself a double-lined SB caught with the components 
at a relatively large velocity (close physical) separation.  Our HIRES spectrum exhibits strong, double-peaked H$\alpha$ 
emission with dual self-absorption.  The FEROS spectrum shows double-peaked \ion{Ca}{2} H and K emission, strong H$\alpha$ emission with 
dual self-absorption, dual reversed-emission cores in each of the \ion{Ca}{2} IR triplet and the Na~D lines, \ion{He}{1}~$\lambda$5876 emission, 
and single-peaked H$\beta$, H$\gamma$, and H$\delta$ emission. Similarly, the ESPaDOnS spectrum displays double-peaked 
\ion{Ca}{2} H and K emission, strong H$\alpha$ emission with dual self-absorption, dual reversed-emission cores in each of the \ion{Ca}{2} IR triplet 
and the Na~D lines, dual \ion{He}{1}~$\lambda$5876 emission, and dual-peaked H$\beta$, H$\gamma$, H$\delta$ and H$\epsilon$ emission. 

We used used a cross-correlation (CC) package from \citet{Bender:2008xk} to determine the RV 
of the primary (Aa) and secondary (Ab) at each epoch, systemic velocity, and mass ratio of the system.
Following the methods of \citet{Mazeh:2002wv}, we performed cross-correlations between the science spectrum and template spectra 
to measure the RV of each component.  The templates can be observed standards or models, and the routine automatically accounts for
the relative normalization of the spectra. 

The HIRES, FEROS, and ESPaDOnS spectra were analyzed in a similar fashion.  The echelle orders were first trimmed to avoid telluric features,
normalized, and, when necessary, the continua were flattened by fitting and dividing by a polynomial.  A set of K5--M4 RV templates
were selected from \citet{Prato:2002hp} and observed using FEROS.  These same templates were used for the HIRES and ESPaDOnS analysis.
We attempted CC with single, binary, and rotationally-broadened templates.  For our FEROS data, 
the strongest CC power ($\sim$90\%) was achieved for the primary and secondary using an M2.5+M2.5 system with $RV_P$=15.7~$\pm$~1 km~s$^{-1}$ 
and $v_P$sin$i_P$ = 14~$\pm$~2~km~s$^{-1}$  and $RV_S$ = --18.2~$\pm$~1 km~s$^{-1}$ and $v_S$sin$i_S$ = 12~$\pm$~2 km~s$^{-1}$. 
The RV uncertainties are dominated by the use of observed template spectra, and the projected rotational velocity uncertainties reflect the
velocity increments used to generate the suite of broadened templates.
For our HIRES data we measure CC peaks at $RV_P$=50.9~$\pm$~1 km~s$^{-1}$ and $RV_S$ = --51.9~$\pm$~1 km~s$^{-1}$, and
for our ESPaDOnS data we find $RV_P$=69.5~$\pm$~1 km~s$^{-1}$ and $RV_S$ = --75.5~$\pm$~1 km~s$^{-1}$.
The range of flux ratios we infer are between 0.5--0.75 at $\approx$6400~\AA. 

With three epochs of RV data for each component, the mass ratio of the system ($q$) and systemic RV ($\gamma$) can be determined 
following the methods of \citet{Wilson:1941ev}.  Figure~\ref{fig:pycsb2} shows the component velocities plotted against each other; the mass ratio of the 
system is the negative of the slope of the line.  We find a nearly equal mass ratio of $q$~=~0.95~$\pm$~0.01 
and systemic velocity $\gamma$~=~--0.3~$\pm$~1.0 km~s$^{-1}$.

The large velocity semi-amplitude of PYC~J11519+0731~Aab implies a short orbital period (at most a few days), 
so the high energy emission is most likely a result of fast rotation from tidal synchronization rather than youth 
(\citealt{Torres:2002eq}; \citealt{Torres:2003hp}; \citealt{Shkolnik:2010ba}; \citealt{Kraus:2011ju}).
Further RV and photometric monitoring could provide a spectroscopic orbit and rotationally-modulated period, which would
give a better indication of whether tidal spin up could be responsible for the observed activity.

\subsubsection{Age and Kinematics of PYC J11519+0731 Aab}

In general, \ion{Li}{1} absorption at 6708~\AA \ from either component of PYC J11519+0731Aab would provide strong, 
independent evidence of youth. The top panels of Figure~\ref{fig:pychalpha} show the  \ion{Li}{1} region in our three high-resolution spectra.
Our HIRES spectrum shows an absorption line that is close to the predicted position of  \ion{Li}{1} in the primary. However, in 
the other epochs there is no convincing detection of  \ion{Li}{1} in either component.  At the 12--25~Myr age of the $\beta$~Pic 
moving group (\citealt{Yee:2010fh}; \citealt{Binks:2014gd}; \citealt{Mamajek:2014bf}),  lithium is not typically detected for spectral types of 
M2; only at younger ages of $\lesssim$10~Myr is photospheric lithium still present (\citealt{Mentuch:2008gb}).
The lack of strong lithium in either component sets a lower
age limit of $\sim$10~Myr for the system and is therefore consistent with the age of $\beta$~Pic.

Despite lacking a parallax, our systemic RV enables constraints on the kinematics of PYC J11519+0731 Aab.
Figure~\ref{fig:pycuvw} shows the range of $UVW$ heliocentric space velocities (based on the procedure
from \citealt{Johnson:1987ji} with the coordinate transformation matrix from \citealt{Murray:1989vh})
and $XYZ$ galactic positions for PYC~J11519+0731 compared to members of young moving groups 
from \citet{Torres:2008vq}.\footnote{Note that 
the proper motion of PYC~J11519+0731 is incorrectly listed in several all-sky catalogs.  USNO-B (\citealt{Monet:2003bw}) and
the updated PPMXL Catalog (\citealt{Roeser:2010cr}) list proper motions of $\mu_{\alpha}$cos$\delta$ = --112~mas~yr$^{-1}$ and 
$\mu_{\delta}$ = 122~mas~yr$^{-1}$ and $\mu_{\alpha}$cos$\delta$ = --126~mas~yr$^{-1}$ and 
$\mu_{\delta}$ = 110~mas~yr$^{-1}$, respectively.  However, visual inspection of older Digitized Sky Survey and 
2MASS images clearly shows movement in the southwest direction.  We therefore adopt the original PPMXL
proper motion of $\mu_{\alpha}$cos$\delta$ = --142~$\pm$~19 mas~yr$^{-1}$ and $\mu_{\delta}$ = --89~$\pm$~18 ~mas~yr$^{-1}$.}
The systemic RV we measure ($\gamma$~=~--0.3~$\pm$~1.0 km~s$^{-1}$) agrees with the value of 
--1.5~$\pm$~1.0 km~s$^{-1}$ predicted by \citet{Schlieder:2012gj} assuming $\beta$~Pic membership, and the range of 
$UVW$ velocities is consistent with that group for distances between $\sim$20--30~pc.
However, the physical location of PYC~J11519+0731 disagrees from known members in $Z$ by at least 20~pc.

Group membership can be further tested by placing PYC J11519+0731 on a color-magnitude diagram.  Because they are
still contracting, $\beta$~Pic members are overluminous by $\sim$1~mag in $V$ band compared to field stars 
on the main sequence (e.g., \citealt{Riedel:2014ce}).  
Assuming similar spectral types for PYC J11519+0731 Aa and Ab and a near-IR flux ratio of 0.6~$\pm$~0.1 from our CC analysis
(Section~\ref{sec:pycbin}), the integrated apparent $V$-band magnitude of 12.5~$\pm$~0.1~mag from \citet{Droege:2006eq} 
can be decomposed into individual $V$-band magnitudes of 13.01~$\pm$~0.12~mag and 13.57~$\pm$~0.16~mag for Aa and Ab.
At a distance of 20~pc (30~pc), these correspond to absolute magnitudes of 11.50~$\pm$~0.12~mag (10.62~$\pm$~0.12~mag) and
12.07~$\pm$~0.16~mag (11.19~$\pm$~0.16~mag) for Aa and Ab, respectively.  
At the system $V$--$K$ color of 4.6~$\pm$~0.1~mag, $\beta$~Pic members are expected to have $M_V$$\approx$9.5~mag 
(\citealt{Riedel:2014ce}).  Instead, both components of PYC J11519+0731~Aab appear consistent with the main sequence given 
the range of plausible distances assuming group membership.  It is therefore likely this system is simply an old but active interloper 
with (partially-constrained) kinematics similar to the $\beta$~Pic YMG.

\subsubsection{PYC J11519+0731 B}

In 2012 we imaged PYC J11519+0731 as part of our PALMS survey.  
As expected, the close SB2 pair PYC J11519+0731~Aab was not resolved in our NIR images, implying a separation less than about one
PSF FWHM (40.4~$\pm$~0.8~mas in $H$ band, or $\approx$1.5~AU at 37~pc).  Indeed, two 0.2~\Msun \ stars
on circular, edge-on orbits would need to reside at separations of 0.03~AU to match the maximum observed RV semi-amplitude. 
We did, however, 
identify a moderate-contrast ($\Delta H$=5.4~$\pm$~0.2~mag) point source 
located at $\approx$0$\farcs$5 from PYC J11519+0731~Aab and immediately obtained unsaturated frames in $Y$, $J$, $H$, $K_S$, and $L'$ 
(Figure~\ref{fig:pycimgs} and Table~\ref{tab:astrometry}).  The companion's red colors point to a cool temperature and late-M spectral type.  
To search for additional companions in the system we obtained deeper $H$-band coronagraphic images in 2013 (20 frames each with 
30-sec integration times) in ADI mode.

A single image in our ADI sequence is shown alongside the PSF-subtracted and coadded version using the entire sequence
in Figure~\ref{fig:pycloci}.  The companion is visible in the single image and is similar in appearance to the quasi-static speckles,
but stands out prominently with a peak signal-to-noise ratio of 370 in our PSF-subtracted image.
The PSF subtraction was carried out in a similar fashion as the deep imaging in \citet{Bowler:2015ja}, except
here we have made use of 100 reference PSF images for our LOCI reduction (\citealt{Lafreniere:2007bg}) from a PSF library of over 
2000 registered coronagraphic images from our PALMS survey.  Reference images were selected as having the lowest residuals
between 0$\farcs$3--2$''$ after scaling and subtracting the library template to each science image.  Local masking was applied at
the location of PYC J11519+0731~B to minimize biasing the LOCI coefficients.

We obtained three additional epochs of astrometry with Keck/NIRC2 between 2012 and 2014, and one
with Gemini-S/NICI in 2013 (Table~\ref{tab:astrometry}).  Compared to the expected motion of
a stationary object (Figure~\ref{fig:pycback}), our astrometry unambiguously shows that PYC J11519+0731~B is 
comoving with the primary Aab.  We also find clear signatures of orbital motion.  The reduced chi-squared ($\chi^2_{\nu}$) value 
of a constant (linear) fit to the separation measurements is 7.7 (1.0), and a constant (linear) fit to the PA is 268 (3.5).
The linear fits correspond to changes of +9.7~$\pm$~1.2 mas~yr$^{-1}$ and --2.04~$\pm$~0.04 $^{\circ}$~yr$^{-1}$ for
the separation and PA, respectively.  Curvature in the orbit should be evident within a few years.

Our $K$-band Keck/OSIRIS spectrum of PYC J11519+0731~B is shown in Figure~\ref{fig:pycspec} compared
to field templates from the IRTF Spectral Library (\citealt{Cushing:2005ed}; \citealt{Rayner:2009ki}) and 
 low-gravity (VL-G) templates from  \citet{Allers:2013hk}.  PYC J11519+0731~B appears most similar to
 the field M8 template in overall shape and the depth of CO and \ion{Na}{1} features.  
 The 2.29~$\mu$m CO absorption is weaker and the $\sim$2~$\mu$m steam absorption is stronger in the very low gravity objects 
 compared to PYC~J11519+0731~B.  In addition, the 2.207~$\mu$m \ion{Na}{1}  doublet, which is sensitive to
surface gravity (and hence age), is weaker in \emph{all} very low-gravity templates.  This is perhaps the strongest indication that
PYC~J11519+0731~B is older than $\sim$30~Myr.
Because of the limited spectral coverage, we adopt a spectral type of M8 with uncertainty of $\pm$1 subtype.

The mass of PYC~J11519+0731~B depends on the distance and, more importantly, on the system age.  
The photometric distance to PYC~J11519+0731~B using the $H$-band absolute magnitude-spectral type
relations from \citet{Dupuy:2012bp} is 37~$\pm$~6~pc.  This incorporates uncertainties in the photometry,
spectral classification, and rms of the relation.  Using the $H$-band bolometric correction from \citet{Liu:2010cw}
we infer a luminosity of log$L$/$L_*$=--3.35~$\pm$0.16 dex.  At ages of \{21~$\pm$~2~Myr, 100~$\pm$~10~Myr,
1.0~$\pm$~0.1~Gyr, 5.0~$\pm$~0.5~Gyr\}, the evolutionary models of \citet{Burrows:1997jq} imply masses
of \{15.3~$\pm$~1.3, 42~$\pm$~7, 89~$\pm$~5, 90~$\pm$~4\} \Mjup \ for PYC~J11519+0731~B.
If the system is younger than $\approx$350~Myr then the companion mass lies below the hydrogen-burning limit.

No other candidate companions were identified in our deep imaging, which extends out to $\approx$4$''$ for complete 
azimuthal sensitivity and to $\approx$8$''$ for partial sensitivity.  The 7~$\sigma$ $H$-band contrast curve shown in Figure~\ref{fig:pycsens}
is measured by computing the rms in 5~pixel (50~mas) annuli as described in \citet{Bowler:2015ja}.  
Unsaturated frames are used at separations below 0$\farcs$3 and our deep ADI data are used at larger radii.
Our final contrast curve in $\Delta$$H$ and field of view coverage (in brackets) is \{4.0, 5.4, 10.6, 12.2, 13.1, 13.8, 13.4, 14.0, 14.2, 14.0, 
13.4\} mag for separations of \{0$\farcs$1, 0$\farcs$2, 0$\farcs$5, 0$\farcs$75, 1$\farcs$0, 1$\farcs$5, 2$''$, 3$''$, 4$''$ [1.0], 5$''$ [0.69], 8$''$ [0.05]\}.
That is, we are fully sensitive to spatial coverage within $\approx$4$''$, but are only sensitive to a fraction of the sky beyond that owing to the square detector.
At ages of \{21~Myr, 100~Myr, 1~Gyr, 5~Gyr\}, we remain highly sensitive ($>$90\%) to companions with masses as low as \{1, 2, 10, 25\}~\Mjup \ 
at 100~AU based on the evolutionary models of \citet{Baraffe:2003bj}.

\subsection{G~180-11 AB (2MASS J15553178+3512028 AB)}

A 1$\farcs$6 companion to the M4.0~$\pm$~0.5 star G~180-11 (a.k.a. GJ 3928) 
was first discovered by \citet{McCarthy:2001gt} and confirmed to be physically bound 
by \citet{Daemgen:2007fg}, \citet{Law:2008fn}, and \citet{Janson:2012dc}.  
Unresolved detections in $ROSAT$ and $GALEX$ together with strong H$\alpha$ emission  
($EW$$\approx$--6 to --8.1~\AA; \citealt{Riaz:2006du}; \citealt{Shkolnik:2009dx}) indicate the primary and/or companion 
are active.  \citet{Riaz:2006du} found a saturated fractional X-ray luminosity of log~$L_X$/$L_\mathrm{bol}$=--2.98~dex 
for the system, pointing to a relatively young age.
\citet{Hartman:2011cq} measured a rotational period of 3.5209 days as part of the HATNet survey.

Recently, \citet{Dittmann:2014cr} measured a parallactic distance of 28~$\pm$~3~pc to G~180-11~AB 
as part of the MEarth transit survey.  At this distance, the primary-companion separation is 45~AU.
Together with the RV of --15.5~$\pm$~0.7~km~s$^{-1}$ from
\citet{Shkolnik:2012cs}, this corresponds to $UVW$ kinematics of 
\{--35~$\pm$~3, --17.6 $\pm$ 1.4, and 8 $\pm$ 2\} km~s$^{-1}$ and $XYZ$ space positions of \{9.9 $\pm$ 1.1, 
14.9 $\pm$ 1.6, and 21.5 $\pm$ 2\} pc.  
\citet{Malo:2013gn, Malo:2014dk} find that G~180-11~AB is a likely member of the Argus moving 
group based on the RV alone, which matches the predicted value assuming group membership, but 
the parallactic distance disagrees with the statistical distance of 13~$\pm$~1~pc so the similar RV is 
probably coincidental.  Indeed, the kinematics
of G~180-11~AB do not appear to be consistent with any known moving groups.

The pair are easily resolved with our Keck/NIRC2 and P60/Robo-AO observations spanning 0.6--2.2~$\mu$m (Figure~\ref{fig:roboao})
Compared with previous astrometry for this system (see \citealt{Janson:2012dc} for a summary), our new data show
slight outward motion by $\approx$6~mas~yr$^{-1}$ and P.A. change of $\approx$0$\fdg$8~yr$^{-1}$.  The
orbital motion is quite slow so it will take decades to complete an appreciable portion of its orbit.

\citet{Daemgen:2007fg} infer a spectral type of M7~$\pm$~2 for G~180-11~B from its $K$-band contrast ratio, 
\citet{Cruz:2003fi} deduce an $\sim$M6 classification, and \citet{Law:2008fn} estimate a spectral type of M6.5 from
its optical colors.  In 2013 we obtained resolved near-infrared ($JHK$) spectra of G~180-11~B with Keck/OSIRIS.
Compared to templates from the IRTF Spectral Library (\citealt{Rayner:2009ki}) in Figure~\ref{fig:g180hgcomp},
G~180-11~B best resembles M5--M6 field objects from 1.2--2.4~$\mu$m and in individually normalized
$J$, $H$, and $K$ bands with a slightly better match to the M6 spectrum.  We therefore adopt a near-infrared 
spectral type of M6.0~$\pm$~0.5, in agreement with previous estimates from colors.
Although the 2.21~$\mu$m \ion{Na}{1} doublet line is equally strong in G~180-11~B as in the field templates,
the 1.244 and 1.253~$\mu$m \ion{K}{1} lines are noticeably weaker, pointing to a low surface gravity in line
with an age younger than $\sim$200~Myr.  

Based on the parallactic distance and an $H$-band bolometric correction from \citet{Liu:2010cw},
we measure a luminosity of log~$L$/$L_{\odot}$ = --2.48~$\pm$~0.13~dex  for G~180-11~B.
At ages of 50~$\pm$~5 Myr, 100~$\pm$~10~Myr, and 200~$\pm$~20 Myr, this corresponds to 
masses of 88~$\pm$~18~\Mjup, 120 $\pm$ 20~\Mjup, and 150 $\pm$ 20 \Mjup \ using the \citet{Burrows:1997jq} evolutionary models.  
Unless this system is younger than $\sim$40~Myr, G~180-11~B is a low-mass star instead of a brown dwarf.

\subsection{2MASS~J15594729+4403595 AB} 

2MASS~J15594729+4403595 is a young active early-M dwarf with a comoving substellar companion
first identified by \citet[see \citealt{Bowler:2015ja} for details]{Janson:2012dc} at a projected separation of 5$\farcs$6 (150~AU).
The companion was easily recovered with our Robo-AO imaging in $i'$ and $z'$ bands (Figure~\ref{fig:roboao}).
Our SNIFS 4000--9000~\AA \ optical spectrum of the primary in Figure~\ref{fig:pri_optspec} reveals weak H$\alpha$ emission ($EW$=--2.3~\AA)
and is well-matched with an M1.5V spectrum, created by averaging the M1 and M2 templates from \citet{Bochanski:2007it}.  
Our own RV of --19.6~$\pm$~0.6 km~s$^{-1}$ from ESPaDOnS disagrees with the value of --15.8~$\pm$~0.5 km~s$^{-1}$ 
from \citet{Malo:2014dk} by 4~km~s$^{-1}$, a 4.9~$\sigma$ difference (Table~\ref{tab:rvs}),
raising the possibility that the primary is an SB1.
Although the system does not appear to coincide with any known moving groups, the intermediate-gravity near-infrared spectrum of 
2MASS~J15594729+4403595~B implies a young age of $\approx$50--200~Myr (\citealt{Bowler:2015ja}).  

Our spatially-resolved Keck/ESI optical spectrum of 2MASS~J15594729+4403595~B is shown in Figure~\ref{fig:twom1559_optspec}.
2MASS~J15594729+4403595~B exhibits modest H$\alpha$ emission ($EW$=--6.4~\AA) and absorption features typical of late-M dwarfs.  
Compared to templates from Bochanski et al., 2MASS~J15594729+4403595~B fits the M7.0 and M7.5 objects
reasonably well.  The index-based classification from \texttt{Hammer} (\citealt{Covey:2007bj}) is M7.  Altogether we adopt an optical classification of M7.5~$\pm$~0.5.
Note, however, that the best match to the entire 0.5--2.5~$\mu$m spectrum is slightly later (M8; Figure~\ref{fig:twom1559_allplot}),
which is common with visual classification of young brown dwarfs (\citealt{Allers:2013hk}).

As shown in Figure~\ref{fig:twom1559_allplot}, we find strong \ion{Li}{1} $\lambda$6708 absorption in the companion ($EW$ = 0.71~\AA).  
The detection of lithium provides independent constraints on the mass and age of the companion through
the ``lithium test'' (\citealt{Rebolo:1992hn}; \citealt{Rebolo:1996co}; \citealt{Kirkpatrick:2008ec}).  
\citet{Basri:1998tg} shows that an object with \ion{Li}{1} absorption with a luminosity below log~($L$/$L_{\odot}$) = --3.05~dex
or \Teff \ $<$ 2670~K (spectral type $\gtrsim$M7) is unambiguously substellar and young ($\lesssim$200~Myr).  
This is consistent with the the intermediate surface gravity and mass of 43~$\pm$~9~\Mjup \ for 
2MASS~J15594729+4403595~B found in \citet{Bowler:2015ja}.

\subsection{GJ 4378 Aa12b (2MASS J23572056--1258487 Aa12b)}  

\subsubsection{The wide binary GJ 4378 A and GJ 4379 B}

GJ~4378~A (LP~704-15, G~273-186) and GJ~4379~B (2MASS~J23571934--1258406, 
G~273-185, LP~704-14) form a 
pair of comoving mid-M dwarfs separated by 20$''$ (\citealt{Giclas:1975un}).
Both stars have similar reported $V$-band (13.0~mag from APASS/UCAC4, \citealt{Zacharias:2013cf}) and $r'$-band 
(12.4~mag from CMC15, \citealt{Muinos:2014ew}) magnitudes,
but GJ~4378~A is brighter in $JHK$ by $\sim$0.5~mag (from 2MASS, \citealt{Skrutskie:2006hl}).
GJ~4378~A is classified by \citet{Reid:2004be}, \citet{Reid:1995kw}, and \citet{Riaz:2006du} as M4, M4, and M3,
respectively, from optical spectroscopy.  The same studies classify GJ~4379~B as
M4, M3, M4.  We find similar spectral types of M3.5 for GJ~4379~B and M4.5 for GJ~4378~A from our own
low-resolution optical spectra (Figure~\ref{fig:pri_optspec}).  Although GJ~4378~A is slightly brighter, its
spectral type is slightly later.  Our moderate-resolution SOAR spectra of the pair are shown in Figure~\ref{fig:pri_medresspec}.
GJ~4378~A exhibits H$\alpha$ emission ($EW$=--4.5~\AA) while none is apparent in the companion GJ~4379~B.

The system coincides with a $ROSAT$ Bright Source Catalog detection (1RXS~J125720.0--125852), 
but it is unclear which component is the counterpart because of the $ROSAT$ PSPC's large positional uncertainty of $\sim$13$''$ (1~$\sigma$;
\citealt{Voges:1999ws}).  GJ~4378~A
was detected in both $GALEX$ in $FUV$ and $NUV$ bandpasses, while GJ~4379~B was detected in $NUV$ alone.
The $NUV$--$J$ color, which traces UV excess from chromospheric emission, is 10.58~$\pm$~0.05~mag
and 13.17~$\pm$~0.19~mag for the A and B, respectively.  The blue $NUV$--$J$ of GJ~4378~A is consistent with 
activity levels of YMG members (10--150~Myr), but the red color of GJ~4379~B is more consistent with
older M dwarfs in the Hyades and the field (\citealt{Rodriguez:2013fv}).  

\citet{Malo:2013gn} identify GJ~4378~A as a candidate member of the $\sim$40~Myr Argus association based
on its proper motion and activity.  Assuming group membership, they predict a kinematic distance of 25~$\pm$~1~pc and RV
of --1.0~$\pm$~0.8~km~s$^{-1}$.  In 2009 we obtained echelle spectroscopy of GJ~4378~A and GJ~4379~B 
with the du~Pont telescope.
Cross-correlation analysis of our du~Pont spectrum of GJ~4378~A shows it is a near-equal flux SB2 with component 
velocities of --7.0~$\pm$~0.8~km~s$^{-1}$ for the primary
(GJ~4378~Aa1) and 49.2~$\pm$~1.1~km~s$^{-1}$ for the companion (GJ~4378~Aa2)\footnote{To avoid confusion,
we adopt the following nomenclature for this hierarchical quadruple system: we retain the names of the 20$''$ pair GJ~4378~A 
and GJ~4378~B, GJ~4378~A is a $\approx$0$\farcs$5 visual binary (see \ref{sec:gj4378ab}) with GJ~4378~Aa and GJ~4378~Ab components,
and GJ~4378~Aa is an SB2 with components GJ~4378~Aa1 and GJ~4378~Aa2.}.
Assuming the pair are equal mass, the systemic RV of GJ~4378~Aa12 is 21.1~$\pm$~0.7 km s$^{-1}$.
We measure an RV of 20.6~$\pm$~0.4~km~s$^{-1}$ for GJ~4379~B and find no signs it is a spectroscopic binary.
Assuming the RV for GJ~4379~B represents the systemic velocity, the range of $UVW$ values for distances of 20--60~pc
do not coincide with any known moving groups.  

Despite their similar spectral types and presumed coevality, GJ~4378~A and GJ~4379~B appear to have very different 
chromospheric activity levels.  Like PYC J11519+0731~Aab (Section~\ref{sec:pyc11519}), GJ~4378~Aa12 is an SB2 with enhanced
UV emission (and the likely source of the X-ray emission) probably induced by tidally locked stars instead of intrinsic fast rotation from youth.
On the other hand, GJ~4379~B appears to be single and is therefore a better tool to diagnose the system age.
Indeed, neither our spectroscopy nor shallow Keck/AO $K_S$-band imaging of GJ~4379~B on UT 2014 August 04 
shows any signs of a close companion.
\citet{West:2008eq} find a characteristic activity lifetime of 4.5$^{+0.5}_{-1.0}$~Gyr for M4 stars as traced 
by the presence of H$\alpha$ emission.  The lack of H$\alpha$ emission in GJ~4379~B therefore suggests an old age of $\gtrsim$4~Gyr 
for the system.

\subsubsection{GJ~4378~Ab}{\label{sec:gj4378ab}

We imaged GJ~4378~A in Oct 2012 with IRCS on Subaru, easily resolving it into close (0$\farcs$47) 
modest-contrast ($\Delta$$H$=3.1~mag) binary.  Astrometric monitoring throughout 2013 and 2014 confirm the pair are
comoving (Figure~\ref{fig:gj4378backtracks}) and undergoing significant orbital motion of 0$\farcs$0564~$\pm$~0$\farcs$0006 yr$^{-1}$ 
in separation and --1$\fdg$06~$\pm$~0$\fdg$02 yr$^{-1}$ in PA.

In August 2013 we obtained 1--4~$\mu$m Keck/NIRC2 AO images of the system (Figure~\ref{fig:gj4378imgs}).
Decomposed photometry of the pair presented in Table~\ref{tab:astrometry} imply the companion has red colors consistent
with a late-M dwarf.  
This is confirmed with our $H$- and $K$-band OSIRIS spectroscopy of GJ 4378 Ab (Figure~\ref{fig:gj4378spec}),
which shows atomic and molecular absorption features characteristic of late-M/early-L dwarfs.
Compared to field late-M dwarfs from \citet{Cushing:2005ed}, the overall 1.5--2.4~$\mu$m spectrum is 
most similar to the M9V template.
The $H$-band spectrum alone provides a reasonable match to the M9V template, while the $H$-band 
spectrum is an excellent match to the M8V template.  Altogether we adopt a spectral type of M8~$\pm$~1 for GJ~4378~Ab.
Interestingly, the $H$-band spectral shape appears 
somewhat more angular than the field objects, which would imply a young age ($<$150~Myr) for
the system (\citealt{Allers:2013hk}) and appears to contradict the old age inferred from the lack of H$\alpha$ emission in 
the wide companion GJ~4379~B.
Moderate-resolution $J$-band spectroscopy of GJ~4378~Ab would help address the system age
since the 1.1--1.3~$\mu$m region possesses a number of gravity-sensitive features.

Using $H$-band spectral type-absolute magnitude relations from \citet{Dupuy:2012bp} and $H$-band bolometric
correction from \citet{Liu:2010cw}, we infer a spectrophotometric distance of 14~$\pm$~2~pc for GJ~4378~Ab and
a luminosity of log($L_\mathrm{bol}$/$L_*$)=--3.35~$\pm$0.14 dex.  This agrees with the photometric distance 
of 12~pc to GJ~4378~Aa from \citet{Lepine:2011gl}.  
Based on the evolutionary models of \citet{Burrows:1997jq}, the luminosity of GJ~4378~Ab 
implies a mass of 90~$\pm$~3~\Mjup \ for an age of 5~$\pm$~2~Gyr with little dependence of 
age beyond $\approx$1~Gyr.

\section{Discussion and Conclusions}

The high density and sheer number of M dwarfs in the solar neighborhood imply that many old stars
will share similar $UVW$ space motions with nearby YMGs.  The task of sifting through this haystack of potential
members is made even more difficult when only an RV or parallax is available, and in most 
cases neither of these are.
We have undertaken a detailed kinematic and spectroscopic analysis of 13 K and M-type systems 
hosting low-mass companions to clarify their ages, companion masses, and possible associations with YMGs (Figure~\ref{fig:benchmarks}).
Eleven of the primary stars have been previously identified as
being young in the literature, eight of which have been listed as candidate members of YMGs.
Three of the late-type companions to moving group candidates were discovered during our PALMS
AO imaging survey and are new in this work
(2MASS~J06475229--2523304~B, PYC J11519+0731~B, and GJ~4378~Ab).

Only two of the systems appear to be likely members
of YMGs (2MASS J01225093--2439505 AB and 2MASS J02155892--0929121 AabBC).  Another four 
show signs of youth but are probably not kinematically associated with any known moving groups 
(1RXS~J034231.8+121622 AB, 2MASS J08540240--3051366~AB, G~180-11 AB, and 2MASS~J15594729+4403595 AB).  
With the exception of 2MASS~J11240434+3808108~B, whose kinematics are consistent with the $\sim$500~Myr
Ursa Major moving group, the remaining systems with low-mass companions are probably old.
In particular, we have shown that the host stars to 
2MASS~J06475229--2523304~B, PYC J11519+0731~B, and GJ~4378~Ab are close spectroscopic binaries
with little or no evidence that they belong to nearby young moving groups.  This underscores the need for 
follow-up RV (and ideally parallax) measurements for new proposed members of YMGs 
to identify old M+M SB1s and SB2s that can masquerade as young stars.

2MASS J02155892--0929121 AabBC is likely to be an especially useful system in the future.  If its systemic RV
and parallax confirm it is a member of Tuc-Hor, which is suggested by the mean (4.0~$\pm$~0.15 km s$^{-1}$) 
and range (--0.6 to 10.0 km~s$^{-1}$) of RV measurements, 
2MASS J02155892--0929121 will offer an excellent opportunity to measure dynamical masses
for the Aab components and, eventually, the B component.  Since the outer companion (``C'') is a young brown dwarf,
this quadruple system offers the opportunity to test pre-main sequence evolutionary models across the substellar boundary.
Regardless of any possible age spread in Tuc-Hor itself, the components of this system were probably formed simultaneously,
removing any ambiguity about possible age differences.  This is analogous to the well-studied young 
high-order benchmarks LkCa~3 (\citealt{Torres:2013gm}), HD~98800 (e.g., \citealt{Prato:2001co}), and especially
GG~Tau (\citealt{White:1999bt}), which also contains a substellar member and recently gained a fifth component 
from sparse aperture masking interferometry (\citealt{DiFolco:2014cma}).

Four systems in our sample are examples of close spectroscopic binaries on what are likely tidally-circularized, few-day orbits with 
nearby lower-mass tertiaries that we have directly resolved with AO (G~160-54~AabBC, 
2MASS~J06475229--2523304~AabB, PYC~J11519+0731~AabB, GJ~4378~Aa12b).  Spectroscopic binaries do not form \emph{in situ} but instead must lose angular momentum
either through gas drag in a viscous molecular cloud core during the protostellar phase (e.g., \citealt{Gorti:1996tj}; \citealt{Stahler:2010ia}) 
or through dynamical 
interactions with a third body (e.g., \citealt{Eggleton:2001te}).  These close hierarchical triples are consistent with this latter scenario in which the tertiary 
remained relatively close in after a dynamical encounter with the (now tidally locked) secondary star instead of being ejected to a wide orbit or completely out of the system 
(\citealt{Tokovinin:2006ec}; \citealt{Reipurth:2010eh}; \citealt{Allen:2012dn}; \citealt{Reipurth:2013kp}).  
These tertiaries are expected to have relatively high eccentricities, which can eventually be tested 
with continued astrometric monitoring.

Regardless of their ages, many of these systems are useful for calibrating empirical metallicity relationships 
at the bottom of the main sequence.  Recently \citet{Mann:2014bza} presented an empirical method to measure the metallicities
of late-M dwarfs with $K$-band spectra by leveraging the metallicity of wide, earlier-type binary companions for calibration.
Although this was the largest such study to date, only a handful of late-M dwarfs in binaries were used (7 systems beyond M7.5),
so building larger samples of these benchmark systems will help better constrain these relationships. 

$Gaia$ will soon provide precise parallaxes for the currently unknown 
stellar members of YMGs, but its estimated 1--15~km~s$^{-1}$ RV precision is generally not adequate
to differentiate moving groups with $UVW$ space velocities only a few km~s$^{-1}$ apart
like TWA, Tuc-Hor, and Carina (\citealt{Lindegren:2008jf}; \citealt{Eyer:2015wj}).  
Ground-based RVs with $<$1~km~s$^{-1}$ precisions will therefore continue to play
a critical role in confirming this nearby population of young stars in the foreseeable future.

\acknowledgments

We thank the referee for their helpful suggestions and Niall Deacon for obtaining 
some of the IRTF observations presented here.  
M.C.L. has been supported by NASA grant NNX11AC31G and NSF grant AST09-09222.
This paper is based on observations at Cerro Tololo Inter-American Observatory, 
National Optical Astronomy Observatory (NOAO Prop. ID: 2014B-0083; PI: Bowler),
which is operated by the Association of Universities for Research in Astronomy (AURA) 
under a cooperative agreement with the National Science Foundation.
This research is also based in part on observations obtained at the Gemini Observatory, which is operated by the 
Association of Universities for Research in Astronomy (AURA) under a cooperative agreement with the NSF on 
behalf of the Gemini partnership: the National Science Foundation (United States), the Science and Technology 
Facilities Council (United Kingdom), the National Research Council (Canada), CONICYT (Chile), the Australian 
Research Council (Australia), CNPq (Brazil) and CONICET (Argentina).
The Robo-AO system was developed by collaborating partner institutions, the California Institute of Technology 
and the Inter-University Centre for Astronomy and Astrophysics, and with the support of the National Science Foundation 
under Grant Nos. AST-0906060, AST-0960343 and AST-1207891, the Mt. Cuba Astronomical Foundation and by a gift 
from Samuel Oschin. Ongoing science operation support of Robo-AO is provided by the California Institute of Technology 
and the University of Hawai`i. C.B. acknowledges support from the Alfred P. Sloan Foundation.
B.T.M. is supports by the National Science Foundation Graduate Research Fellowship under Grant No. DGE1144469.
We utilized data products from the Two Micron All Sky Survey, which is a joint project of the University of 
Massachusetts and the Infrared Processing and Analysis Center/California Institute of Technology, funded by the 
National Aeronautics and Space Administration and the National Science Foundation.
 NASA's Astrophysics Data System Bibliographic Services together with the VizieR catalogue access tool and SIMBAD database 
operated at CDS, Strasbourg, France, were invaluable resources for this work.
This research has made use of the Washington Double Star Catalog maintained at the U.S. Naval Observatory.
Finally, mahalo nui loa to the kama`\={a}ina of Hawai`i for their support of Keck and the Maunakea observatories.
We are grateful to conduct observations from this mountain.

\facility{{\it Facilities}: 
\facility{Keck:II (NIRC2, ESI)}, 
\facility{Subaru (IRCS)},
\facility{Gemini:South (NICI)},
\facility{PO:1.5m (Robo-AO)},
\facility{Keck:I (OSIRIS, HIRES)},
\facility{IRTF (SpeX)},
\facility{Mayall (RC-Spec)},
\facility{SOAR (Goodman)},
\facility{UH:2.2m (SNIFS)},
\facility{CFHT (ESPaDOnS)},
\facility{Du Pont (Echelle)},
\facility{Max Planck:2.2m (FEROS)},
\facility{CTIO:1.5m (CHIRON)}
}

\newpage

\bibliographystyle{apj}
\bibliography{palms5_astroph.bbl}

\clearpage

\begin{deluxetable}{lccccccccccccc}
\tabletypesize{\scriptsize}
\rotate
\setlength{ \tabcolsep } {.12cm} 
\tablewidth{0pt}
\tablecolumns{14}
\tablecaption{Target Properties\label{tab:targets}}
\tablehead{
   \colhead{Name} & \colhead{$\alpha_\mathrm{J2000}$} & \colhead{$\delta_\mathrm{J2000}$}  & \colhead{$\mu_{\alpha}$cos$\delta$}  & \colhead{$\mu_{\delta}$}  & \colhead{$\mu$}      & \colhead{SpT}         & \colhead{SpT} & 
  \colhead{Distance} &  \colhead{Dist.} &  \colhead{Dist.} & \colhead{YMG?\tablenotemark{d}} & \colhead{Age}   & \colhead{Age} \\
   \colhead{}     & \colhead{}                        & \colhead{}                         & \colhead{(mas yr$^{-1}$)}            & \colhead{(mas yr$^{-1}$)} &  \colhead{Ref.}           & \colhead{ ($\pm$0.5)} &  \colhead{Ref.}       & 
  \colhead{(pc)}     &  \colhead{Method}       &   \colhead{Ref.}        & \colhead{}     & \colhead{(Myr)} & \colhead{Ref.} 
        }
\startdata
2MASS~J01225093--2439505 AB          &  01 22 50.93   & --24 39 50.5     &  120.2~$\pm$~1.9  &  --120.3~$\pm$~1.7    &  Z13    &  M4.0     & TW   & 36~$\pm$~4     & Phot  & B13  & AB~Dor?   & 120~$\pm$~10  & B13, M13  \\
2MASS~J02155892--0929121 AabBC  &  02 15 58.92   & --09 29 12.1     &   96.6 $\pm$ 1.9      &  --46.5 $\pm$ 2.6         &  Z13    &  M3.5     & TW   & 26 $\pm$ 10     & Phot  & R06   &  Tuc-Hor? &  $\sim$30--40  &  M13, B15  \\
2MASS~J02594789--0913109 AB          & 02 59 47.89    & --09 13 10.9    &  296~$\pm$~8         &  --527~$\pm$~ 8           &  Z13    &  M4.0     & R07  & 36~$\pm$~13   & Phot  &  J12  &  $\cdots$  &  $\gtrsim$4000  &  TW   \\
1RXS~J034231.8+121622 AB                 & 03 42 31.80   & +12 16 22.5      &  196.8 $\pm$ 2.2     & --16.3 $\pm$ 3.8           &  Z13    &  M4.0     & R06  & 23.9 $\pm$ 1.1 &  $\pi$ & D14  &  $\cdots$ & 60--300 &  Sh09        \\
HD 23514 AB                                               & 03 46 38.39   &  +22 55 11.2     & 20.5 $\pm$ 0.4        & --42.6 $\pm$ 0.5           &  Z13    &  F5V       &  G01  &  136.2~$\pm$~1.2 &  Cluster  & Me14 & Pleiades  &  120~$\pm$~10  &  S98  \\
G~160-54 AabBC                                        & 04 13 45.85    & --05 09 04.9   & 182.0 $\pm$ 8.0       &  --112.5 $\pm$ 8.0        &  Z13    &   M4.5    &  TW   & 21 $\pm$ 9        &  Phot    &   L11    &  $\cdots$  &  $\gtrsim$1000  &  TW   \\
2MASS J05464932--0757427 AB           & 05 46 49.32    & --07 57 42.7   & 127.9 $\pm$ 2.2        &  --53.8 $\pm$ 2.1          & Z13     &  M3.0     & R07    &  45                     &  Phot  & R07  &   $\cdots$  &  $\gtrsim$1000  &  TW  \\
2MASS~J06475229--2523304 AB          & 06 47 52.29    & --25 23 30.4   & 22.7 $\pm$ 0.8          &  --72.0 $\pm$ 1.1         &  Z13     &  K7.0      & TW     &  240 $\pm$ 40   &  Phot   & TW  &  $\cdots$  &  $\cdots$ & $\cdots$  \\
2MASS J08540240--3051366 AB           & 08 54 02.40    & --30 51 36.6   & --281.6 $\pm$ 5.7      & --8.9 $\pm$ 5.7            &  R10    &  M4.0     & R06    &  10 $\pm$ 4    & Phot & R06  &  $\cdots$  &  $\lesssim$200?  &  C10     \\
2MASS~J11240434+3808108 AB          & 11 24 04.34    &  +38 08 10.8   & 121.7 $\pm$2.6        &  --12.4 $\pm$ 2.6           & Z13     &   M4.5    & R07    &  20.3 $\pm$ 1.3 & Phot  &  B15  &  UMa?  &  500 $\pm$ 100 &  Sh12 \\
PYC J11519+0731 AabB                          & 11 51 56.81    & +07 31 26.2   & --142~$\pm$~19        & --89~$\pm$~18             & R08     &  M2.0    &  TW     &    37~$\pm$~6  &  Phot  &  TW     &  $\cdots$          &   $>$10                &  TW    \\
G 180-11 AB                                                 & 15 55 31.78    &  +35 12 02.8   &  --233~$\pm$~8.0    & 150.0~$\pm$~8.0          &  Z13    &  M4.0    &  R06   &  28~$\pm$~3   &  $\pi$  &   D14  &  $\cdots$         &  $\lesssim$200   &  TW   \\
2MASS~J15594729+4403595 AB          & 15 59 47.29    & +44 03 59.5    &  --70.7~$\pm$~0.9   &  --8.9~$\pm$~0.6            &  Z13    &  M1.5    &  TW    &  27~$\pm$~2    &  Phot   &  B15   &  $\cdots$        &  50--200    &  B15  \\
GJ 4378 Aa12b                                           & 23 57 20.56    & --12 58 48.7   &   208.3 $\pm$ 8.0      &  25.2 $\pm$ 8.0              &  Z13    &  M4.5   &  TW      &  14 $\pm$ 2       &  Phot  & TW    &  $\cdots$                 &  $\gtrsim$4000   &  TW  \\
\enddata
\tablerefs{
B13=\citet{Bowler:2013ek};
B15 = \citet{Bowler:2015ja};
D14=\citet{Dittmann:2014cr};
C10 = \citet{Chauvin:2010hm};
G01=\citet{Gray:2001uo};
H96=\citet{Hawley:1996hg};
J12=\citet{Janson:2012dc}; 
L11=\citet{Lepine:2011gl};
L13=\citealt{Lepine:2013hc};
M13=\citealt{Malo:2013gn}; 
Me14=\citealt{Melis:2014id};
R04=\citealt{Reid:2004be};
R06=\citet{Riaz:2006du};
R07=\citet{Reid:2007gl};
R08=\citet{Roser:2008bk};
R10=\citet{Roeser:2010cr};
S98=\citet{Stauffer:1998kt};
Sh09=\citet{Shkolnik:2009dx};
Sh12=\citet{Shkolnik:2012cs};
TW=This work;
Z05=\citet{Zacharias:2005tw};
Z13=UCAC4 (\citet{Zacharias:2013cf})
}
\end{deluxetable}
\clearpage

\begin{deluxetable}{lcccccccccccc}
\tabletypesize{\scriptsize}
\tablewidth{0pt}
\tablecolumns{13}
\tablecaption{Companion Properties\label{tab:comp}}
\tablehead{
        \colhead{Name}      &  \colhead{Ang. Sep}  &  \colhead{Proj. Sep}   &  \colhead{SpT}   &   \colhead{YMG?}  &   \colhead{Age}   &    \colhead{Mass}   & \colhead{Discovery}  \\
        \colhead{}                  &  \colhead{($''$)}        &  \colhead{(AU)}          &    \colhead{}         &    \colhead{}          &      \colhead{(Myr)}  & \colhead{(\Mjup)}  &  \colhead{Reference}     
        }   
\startdata
2MASS~J01225093--2439505 B          &  1.5   &   52 $\pm$ 6     &  L5 $\pm$ 1  &  AB~Dor?     & 120~$\pm$~10  & 13--25  & B13 \\
2MASS~J02155892--0929121 C          &  3.5   &    90 $\pm$ 10   &  M7 $\pm$ 1  & Tuc-Hor? &  30--40  &  42 $\pm$ 15 & B10, TW \\
2MASS~J02594789--0913109 B          &  0.6   &    22 $\pm$ 8   &  [M9.5]\tablenotemark{a}    &  $\cdots$      &  $\gtrsim$4000  &  87 $\pm$ 13 & J12, TW  \\
1RXS~J034231.8+121622 B                 & 0.8    &   19.1  $\pm$ 0.9   &  L0 $\pm$ 1   &  $\cdots$     & 60--300                &  35 $\pm$ 8  & B15   \\
HD 23514 B                                               & 2.6    &  360 $\pm$ 3     & M8 $\pm$ 1   & Pleiades      &  120~$\pm$~10  &  60 $\pm$ 10 & R12 \\
G~160-54 C                                               & 3.3    &   70 $\pm$ 30    & M7.0 $\pm$ 0.5 &  $\cdots$  &  $\gtrsim$1000  &  85~$\pm$~16 & B15   \\
2MASS J05464932--0757427 B           & 2.8    &   130   & [M/L]\tablenotemark{a}   &   $\cdots$      &  $\gtrsim$1000  &  87~$\pm$~3 &  J12 \\
2MASS~J06475229--2523304 B          & 1.1    &    260 $\pm$ 40  & M3.0 $\pm$ 0.5 &  $\cdots$  &  $\cdots$ & $\cdots$  & TW  \\
2MASS J08540240--3051366 B           & 1.7    &    17 $\pm$ 7  & M5.0 $\pm$ 0.5  &  $\cdots$  &  $\lesssim$200?  &  $\sim$40 $\pm$ 20  &  C10  \\
2MASS~J11240434+3808108 B          & 8.3    &   170 $\pm$ 11   & M9.5 $\pm$0.5 &   UMa?      &  500 $\pm$ 100 &  81 $\pm$ 5 &  C03 \\
PYC J11519+0731 B                              & 0.5    &    18 $\pm$ 3   & M8 $\pm$ 1       &  $\cdots$   &   $>$10                &  15--90  &  TW  \\
G 180-11 B                                                & 1.6    &    45 $\pm$ 5   &  M6.0 $\pm$ 0.5  &  $\cdots$    &  $\lesssim$200   &  90--150 & M01, D07   \\
2MASS~J15594729+4403595 B          & 5.6    & 150 $\pm$ 11   &  M7.5 $\pm$ 0.5   &  $\cdots$   &  50--200    &  43 $\pm$ 9 &  J12, B15 \\
GJ 4378 Ab                                               & 0.5    &     7 $\pm$ 1     &   M8 $\pm$ 1         &  $\cdots$      &  $\gtrsim$4000   &  90 $\pm$ 3 & TW   \\
\enddata
\tablenotetext{a}{Estimated spectral type based on colors.}
\tablerefs{
B10=\citet{Bergfors:2010hm};  
B13=\citet{Bowler:2013ek}; 
B15=\citet{Bowler:2015ja}; 
C03=\citet{Close:2003ie};
C10=\citet{Chauvin:2010hm}; 
D07=\citet{Daemgen:2007fg}; 
J12=\citet{Janson:2012dc}; 
M01=\citet{McCarthy:2001gt};
R12=\citet{Rodriguez:2012ef};
TW=This work
}
\end{deluxetable}

\clearpage

\begin{deluxetable}{lcccccccccc}
\tabletypesize{\scriptsize}
\rotate
\setlength{ \tabcolsep } {.1cm} 
\tablewidth{0pt}
\tablecolumns{11}
\tablecaption{Adaptive Optics Imaging Observations\label{tab:astrometry}}
\tablehead{
      \colhead{Name} &  \colhead{Date}  &  \colhead{Epoch}   &   \colhead{Telescope/}  &  \colhead{Filter}    & \colhead{$N$ $\times$ Coadds $\times$ $t_\mathrm{exp}$} & \colhead{Separation}    &    \colhead{P.A.}   &   \colhead{$\Delta$mag}  &  \colhead{$m_\mathrm{A}$\tablenotemark{a}}  &  \colhead{$m_\mathrm{B}$\tablenotemark{b}}   \\
                                     &  \colhead{(UT)}  &  \colhead{(UT)}   &   \colhead{Instrument}     &                          &     \colhead{(s)}     &        \colhead{(mas)}    &    \colhead{($^{\circ}$)}   &   &  \colhead{(mag)}   &    \colhead{(mag)}   
        }   
\startdata
2M01225093--2439505 AB &  2013 Aug 17   &  2013.6258 & Keck~II/NIRC2  &  $Y$  & 14 $\times$ 20 $\times$ 2  &  1448 $\pm$ 3  & 216.52 $\pm$ 0.09   & 7.6 $\pm$ 0.3  & 10.61 $\pm$ 0.03  &   18.2 $\pm$  0.3   \\  
2M02155892--0929121 AB  &  2012 Aug 23   &  2012.6451 & Keck~II/NIRC2  &  $K_S$  & 20 $\times$ 100 $\times$ 0.028  &  575.5 $\pm$ 0.5  & 289.98 $\pm$ 0.03   & 1.17 $\pm$ 0.02  & 7.86 $\pm$ 0.02  &   9.03 $\pm$  0.03   \\  
                                                    &  2012 Oct 12   & 2012.7816  &  Subaru/IRCS    & $K$  &  5 $\times$ 5 $\times$ 0.5 &  575 $\pm$ 2  &  289.9 $\pm$ 0.5   &   1.16~$\pm$~0.02  & 7.84 $\pm$ 0.03\tablenotemark{c}  &   9.00 $\pm$  0.03\tablenotemark{c}  \\
                                                    &  2013 Aug 17  & 2013.6259  &  Keck~II/NIRC2  &  $K_S$  & 10 $\times$ 10 $\times$ 0.2  &  576.35 $\pm$ 0.11  & 289.351 $\pm$ 0.011 & 1.112 $\pm$ 0.006 &   7.88 $\pm$ 0.02  &   8.99 $\pm$  0.02 \\
2M02155892--0929121 AC  &  2012 Aug 23  &  2012.6451 & Keck~II/NIRC2  &  $K_S$+cor600  &  5 $\times$ 1 $\times$ 2.9  & 3430 $\pm$ 3     &  299.4 $\pm$ 0.2    &   4.1 $\pm$ 0.2  & 7.57 $\pm$ 0.02  &  11.7 $\pm$  0.2 \\
                                                    &  2012 Oct 12    & 2012.7816  &  Subaru/IRCS    & $K$       & 4 $\times$ 5 $\times$ 0.5    &  3370 $\pm$ 80 &  299.57 $\pm$ 0.17  &   4.50 $\pm$ 0.16  & 7.54 $\pm$ 0.03\tablenotemark{c}  &  12.04 $\pm$  0.16\tablenotemark{c} \\
2M02155892--0929121 Aab  & 2012 Aug 23   &  2012.6451 & Keck~II/NIRC2  &  $K_S$  & 20 $\times$ 100 $\times$ 0.028  &  42 $\pm$ 7  & 112 $\pm$ 2   & 1.2 $\pm$ 0.3      & 7.86 $\pm$ 0.08  &   9.1 $\pm$ 0.2   \\  
                                                      & 2013 Aug 17  & 2013.6259  &  Keck~II/NIRC2  &  $K_S$  & 10 $\times$ 10 $\times$ 0.2         &  42 $\pm$ 7  & 308 $\pm$ 5   & 1.17 $\pm$ 0.05 &   7.86 $\pm$ 0.03  &   9.03 $\pm$ 0.04 \\
2M02594789--0913109 AB     &  2013 Aug 17   & 2013.6259  & Keck~II/NIRC2  &  $K_S$  & 10 $\times$ 10 $\times$ 0.2   &  648.6 $\pm$ 0.3  & 320.53 $\pm$ 0.01   & 2.70 $\pm$ 0.02 &  10.22 $\pm$ 0.02  &  12.92 $\pm$  0.03 \\  
1RXS~J034231.8+121622 AB  &  2007 Dec 14  &  2007.9513 & Keck~II/NIRC2  &  $K_S$     &  10 $\times$ 30 $\times$ 1.5         & 883.0 $\pm$ 0.2           &  17.58 $\pm$ 0.09         &   3.62 $\pm$ 0.04 &  9.313 $\pm$ 0.018 &  12.93 $\pm$  0.04 \\
                                                         &  2010 Aug 29  &  2010.6585 & Gemini-S/NICI  &  $H$          &  8 $\times$ 1 $\times$ 60               & 851 $\pm$ 3                 &  18.7 $\pm$ 0.1              &   3.84 $\pm$ 0.06 &  9.580 $\pm$ 0.019 &  13.42 $\pm$  0.06 \\
2M05464932--0757427 AB  &  2013 Jan 17   & 2013.0448  & Keck~II/NIRC2  &  $K_S$  & 10 $\times$ 10 $\times$ 1.0    &  2877.3 $\pm$ 1.2  & 114.48 $\pm$ 0.03   & 3.86 $\pm$ 0.03 &   9.81 $\pm$ 0.02  &  13.67 $\pm$  0.04 \\  
2M06475229--2523304 AB  & 2012 Oct 13  &  2012.7847  &  Subaru/IRCS    & $J$  &  5 $\times$ 5 $\times$ 0.4      &  1116 $\pm$ 8  &  28.9 $\pm$ 0.5  &  5.22 $\pm$ 0.09  & 8.36 $\pm$ 0.02 &  13.58 $\pm$ 0.09 \\
                                                    & 2012 Oct 13  &  2012.7847  &  Subaru/IRCS    & $H$  &  4 $\times$ 1 $\times$ 10.0  &   1071 $\pm$ 7  &  28.7 $\pm$ 0.3  &  4.89 $\pm$ 0.12  & 7.77 $\pm$ 0.04 & 12.66 $\pm$ 0.13 \\
                                                    & 2012 Oct 13  &  2012.7847  &  Subaru/IRCS    & $K$  &  4 $\times$ 10 $\times$ 0.26 &  1089 $\pm$ 23 &  28.8 $\pm$ 0.2  &  5.1 $\pm$ 0.5     & 7.53 $\pm$ 0.02\tablenotemark{c} & 12.6 $\pm$ 0.5\tablenotemark{c} \\
                                                    & 2013 Jan 18  &  2013.0478  &  Keck~II/NIRC2   & $K$  &  5 $\times$ 100 $\times$ 0.106 &  1086.6 $\pm$ 1.2 &  28.08 $\pm$ 0.07  &  5.538 $\pm$ 0.012     & 7.53 $\pm$ 0.02\tablenotemark{c} & 13.07 $\pm$ 0.03\tablenotemark{c} \\
                                                    & 2014 Nov 8   &  2014.8531  &  Keck~II/NIRC2   & $Ks$  &  9 $\times$ 10 $\times$ 0.1    &  1088   $\pm$ 2 &  28.10 $\pm$ 0.09  &  5.33 $\pm$ 0.16     & 7.55 $\pm$ 0.02 & 12.88 $\pm$ 0.16 \\
                                                    & 2014 Nov 8   &  2014.8531  &  Keck~II/NIRC2   & $Ks$+cor600 &  10 $\times$ 1 $\times$ 30    &  1089.9   $\pm$ 0.7 &  27.80 $\pm$ 0.03  &   $\cdots$     & $\cdots$ & $\cdots$   \\
2M08540240--3051366 AB  &  2013 Jan 18   &  2013.0478 & Keck~II/NIRC2  &  $J$     &  4 $\times$ 100 $\times$ 0.053 & 1741 $\pm$ 3           &  159.08 $\pm$ 0.03    &   3.00 $\pm$ 0.07     &   9.08 $\pm$ 0.03  &  12.08 $\pm$ 0.07  \\
                                                    &  2013 Jan 18   &  2013.0478 & Keck~II/NIRC2  &  $H$     &  4 $\times$ 100 $\times$ 0.053 & 1741.2 $\pm$ 1.3   &  159.07 $\pm$ 0.08    &   2.97 $\pm$ 0.03     &   8.46 $\pm$ 0.04  &  11.43 $\pm$ 0.05  \\
                                                    &  2013 Jan 18   &  2013.0478 & Keck~II/NIRC2  &  $K$     &  4 $\times$ 100 $\times$ 0.181 & 1740.9 $\pm$ 1.8   &  159.07 $\pm$ 0.05    &   2.848 $\pm$ 0.008 &   8.142 $\pm$ 0.03\tablenotemark{c}  &  10.99 $\pm$ 0.03\tablenotemark{c}  \\
2M11240434+3808108 AB  & 2014 Jun 13    & 2014.4463 & P60/Robo-AO    & $i'$        & 1 $\times$ 1 $\times$ 60             &  8250 $\pm$ 90     &  131.2 $\pm$ 0.5         &   4.63 $\pm$ 0.05   &  12.554 $\pm$ 0.001\tablenotemark{f} & 17.18 $\pm$ 0.05\tablenotemark{f} \\
                                                    & 2014 Jun 13    & 2014.4463 & P60/Robo-AO    & $z'$        & 1 $\times$ 1 $\times$ 60             &  8270 $\pm$ 90     &  131.1 $\pm$ 0.5         &   3.90 $\pm$ 0.03   &13.985 $\pm$ 0.013\tablenotemark{f} & 17.88 $\pm$ 0.03\tablenotemark{f}  \\
PYC 11519+0731 AabB        &  2012 May 22  &  2012.3896 & Keck~II/NIRC2  &  $Y$     &  11 $\times$ 5 $\times$ 0.5         & 496 $\pm$ 3           &  110.6 $\pm$ 0.3         &   5.3 $\pm$ 0.3 &   9.31 $\pm$ 0.03\tablenotemark{d}   &  14.6  $\pm$  0.3\tablenotemark{d}  \\ 
                                                    &  2012 May 22  &  2012.3896 & Keck~II/NIRC2  &  $J$     &  20 $\times$ 100 $\times$ 0.05  & 497 $\pm$ 2            &  110.4 $\pm$ 0.2         &   5.40 $\pm$ 0.15 &   8.82 $\pm$ 0.03  &  14.22 $\pm$  0.15 \\ 
                                                    &  2012 May 22  &  2012.3896 & Keck~II/NIRC2  &  $H$     &  22 $\times$ 10 $\times$ 0.028  & 496.7 $\pm$ 1.4     &  110.4 $\pm$ 0.2        &   5.4 $\pm$ 0.2  &   8.14 $\pm$ 0.04  &  13.5  $\pm$  0.2  \\
                                                    &  2012 May 22  &  2012.3896 & Keck~II/NIRC2  &  $K_S$ &  17 $\times$ 100 $\times$ 0.028 & 497.1 $\pm$ 1.1   &  110.46 $\pm$ 0.09    &   5.23 $\pm$ 0.09 &   7.90 $\pm$ 0.03  &  13.13 $\pm$  0.10 \\
                                                    &  2012 May 22  &  2012.3896 & Keck~II/NIRC2  &  $L'$      &  10 $\times$ 10 $\times$ 0.11     & 500 $\pm$ 3          &  110.4 $\pm$ 0.2        &   4.97 $\pm$ 0.13 &   7.80 $\pm$ 0.11\tablenotemark{e}   &  12.77 $\pm$  0.17\tablenotemark{e}  \\
                                                    &  2012 Jun 24   &  2012.4798 & Keck~II/NIRC2  &  $H$      &  20 $\times$ 100 $\times$ 0.028 & 498.2 $\pm$ 2.3  &  110.21 $\pm$ 0.15    &   5.18 $\pm$ 0.09 &   8.14 $\pm$ 0.04  &  13.32 $\pm$  0.10 \\
                                                    &  2012 Jun 24   &  2012.4798 & Keck~II/NIRC2  &  $K_S$  &  19 $\times$ 100 $\times$ 0.028 & 499.2 $\pm$ 0.5  &  110.30 $\pm$ 0.10    &   5.08 $\pm$ 0.12 &   7.90 $\pm$ 0.05  &  12.98 $\pm$  0.13 \\
                                                    &  2013 Jan 18   &  2013.0479 & Keck~II/NIRC2  &  $H$+cor600  &  20 $\times$ 1 $\times$ 30 & 500.9 $\pm$ 1.8  &  108.76 $\pm$ 0.05    &           $\cdots$   & $\cdots$ & $\cdots$   \\
                                                    &  2013 Apr 19  &  2013.2957 & Gemini-S/NICI  &  $H$          &  10 $\times$ 10 $\times$ 1.14      & 517 $\pm$ 10       &  108.3 $\pm$ 0.5        &   6.0 $\pm$ 0.2  &  8.14 $\pm$ 0.04 &  14.1 $\pm$  0.2 \\
                                                    &  2013 Apr 19  &  2013.2957 & Gemini-S/NICI  &  $K_S$      &  10 $\times$ 10 $\times$ 1.14      & 511 $\pm$ 10       &  107.9 $\pm$ 0.5        &   5.4 $\pm$ 0.2  &  7.89 $\pm$ 0.04 &  13.3 $\pm$  0.2 \\
                                                    &  2013 Dec 18   &  2013.9628 & Keck~II/NIRC2  &  $K_S$+cor600 & 9 $\times$ 2 $\times$ 10 & 514 $\pm$ 2        &  107.15 $\pm$ 0.04    &           $\cdots$  & $\cdots$ & $\cdots$     \\
G 180-11 AB                             &  2013 Feb 04  &  2013.0949 & Keck~II/NIRC2  &  $Y$     &  10 $\times$ 10 $\times$ 0.5  & 1620.4 $\pm$ 1.3     &  254.98 $\pm$ 0.03    &   2.089 $\pm$ 0.008  &   9.60 $\pm$ 0.02\tablenotemark{d}   &  11.70 $\pm$  0.02\tablenotemark{d}  \\
                                                    &  2013 Feb 04  &  2013.0949 & Keck~II/NIRC2  &  $J$     &  10 $\times$ 10 $\times$ 0.2   & 1619.9 $\pm$ 1.2     &  254.92 $\pm$ 0.05    &   2.063 $\pm$ 0.010 &   9.08 $\pm$ 0.02  &  11.14 $\pm$  0.02 \\
                                                    &  2013 Feb 04  &  2013.0949 & Keck~II/NIRC2  &  $H$     &  19 $\times$ 10 $\times$ 0.028 & 1619.6 $\pm$ 1.4    &  254.97 $\pm$ 0.03    &   2.08 $\pm$ 0.02   &   8.41 $\pm$ 0.03  &  10.49 $\pm$  0.04 \\
                                                    &  2013 Feb 04  &  2013.0949 & Keck~II/NIRC2  &  $K_S$   &  15 $\times$ 10 $\times$ 0.028 & 1620.5 $\pm$ 1.3  &  254.93 $\pm$ 0.04    &   2.06 $\pm$ 0.03  &   8.19 $\pm$ 0.03  &  10.25 $\pm$  0.04 \\
                                                    & 2014 Jun 13    & 2014.4463 & P60/Robo-AO    & $r'$        & 1 $\times$ 1 $\times$ 60             &   1610 $\pm$ 50     &  253.2 $\pm$ 1.6         &   2.18 $\pm$ 0.03   &  13.544 $\pm$ 0.004\tablenotemark{f}  &  15.724 $\pm$ 0.027\tablenotemark{f} \\
                                                    & 2014 Jun 13    & 2014.4463 & P60/Robo-AO    & $i'$        & 1 $\times$ 1 $\times$ 60             &   1640 $\pm$ 50     &  253.9 $\pm$ 1.6         &   2.09 $\pm$ 0.02   &  11.749 $\pm$ 0.003\tablenotemark{f}  &  13.839 $\pm$ 0.017\tablenotemark{f} \\
                                                    & 2014 Jun 13    & 2014.4463 & P60/Robo-AO    & $z'$        & 1 $\times$ 1 $\times$ 60             &  1630 $\pm$ 50     &  253.8 $\pm$ 1.6         &   2.03 $\pm$ 0.02   &  11.142 $\pm$ 0.005\tablenotemark{f}  &  13.172 $\pm$ 0.018\tablenotemark{f}\\
2M15594729+4403595~AB & 2014 Jun 13 & 2014.4463 & P60/Robo-AO & $i'$           & 1 $\times$ 1 $\times$ 60                & 5670 $\pm$ 70             &  284.4 $\pm$ 0.6           &   6.37 $\pm$ 0.07  & 10.433 $\pm$ 0.001\tablenotemark{f} & 16.80 $\pm$ 0.07\tablenotemark{f} \\
                                                    & 2014 Jun 13 & 2014.4463 & P60/Robo-AO & $z'$           & 1 $\times$ 1 $\times$ 60                &  $\cdots$                       &  $\cdots$                          & 5.55 $\pm$ 0.05   & 10.147 $\pm$ 0.001\tablenotemark{f} & 15.70 $\pm$ 0.05\tablenotemark{f} \\
GJ 4378 Aab                            &  2012 Oct 13   & 2012.7842  &  Subaru/IRCS    & $J$  &  4 $\times$ 1 $\times$ 30.0       &  473.4 $\pm$ 1.2  &  348.20 $\pm$ 0.18   &   2.907~$\pm$~0.011 &   8.71 $\pm$ 0.02  &  11.62 $\pm$  0.02 \\
                                                    &  2012 Oct 13   & 2012.7842  &  Subaru/IRCS    & $H$  &  5 $\times$ 1 $\times$ 8.0       &  475.6 $\pm$ 1.8  &  348.3 $\pm$ 0.16   &   3.14~$\pm$~0.07 &   8.13 $\pm$ 0.03  &  11.27 $\pm$  0.07 \\
                                                    &  2012 Oct 13   & 2012.7842  &  Subaru/IRCS    & $K$  &  5 $\times$ 30 $\times$ 0.145 &  475.3 $\pm$ 1.7  &  348.4 $\pm$ 0.10   &   3.02~$\pm$~0.03 &   7.93 $\pm$ 0.05\tablenotemark{c}   &  10.95 $\pm$  0.06\tablenotemark{c}  \\
                                                    &  2013 Jan 18  & 2013.0471  &  Keck~II/NIRC2  &  $K$  & 13 $\times$ 10 $\times$ 0.053  &  492.8 $\pm$ 1.5  & 347.85 $\pm$ 0.13 & 3.06 $\pm$ 0.03  &   7.92 $\pm$ 0.05\tablenotemark{c}   &  10.98 $\pm$  0.06\tablenotemark{c}  \\
                                                    &  2013 Aug 17  & 2013.6258  &  Keck~II/NIRC2  &  $Y$    & 10 $\times$ 5 $\times$ 0.5  &  526.7 $\pm$ 0.4  & 347.22 $\pm$ 0.03 & 3.353 $\pm$ 0.014  &   9.21 $\pm$ 0.02\tablenotemark{d}   &  12.56 $\pm$  0.03\tablenotemark{d}  \\
                                                    &  2013 Aug 17  & 2013.6258  &  Keck~II/NIRC2  &  $J$    & 10 $\times$ 10 $\times$ 0.05  &  526.0 $\pm$ 0.3  & 347.22 $\pm$ 0.03 & 3.11 $\pm$ 0.03  &   8.70 $\pm$ 0.02  &  11.81 $\pm$  0.04 \\
                                                    &  2013 Aug 17  & 2013.6258  &  Keck~II/NIRC2  &  $H$    & 10 $\times$ 10 $\times$ 0.05  &  526.1 $\pm$ 0.2  & 347.18 $\pm$ 0.03 & 3.107 $\pm$ 0.019  &   8.13 $\pm$ 0.03  &  11.24 $\pm$  0.03 \\
                                                    &  2013 Aug 17  & 2013.6258  &  Keck~II/NIRC2  &  $K_S$  & 15 $\times$ 10 $\times$ 0.1  &  525.96 $\pm$ 0.16  & 347.15 $\pm$ 0.03 & 3.066 $\pm$ 0.008  & 7.87 $\pm$ 0.03  &  10.94 $\pm$  0.03 \\
                                                    &  2013 Aug 17  & 2013.6258  &  Keck~II/NIRC2  &  $K$    & 10 $\times$ 10 $\times$ 0.1  &  526.06 $\pm$ 0.08  & 347.194 $\pm$ 0.018 & 3.046 $\pm$ 0.018 &   7.93 $\pm$ 0.05\tablenotemark{c}   &  10.97 $\pm$  0.05\tablenotemark{c}   \\
                                                    &  2013 Aug 17  & 2013.6258  &  Keck~II/NIRC2  &  $L'$    & 20 $\times$ 10 $\times$ 0.114  &  525.9 $\pm$ 0.6  & 347.23 $\pm$ 0.04 & 2.819 $\pm$ 0.011 &   7.58 $\pm$ 0.10\tablenotemark{e}   &  10.40 $\pm$  0.11\tablenotemark{e}   \\
                                                    &  2014 Aug 04  & 2014.5901  &  Keck~II/NIRC2  &  $K_S$  & 20 $\times$ 10 $\times$ 0.2  &  577.2 $\pm$ 0.4  & 346.2 $\pm$ 0.02 & 3.06 $\pm$ 0.03 &   7.87 $\pm$ 0.03  &  10.93 $\pm$  0.04  
\enddata
\tablenotetext{a}{Magnitude of the primary ($m_\mathrm{A}$) decomposed from the integrated-light magnitude of the A and B 
components ($m_\mathrm{AB}$) as follows: $m_\mathrm{A}$ = $m_\mathrm{AB}$ + 2.5 log(1 + 10$^{-\Delta \mathrm{mag}/2.5})$.  Uncertainties are derived in a Monte Carlo fashion.
Unless otherwise noted, $J$, $H$, and $K_S$ magnitudes are from 2MASS (\citealt{Skrutskie:2006hl}).}
  \tablenotetext{b}{Magnitude of the companion ($m_\mathrm{B}$) is computed from $\Delta \mathrm{mag}$ and $m_\mathrm{A}$.}
  \tablenotetext{c}{Integrated-light $K_S$-band magnitude is converted to  $K_\mathrm{MKO}$ using relations from \citet{Leggett:2006gg}.}
  \tablenotetext{d}{$Y$-band integrated-light magnitude is derived from the typical $Y-J$ color of dwarfs for the system spectral type from \citet{Rayner:2009ki} and the integrated-light $J$-band magnitude.}
  \tablenotetext{e}{$L'$-band integrated-light magnitude is derived from the typical $K_S-L'$ color of dwarfs for the system spectral type from \citet{Golimowski:2004en} and the integrated-light $K_S$-band magnitude.}
  \tablenotetext{f}{Unresolved $r'$, $i'$, and $z'$ magnitudes are from the Sloan Digital Sky Survey DR9 (\citealt{Ahn:2012ih}).  Note that optical variability of the primary and/or companion 
  can affect the decomposed photometry of one or both components.}
\end{deluxetable}
\clearpage

\begin{deluxetable}{lcccccccc}
\tabletypesize{\scriptsize}
\rotate
\setlength{ \tabcolsep } {.1cm} 
\tablewidth{0pt}
\tablecolumns{9}
\tablecaption{Spectroscopic Observations\label{tab:specobs}}
\tablehead{
        \colhead{Object}   &  \colhead{Date}      &   \colhead{Telescope/}  &  \colhead{Filter}  &  \colhead{Slit Width}  & \colhead{Plate Scale}   &    \colhead{Exp. Time}   &   \colhead{Resolution} & \colhead{Standard\tablenotemark{a}}   \\
                     \colhead{}  &  \colhead{(UT)}   &   \colhead{Instrument}       &  \colhead{}           &  \colhead{($''$)}               & \colhead{(mas pix$^{-1}$)}             &    \colhead{(min)}   & \colhead{(=$\Delta \lambda$/$\lambda$)}  &   \colhead{}             
        }   
\startdata
2MASS~J01225093--2439505 A    &    2014 Jun 28     &   SOAR/Goodman  &  $GG 455$ &   0.46  &  $\cdots$   &    5  & 1800  &  LTT 6248  \\
2MASS~J02155892--0929121 A    &    2012 Dec 28    &    Keck~I/HIRES       &  $\cdots$    &   0.86  &  $\cdots$  &    1   &  45000 &  GJ 908  \\ 
                                                               &    2014 Jun 28     &   SOAR/Goodman  &  $GG 455$ &   0.46  &  $\cdots$   &    4  & 1800  &  LTT 6248  \\
                                                               &    2014 Jun 28     &   SOAR/Goodman  &  $GG 455$ &   0.46  &  $\cdots$   &    5  & 5900  &  LTT 6248  \\
                                                               &    2014 Dec 08    &   CTIO 1.5-m/CHIRON  &  $\cdots$  &  $\cdots$  &  $\cdots$  &  45  &  28000  &  HD 42581 \\
2MASS~J02155892--0929121 C   &    2014 Jan 18     &   IRTF/SpeX-SXD    &  $\cdots$    &   0.5     &  150          &   12  & 1200    &  HD 13936  \\
2MASS~J02594789--0913109 A   &    2012 Dec 28    &    Keck~I/HIRES       &  $\cdots$    &   0.86  &  $\cdots$  &    5   &  45000  &  GJ 908  \\ 
HD 23514 B                                         &    2013 Feb 3      &   Keck~I/OSIRIS    &  $Hbb$          &   $\cdots$  &  50    &   45   & 3800   &  HD 23258  \\
                                                               &    2013 Feb 3      &   Keck~I/OSIRIS    &  $Kbb$           &   $\cdots$  &  50    &    42   & 3800  &   HD 23258 \\
G 160-54 A                                          &    2007 Dec  19   &  CFHT/ESPaDOnS  &  $\cdots$    & 0.53      & $\cdots$ & 16.7 & 68000 & GJ 273    \\ 
                                                               &    2012 Jan 29    &   UH 2.2-m/SNIFS &  $\cdots$        &   $\cdots$  &  400  &   7   & 1300  & HR 1544, G191B2B  \\
G 160-54 C                                          &    2012 Sep 24    &   IRTF/SpeX-SXD &  $\cdots$      &   0.5        &  150  &   24   & 1200  & HD 25792  \\
2MASS~J06475229--2523304 A   &     2013 Apr 30    &  du Pont/Echelle   &  $\cdots$      &  0.75      &  $\cdots$  & 10    & 45000  & GJ 273, GJ 433 \\  
                                                               &    2014 Dec 3      &  UH 2.2-m/SNIFS &  $\cdots$ &   $\cdots$ &  400  &   3.3   & 1300  & Feige 67  \\
                                                               &    2014 Dec 10    &   CTIO 1.5-m/CHIRON  &  $\cdots$  &  $\cdots$  &  $\cdots$  &  15  &  28000  &  HD 217357 \\
2MASS~J06475229--2523304 B   &    2014 Dec 8      &   Keck~I/OSIRIS    &  $Kbb$    & $\cdots$ &  20    &   18  & 3800 &  HD 58886  \\
2MASS J08540240--3051366 A     &     2009 Jun 13    &  du Pont/Echelle   &  $\cdots$      &  0.75      &  $\cdots$  & 15    & 45000  & GJ 908 \\ 
2MASS J08540240--3051366 B     &    2014 Dec 8      &   Keck~I/OSIRIS    &  $Hbb$    & $\cdots$ &  20    &   10  & 3800 &  HD 58886  \\
                                                               &    2014 Dec 8      &   Keck~I/OSIRIS    &  $Kbb$    & $\cdots$ &  20    &   16  & 3800 &  HD 58886  \\
2MASS~J11240434+3808108 B    &    2011 Jan 22     &   IRTF/SpeX-SXD &  $\cdots$ &   0.5        &  150  &   8   & 1200  &  HD 105388  \\
%2MASS~J11240434+3808108 A    &    2011 Jan 22     &   IRTF/SpeX-SXD &  $\cdots$ &   0.3xx    &  150  &        & HD 105388  \\
PYC J11519+0731 Aab                    &    2012 Jun 1       &   IRTF/SpeX-SXD &  $\cdots$ &   0.8        &  150  &   3    &  750  &  HD 97585  \\
                                                               &    2012 Jul 5        &   Keck~I/HIRES     &  $\cdots$ &   0.9         &  $\cdots$  &   3.6    & 4500 & $\cdots$  \\
                                                               &    2013 Feb 22    &   ESO-MPG 2.2-m/FEROS   &  $\cdots$ &   $\cdots$ &  $\cdots$  &  20  &  48000    & $\cdots$  \\
                                                               &    2014 Jun 9      &   CFHT/ESPaDOnS  &  $\cdots$ &   0.53         &  $\cdots$  & 6.7  &  68000    & GJ 625  \\
                                                               &    2014 May 22   &   Mayall/RC-Spectrograph     &  $GG 495$ &   1.5   &  690  &  4  &   2500    & HZ 44  \\
PYC J11519+0731 B                         &    2012 Jun 25    &   Keck~I/OSIRIS &  $Kbb$ &   $\cdots$  &  20    &   40   & 3800   &  HD 116960  \\
G 180-11 B                                          &    2013 Feb 1      &   Keck~I/OSIRIS    &  $Jbb$     & $\cdots$ &  50    &  6   & 3800 & q Her  \\
                                                               &    2013 Feb 1      &   Keck~I/OSIRIS    &  $Hbb$    & $\cdots$ &  50    &  6   & 3800 & q Her  \\
                                                               &    2013 Feb 1      &   Keck~I/OSIRIS    &  $Kbb$    & $\cdots$ &  50    &   6  & 3800 &  q Her  \\
2MASS~J15594729+4403595 A    &    2012 Feb 12    &  CFHT/ESPaDOnS  &  $\cdots$    & 0.53      & $\cdots$ & 5     & 68000 & GJ 628    \\ 
                                                               &    2012 Aug 23    &   UH 2.2-m/SNIFS &  $\cdots$ &   $\cdots$ &  400  &   1.7   & 1300  & HR5501, EG131  \\
2MASS~J15594729+4403595 B    &    2012 Sept 9     &   Keck~II/ESI          &  $\cdots$ &     1.0      &  150  & 40 &  4000   & G191B2B, Feige 67  \\
GJ 4378 A                                            &    2009 Aug 20    &  du Pont/Echelle   &  $\cdots$      &  0.75      &  $\cdots$  & 10    & 45000  & GJ 908, GJ 699 \\ 
                                                               &    2014 Jun 28     &   SOAR/Goodman  &  $GG 455$ &   0.46  &  $\cdots$   &    4  & 1800  &  LTT 6248  \\
                                                               &    2014 Jun 28     &   SOAR/Goodman  &  $GG 455$ &   0.46  &  $\cdots$   &    5  & 5900  &  LTT 6248  \\
GJ 4378 Ab                                          &    2013 Jul 31     &   Keck~I/OSIRIS    &  $Hbb$ &   $\cdots$  &  50    &    28.5  & 3800  &  HD 219833  \\
                                                               &    2013 Jul 31     &   Keck~I/OSIRIS    &  $Kbb$ &   $\cdots$  &  50    &    11    & 3800  &  HD 219833  \\
GJ 4379 B                                            &    2009 Aug 20   &  du Pont/Echelle   &  $\cdots$      &  0.75      &  $\cdots$  & 10    & 45000  & GJ 908, GJ 699 \\ 
                                                               &    2014 Jun 28     &   SOAR/Goodman  &  $GG 455$ &   0.46  &  $\cdots$   &    4  & 1800  &  LTT 6248  \\
                                                               &    2014 Jun 28     &   SOAR/Goodman  &  $GG 455$ &   0.46  &  $\cdots$   &    5  & 5900  &  LTT 6248  \\
\enddata
\tablenotetext{a}{For our high-resolution optical spectra, this refers to the RV standard.  For our near-infrared and low-resolution optical spectra, this
refers to telluric standards.}
\end{deluxetable}

\clearpage

\begin{deluxetable}{lcccccc}
\tabletypesize{\scriptsize}
\setlength{ \tabcolsep } {.12cm} 
\tablewidth{0pt}
\tablecolumns{7}
\tablecaption{H$\alpha$ Emission and Optical Spectral Classification\label{tab:optspec}}
\tablehead{
   \colhead{Name} &  \colhead{H$\alpha$ $EW$\tablenotemark{a}}  & \colhead{TiO5}  & \colhead{CaH1}  & \colhead{CaH2}   & \colhead{CaH3}   & \colhead{Adopted SpT} \\
   &  \colhead{(\AA)} & \colhead{Index} & \colhead{Index} &  \colhead{Index}  &  \colhead{Index} &  \colhead{($\pm$0.5)} 
        }
\startdata
2MASS~J01225093--2439505 A      &    --5.0  &  0.417  &   0.783  &     0.424  &   0.677 &  M4.0  \\  %M4   M4
2MASS~J02155892--0929121 Aab  &     --5.6 &  0.482  &    0.810  &     0.485 &    0.716 & M3.5  \\ %  M3   M3
G~160-54~Aab                                      &    --0.8   &  0.413  &   0.820   &    0.416  &   0.701 &   M4.5 \\  %  M4   M4
2MASS~J06475229--2523304 Aab   &    --0.3  &  0.890  &   1.009  &     0.950  &   0.952 &  K7.0 \\  %M0   K5
PYC J11519+0731 Aab                        &   --2.8   &  0.594   &  0.866   &    0.612  &   0.814 & M2.0   \\  %  M2   M4
2MASS~J15594729+4403595 A       &   --2.3    &  0.685  &   0.877  &     0.648  &   0.821 &  M1.5  \\  % M2   M2
2MASS~J15594729+4403595 B       &    --6.4  &  0.195  &   0.828    &   0.303   &  0.674  &   M7.5  \\  % M7   M7
GJ~4378~Aa12                                     &     --4.5 &   0.393 &    0.769  &     0.408 &    0.670 & M4.5  \\ %  M4   M4
GJ~4379~B                                            &    0.1     &  0.457  &   0.801   &    0.459  &   0.725  &  M3.5  \\ % M3   M3
\enddata
\tablenotetext{a}{Negative values indicate emission.}
\end{deluxetable}
\clearpage

\begin{deluxetable}{lcccc}
\tabletypesize{\scriptsize}
\setlength{ \tabcolsep } {.12cm} 
\tablewidth{0pt}
\tablecolumns{5}
\tablecaption{Radial Velocities\label{tab:rvs}}
\tablehead{
   \colhead{Name} &  \colhead{Component}  & \colhead{UT Date}  & \colhead{RV}   & \colhead{Reference} \\
   \colhead{}    &        &  \colhead{(YYYY-MM-DD)}  & \colhead{km s$^{-1}$}    & \colhead{}    
        }
\startdata
2MASS~J01225093--2439505      &   A    & 2010-09-24  &  11.4~$\pm$~0.2  &  M14 \\
                                                              &  A   &  2011-08-24  &  12.3~$\pm$~2.5  &  K13 \\
                                                              & A   & 2012-12-28  &  9.6~$\pm$~0.7   &  B13 \\
2MASS~J02155892--0929121    &  Aa    & 2007-11-08  &  0.6~$\pm$~5.0   &  K13  \\
                                                            &  Aa   &  2010-07-26   &  0.5~$\pm$~0.3  &  M14 \\
                                                            &  Aa   &  2010-09-19   &  --0.6~$\pm$~0.3  &  M14 \\
                                                            &  Aa   &  2012-02-11   &  8.3~$\pm$~0.3  &  M14 \\ 
                                                            & Aa    & 2012-07-18   &  10.1~$\pm$~0.6  &  K14  \\
                                                            &  Aa   & 2012-12-28   &   6.2~$\pm$~0.4    &  TW \\
                                                            &  Aa   & 2014-12-08    &   8.3~$\pm$~0.6    &  TW  \\
2MASS~J02594789--0913109    &  A      &  2012-12-28  &   4.9~$\pm$~0.5    &  TW \\
1RXS~J034231.8+121622           &   A      &  2005-12-21  &   35.4~$\pm$~0.4  &  S12  \\
2MASS~J06475229--2523304  & Aa       &  $\cdots$         &  --56.5                     &   T06  \\
                                                          & Aa       &  2009-02-28   &  --67 $\pm$ 6        &   K13  \\
                                                          & Aa       &  2011-09-14   &  --24.4 $\pm$ 0.8   &   M14  \\
                                                          & Aa       &   2013-04-30  &  --2.1~$\pm$~0.9  & TW \\
                                                          &  Aa      & 2014-12-10    &  --5.5~$\pm$~0.8    &  TW  \\
2MASS J08540240--3051366   &   A        &  2009-06-13    &   44.5~$\pm$~0.6    &  TW  \\
2MASS~J11240434+3808108 &   A        &  2006-05-11    & --11.5 $\pm$ 0.5   &  S12   \\
                                                      &     B         &  $\cdots$       &  --14 $\pm$ 3    &  R09 \\
PYC J11519+0731                   &    Aa       &  2012-07-05  &   50.9~$\pm$~1.0    &  TW \\   % HIRES
                                                      &     Aa     &  2013-02-22  &  14.7~$\pm$~1.0     &  TW \\  % FEROS
                                                      &      Aa      &  2014-06-09  &   69.5~$\pm$~1.0    &  TW \\  % Espadons
                                                      &     Ab       &   2012-07-05   &   --51.9~$\pm$~1.0    &  TW  \\  % HIRES
                                                      &      Ab      &   2013-02-22   &    --18.2~$\pm$~1.0   &  TW  \\  % FEROS
                                                      &       Ab     &   2014-06-09   &    --75.5~$\pm$~1.0   &  TW  \\  % Espadons
                                                      &  $\gamma$  &    $\cdots$   &  --0.3~$\pm$~1.0  &  TW  \\
G 180-11                                       &    A       & 2006-08-13    &   --15.5~$\pm$~0.7    &  S12   \\
2MASS~J15594729+4403595  &   A       & 2013-03-03  &  --15.8~$\pm$~0.5    &  M14  \\
                                                         &   A        & 2012-02-12   &   --19.6~$\pm$~0.6    &  TW  \\
GJ 4378 A                                     &   a1       & 2009-08-20   &  --7.0~$\pm$~0.8   &  TW  \\
                                                        &   a2       &  2009-08-20   &  49.2~$\pm$~1.1     &  TW  \\
GJ 4379 B                                     &   $\cdots$  & 2009-08-20   &  20.6~$\pm$~0.4         &  TW 
\enddata
\tablerefs{
K13 = \citet{Kordopatis:2013cd};
K14 = \citet{Kraus:2014ur}; 
M14 = \citet{Malo:2014dk}; 
R09 = \citet{Reiners:2009kd}; 
S12 = \citet{Shkolnik:2012cs};
T06 = \citet{Torres:2006bw};
TW = this work.
}
\end{deluxetable}
\clearpage

\begin{deluxetable}{lccccc}
\tabletypesize{\scriptsize}
\setlength{ \tabcolsep } {.12cm} 
\tablewidth{0pt}
\tablecolumns{6}
\tablecaption{1RXS~J034231.8+121622 AB Astrometry\label{tab:rxs0342_ast}}
\tablehead{
   \colhead{Epoch} &  \colhead{Separation}  & \colhead{PA}  & \colhead{Ref}   \\
   & \colhead{(mas)} &  \colhead{($^{\circ}$)}
        }
\startdata
2007.95  &  883.0 $\pm$ 0.2  &  17.58 $\pm$ 0.09  &  TW  \\
2008.63  &  860 $\pm$ 8  &  17.3 $\pm$ 0.4  &  J12 \\
2008.87  &  866 $\pm$ 8  &  17.8 $\pm$ 0.4  &  J12 \\     
2010.66  &  851 $\pm$ 3  &  18.7 $\pm$ 0.1  & TW  \\
2012.02  &  834 $\pm$ 57  &  17.6 $\pm$ 1.7  &  J14 \\
2012.65  &  831 $\pm$ 2  &  18.71 $\pm$ 0.07  &  B15  \\
2013.04  &  822 $\pm$ 8  &  19.1 $\pm$ 0.7  &  B15  
\enddata
\tablerefs{
B15=\citet{Bowler:2015ja};
J12=\citet{Janson:2012dc}; 
J14=\citet{Janson:2014gz};
TW=this work.
}
%\tablenotetext{a}{\citet{Janson:2014gz} list slightly different, but formally consistent astrometry for this epoch compared to \citet{Janson:2012dc}.  %Here we adopt the original values from \citet{Janson:2012dc}.}
\end{deluxetable}
\clearpage

\begin{deluxetable}{lccc}
\tabletypesize{\scriptsize}
\setlength{ \tabcolsep } {.1cm} 
\tablewidth{0pt}
\tablecolumns{4}
\tablecaption{Allers \& Liu 2013 Gravity Indices\label{tab:grav}}
\tablehead{
   \colhead{Parameter} & \colhead{2MASS~J02155892--0929121 C}  &  \colhead{G~160-54 C}  &  \colhead{2MASS~J11240434+3808108  B} 
   }
  \startdata
  
  VO$_z$ Index  &1.023 $\pm$ 0.011 & 1.042 $\pm$  0.007  &  1.069  $\pm$  0.011  \\
  FeH$_z$ Index &  1.082 $\pm$ 0.012  &   1.109 $\pm$  0.010  &1.210 $\pm$ 0.015 \\
  K I$_J$ Index & 1.030 $\pm$ 0.006 &   1.072 $\pm$  0.005 & 1.111 $\pm$  0.007 \\
  $H$-cont Index &1.005 $\pm$ 0.006  &  0.968 $\pm$  0.005  & 0.931 $\pm$ 0.007	 \\
  FeH$_J$ Index & 1.066 $\pm$  0.010  &    1.113 $\pm$ 0.010 & 1.202 $\pm$ 0.019   \\
  $EW$(Na I 1.138 $\mu$m) (\AA) & 6.9 $\pm$ 0.7 &   12.1 $\pm$ 0.2 &   $\cdots$  \\
  $EW$(K I 1.169 $\mu$m) (\AA) & 2.7 $\pm$ 0.5 & 4.3 $\pm$ 0.4  &  5.5 $\pm$ 0.5  \\
  $EW$(K I 1.177 $\mu$m) (\AA) &  4.0 $\pm$ 0.5 & 6.6 $\pm$  0.4 &  8.8 $\pm$ 0.5  \\
  $EW$(K I 1.253 $\mu$m) (\AA) &  2.4 $\pm$ 0.4  &  4.5  $\pm$   0.4 &5.4 $\pm$ 0.4  \\
  Gravity Score\tablenotemark{a} & 1n12 &   0n00 &  0n00  \\
  Gravity Class &  INT-G &   FLD-G  & FLD-G  
\enddata
\tablenotetext{a}{Gravity scores are for the FeH, VO, alkali lines, and $H$-band continuum shape.  See \citet{Allers:2013hk} for details.}
\end{deluxetable}
\clearpage

% Figure 1

\begin{figure}
  \vskip 0.in
  \begin{center}
  \resizebox{5.5in}{!}{\includegraphics{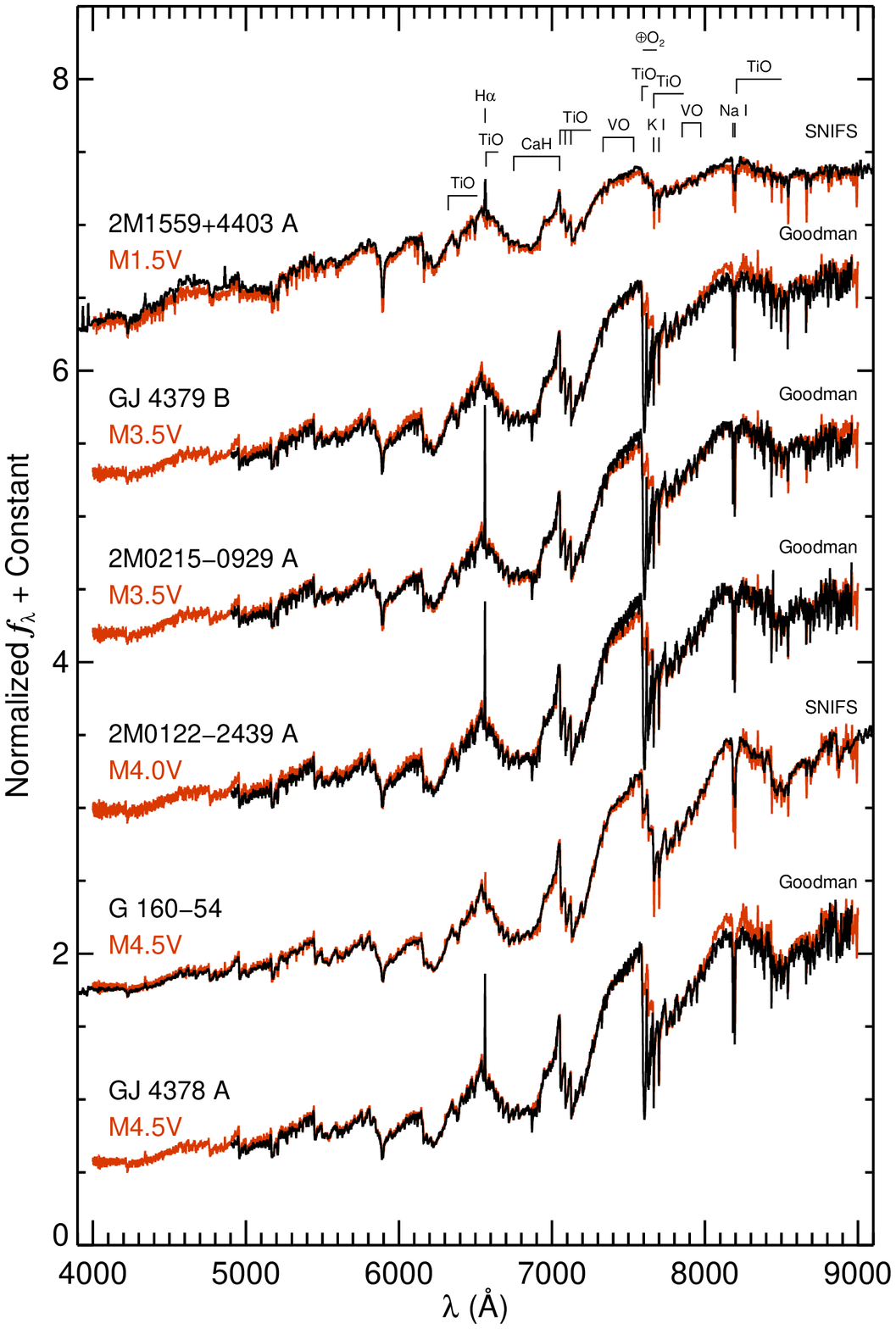}}
  \caption{Low-resolution SOAR/Goodman ($R$$\sim$1800) and UH2.2m/SNIFS ($R$$\sim$1300) optical spectra of 
  host stars to ultracool companions.  Red templates from \citet{Bochanski:2007it} show the best matches based on visual
  and index-based classification with \texttt{Hammer} (Table~\ref{tab:optspec}; \citealt{Covey:2007bj}).    Half-integer subtypes are
  created by averaging two templates together.  All spectra are normalized between 7200--7400~\AA \ and offset by a constant.  
  Although these spectra originate from instruments with different resolving powers and wavelength spans, 
  our spectral types agree with those in the literature (to within $\pm$0.5 subtypes) for nearly all stars 
  which have been previously classified.  \label{fig:pri_optspec} } 
\end{center}
\end{figure}

\clearpage
\newpage

% Figure 2

\begin{figure}
  \vskip -3.in
  \hskip -1 in
  %\begin{center}
  \resizebox{11in}{!}{\includegraphics{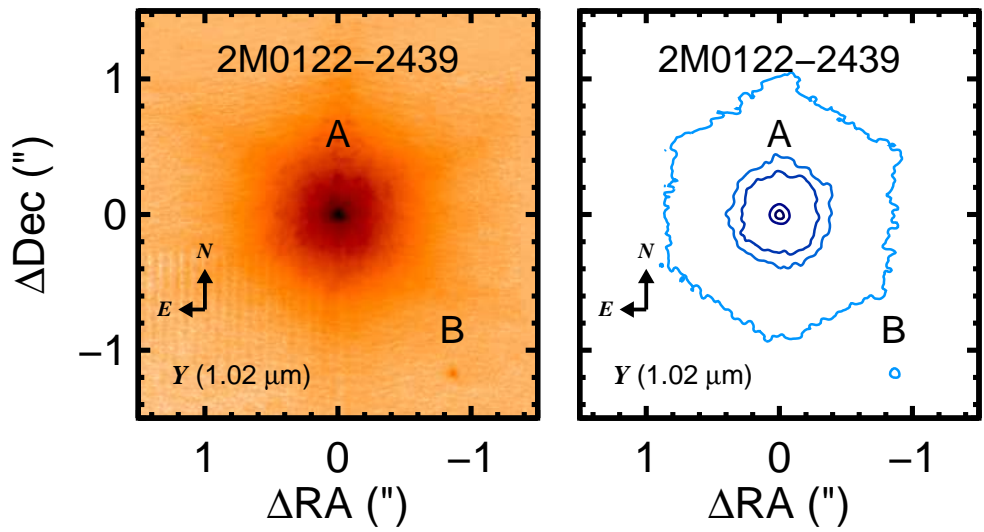}}
  \vskip -10 in
  \caption{Keck/NIRC2 $Y$-band image of 2MASS~J01225093--2439505~B.  Contours represent 0.03\%, 0.5\%, 1\%, 10\%, and 50\% of the peak flux after convolution with
  a Gaussian kernel with a FWHM equal to that of the image PSF.  North is up and east is left.   \label{fig:twom0122y} } 
%\end{center}
\end{figure}

\clearpage
\newpage

% Figure 3

\begin{figure}
  \vskip -2in
  \hskip -.2 in
  \resizebox{7in}{!}{\includegraphics{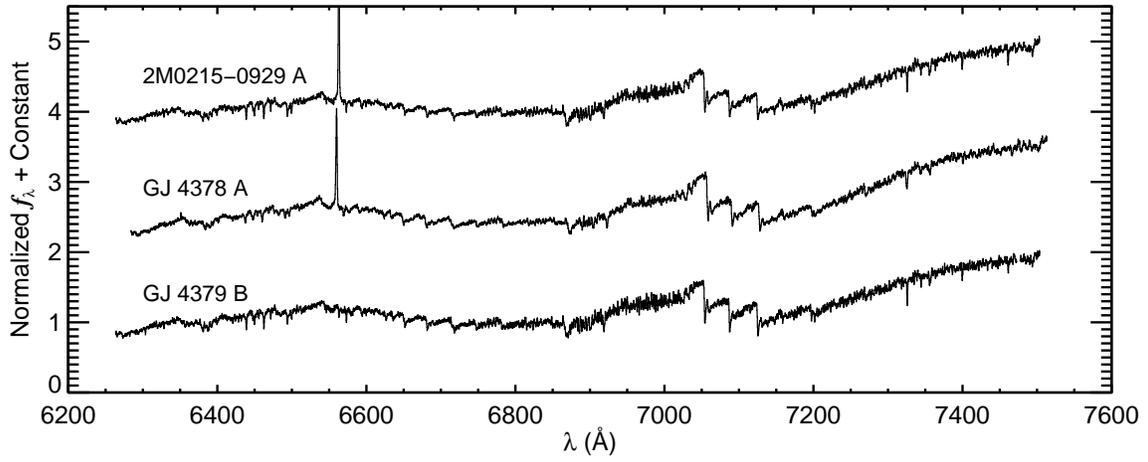}}
  \vskip -1.2in
  \caption{Moderate-resolution ($R$$\sim$5900) spectra  of 2MASS J02155892--0929121~A, GJ~4378~A, and 
  GJ~4379~B from the SOAR/Goodman spectrograph.  None of the targets show signs of \ion{Li}{1}~$\lambda$6708 
  absorption.  All spectra are normalized between 7200--7400~\AA \ and offset by a constant.     \label{fig:pri_medresspec} } 
\end{figure}

\clearpage
\newpage

% Figure 4

\begin{figure}
  \vskip -4.in
  \hskip -1.3in
  \resizebox{9in}{!}{\includegraphics{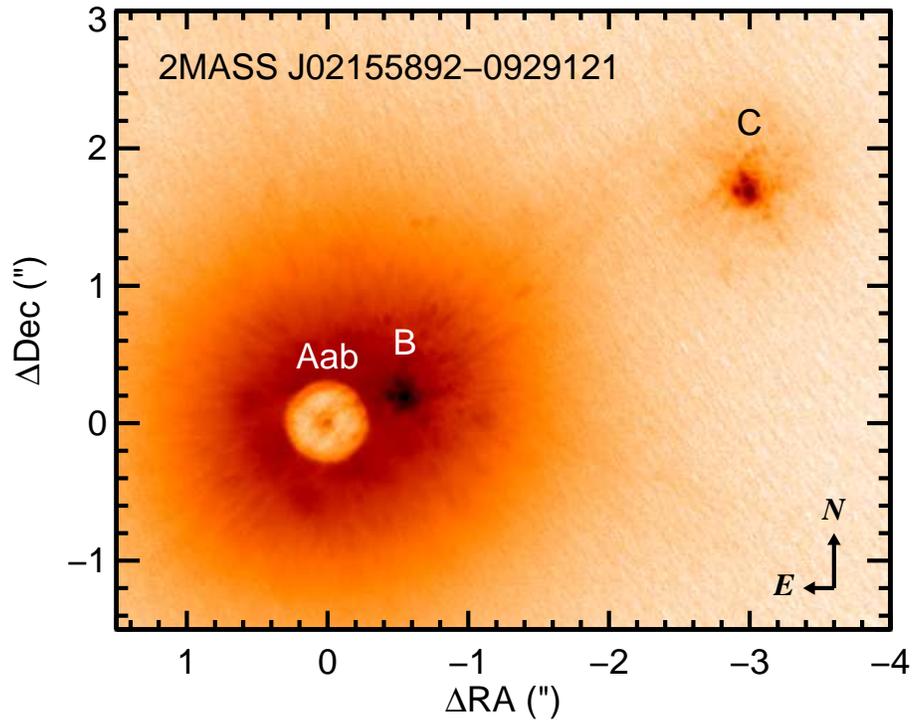}}
  \vskip -5.in
  \caption{Keck/NIRC2 $K_S$-band coadded image of the young 2MASS J02155892--0929121 system.  The primary (``Aab'') is positioned behind the 600~mas partly-opaque coronagraph.  The stellar (``B'') and brown dwarf (``C'') companions are located at separations 0$\farcs$6 and 3$\farcs$5, respectively.   North is up and east is left. \label{fig:twom0215abcnirc2} } 
\end{figure}

\clearpage
\newpage

% Figure 5

\begin{figure}
  \vskip 0.in
  \begin{center}
  \resizebox{7in}{!}{\includegraphics{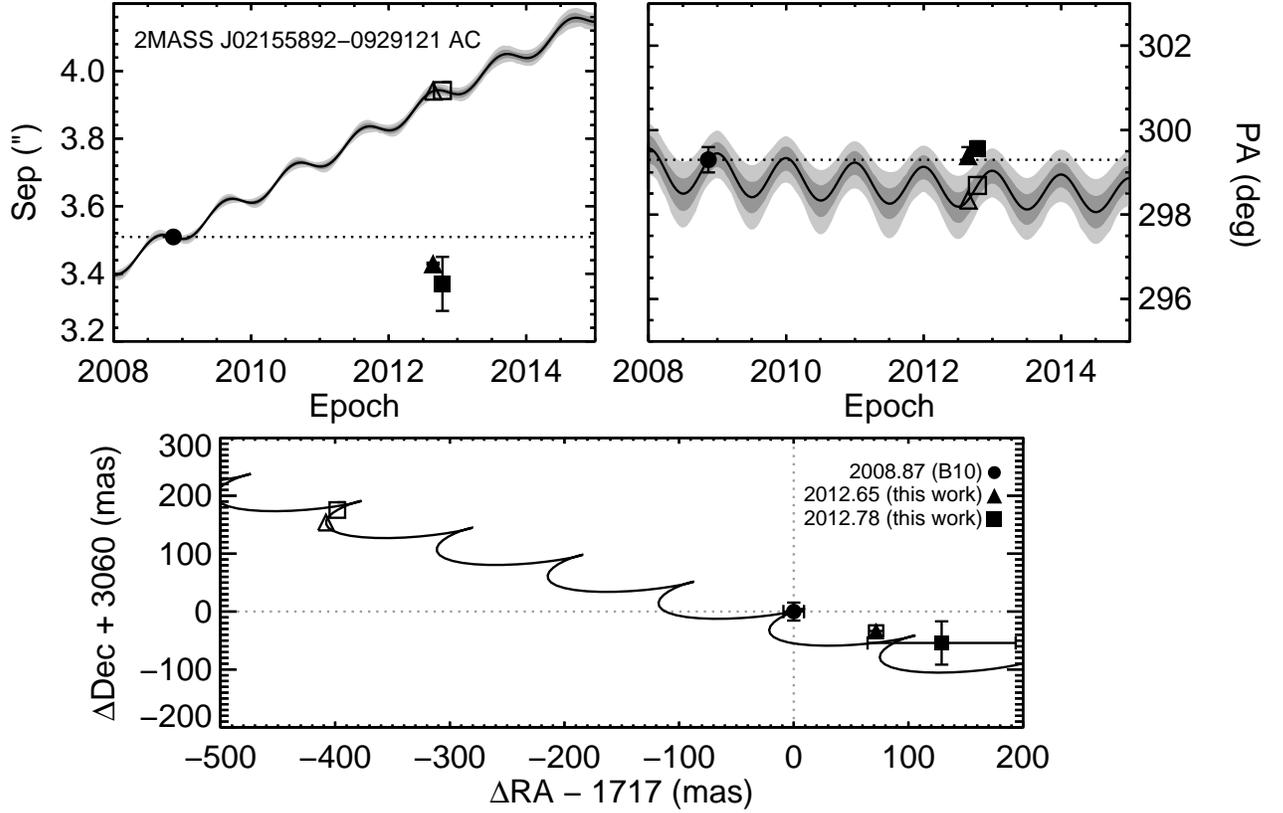}}
  \caption{Astrometry of the substellar companion 2MASS~J02155892--0929121~C between 2008--2014 relative to the 
  primary, which itself is a close 40~mas visual binary.  The solid curve shows the expected relative astrometry of a 
  background object; the gray shaded regions show 1~$\sigma$ and 2~$\sigma$ confidence intervals.  Open symbols represent
  the predicted astrometry for a stationary object at the epochs of the observations.  Compared to 
  astrometry from \citet{Bergfors:2010hm}, 2MASS~J02155892--0929121~C is clearly comoving.  There is some evidence
  of orbital motion in the system, which is likely caused by motion from the Aab primary, Aab-B component, and/or 
  between the Aab-C components of this quadruple system. \label{fig:twom0215back} } 
\end{center}
\end{figure}

\clearpage
\newpage

% Figure 6

\begin{figure}
  \vskip -2.in
  \begin{center}
  \resizebox{6in}{!}{\includegraphics{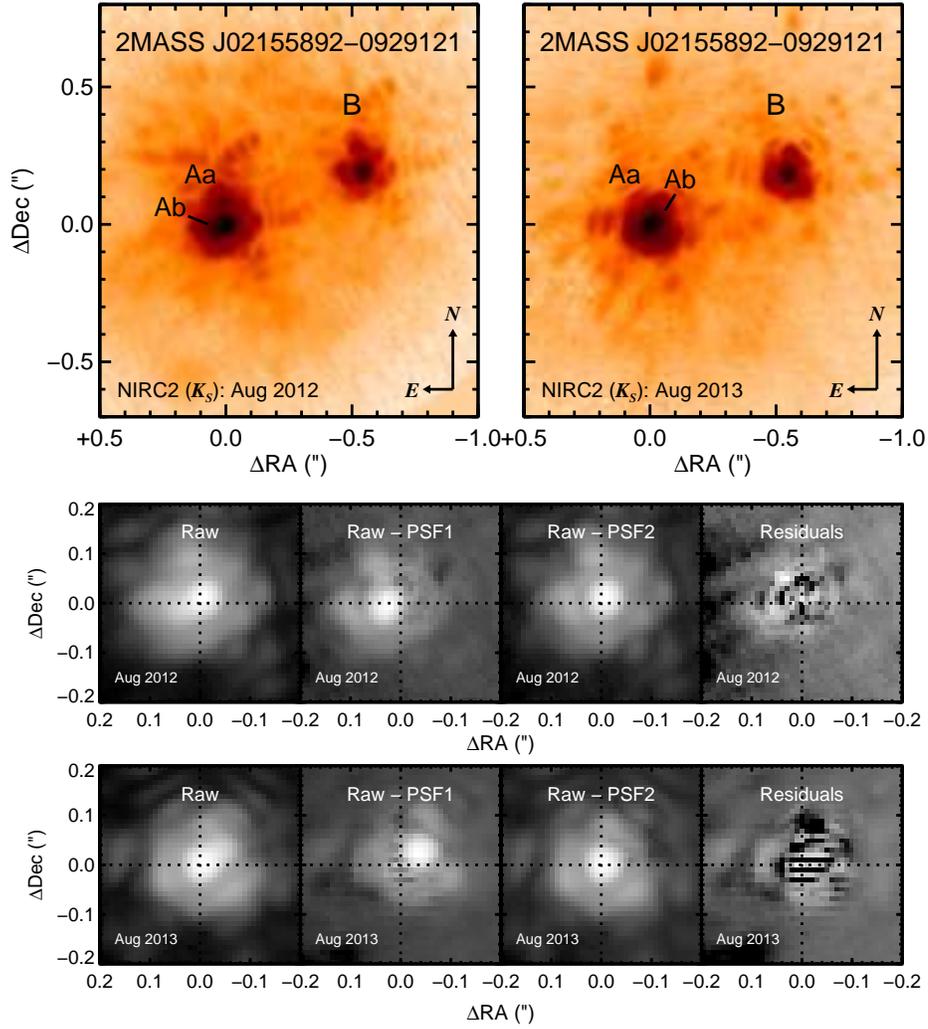}}
  \vskip -2.in
  \caption{NIRC2 images of 2MASS~J02155892--0929121~A and B in 2012 and 2013.  The ``A'' component is slightly elongated compared 
  to ``B'' and is well-fit as a marginally-resolved visual binary utilizing the PSF of ``B''.  
  The bottom two rows from left to right show the raw image of ``A,'' the residuals after jointly fitting a two-star PSF of ``B'' and subtracting the
  the brighter component (``Aa''), the same after subtracting the fainter component (``Ab''), and the residuals after subtracting both.   
  In August 2012, 2MASS~J02155892--0929121~Ab is $\approx$40~mas southeast of Aa, and in August 2013 it is $\approx$40~mas
  northwest of Aa.      \label{fig:twom0215nirc2} } 
\end{center}
\end{figure}

\clearpage
\newpage

% Figure 7

\begin{figure}
  \vskip -2.in
  \hskip -.5 in
  \resizebox{7.5in}{!}{\includegraphics{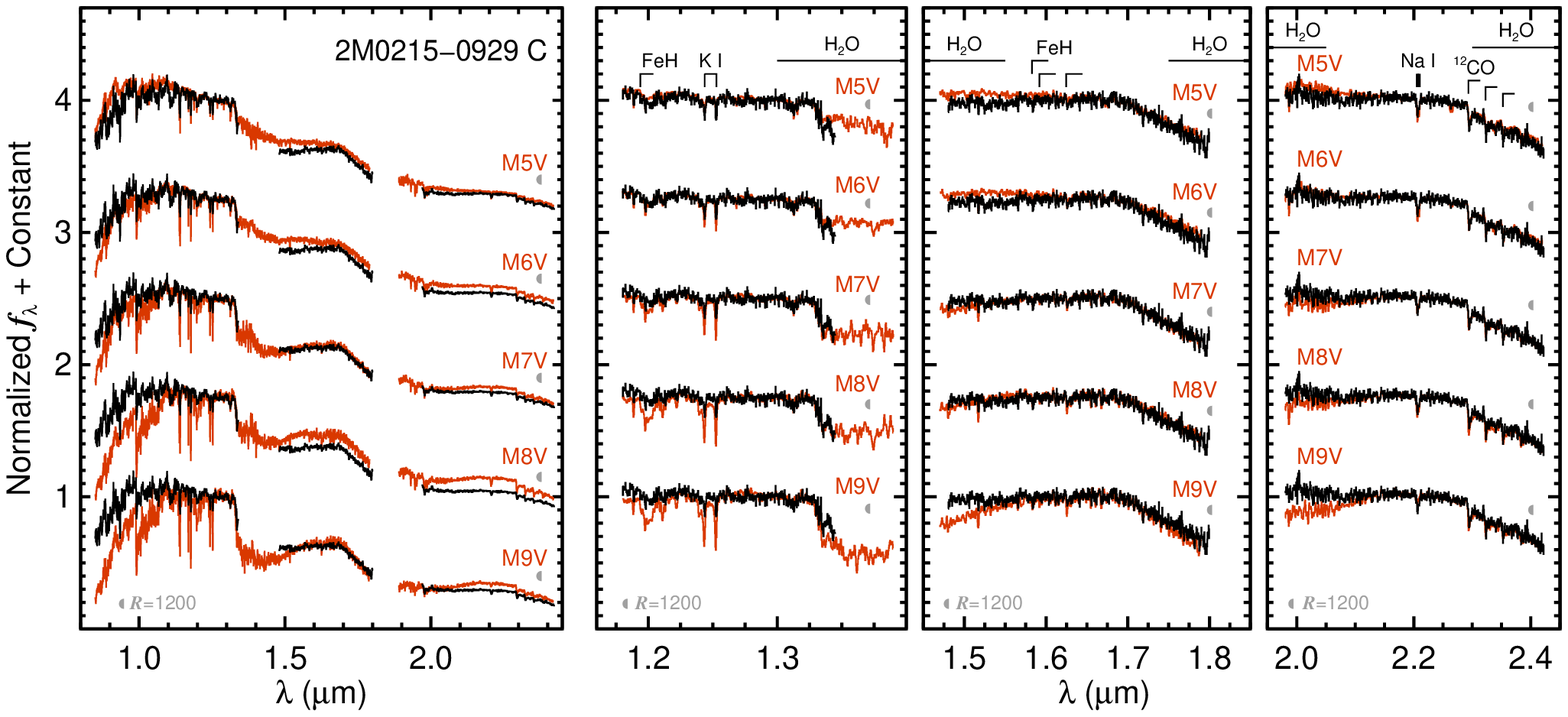}}
  \vskip -1 in
  \caption{IRTF/SpeX 0.8--2.4~$\mu$m spectrum of 2MASS~J02155892--0929121~C (black).  Visual and index-based classification from
   \cite{Allers:2013hk} yields a spectral type of M7~$\pm$~1 and an intermediate surface gravity (``INT-G'').  A reduced surface gravity 
   is evident from the shallow alkali lines and FeH band strength in 2MASS~J02155892--0929121~C compared to 
   field templates (red) from \citet{Cushing:2005ed} and \citet{Rayner:2009ki}.  This implies the companion is quite young 
   (30--150~Myr) and probably substellar, in line with the system's likely association with the Tuc-Hor, $\beta$~Pic, or Columba moving
   groups.  All templates are Gaussian smoothed to match the resolving power of our SpeX spectrum ($R$$\approx$1200).  
   All spectra are normalized and offset by a constant.     \label{fig:twom0215spec} } 
\end{figure}

\clearpage
\newpage

% Figure 8

\begin{figure}
  \vskip -1.in
  \begin{center}
  \resizebox{5in}{!}{\includegraphics{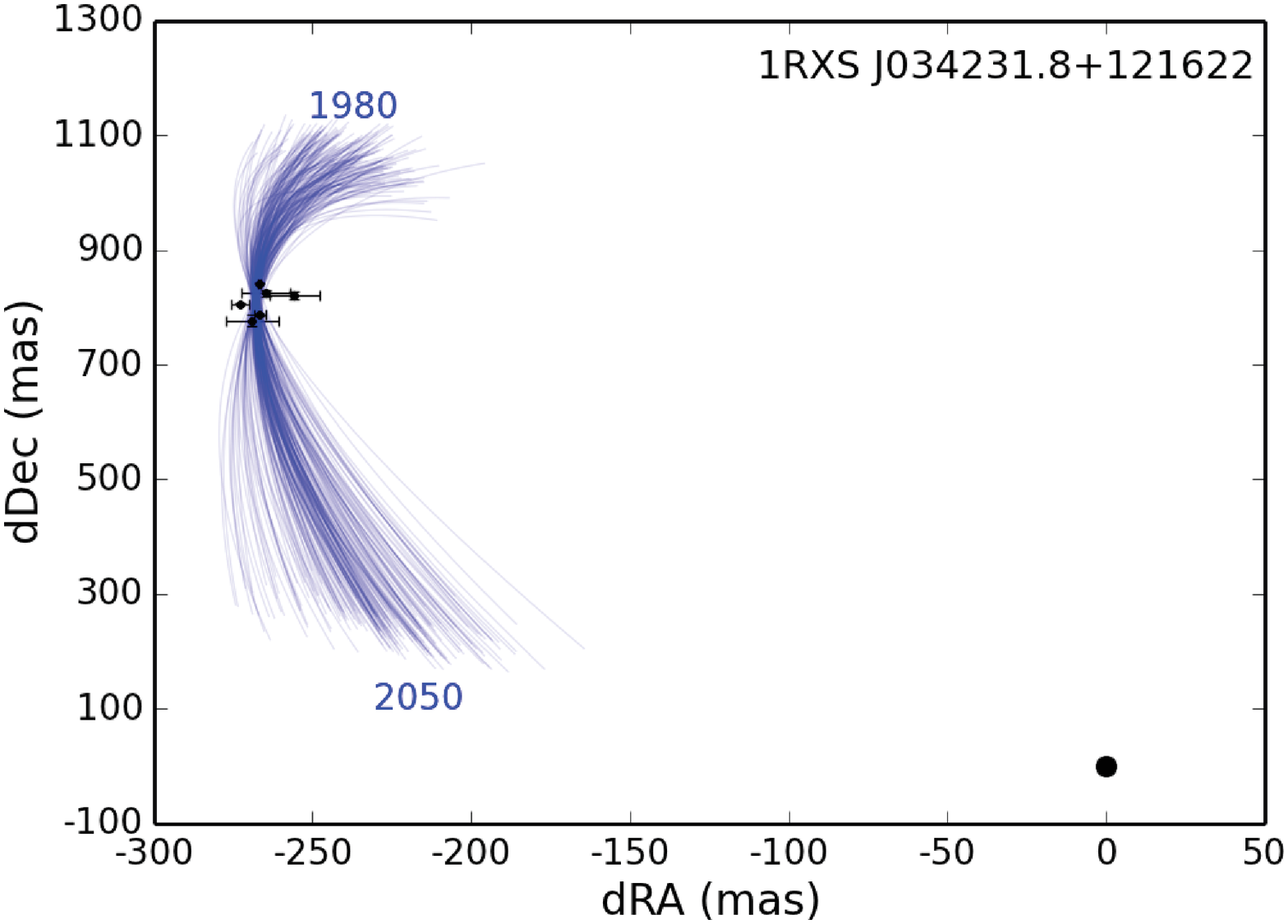}}
  \resizebox{5in}{!}{\includegraphics{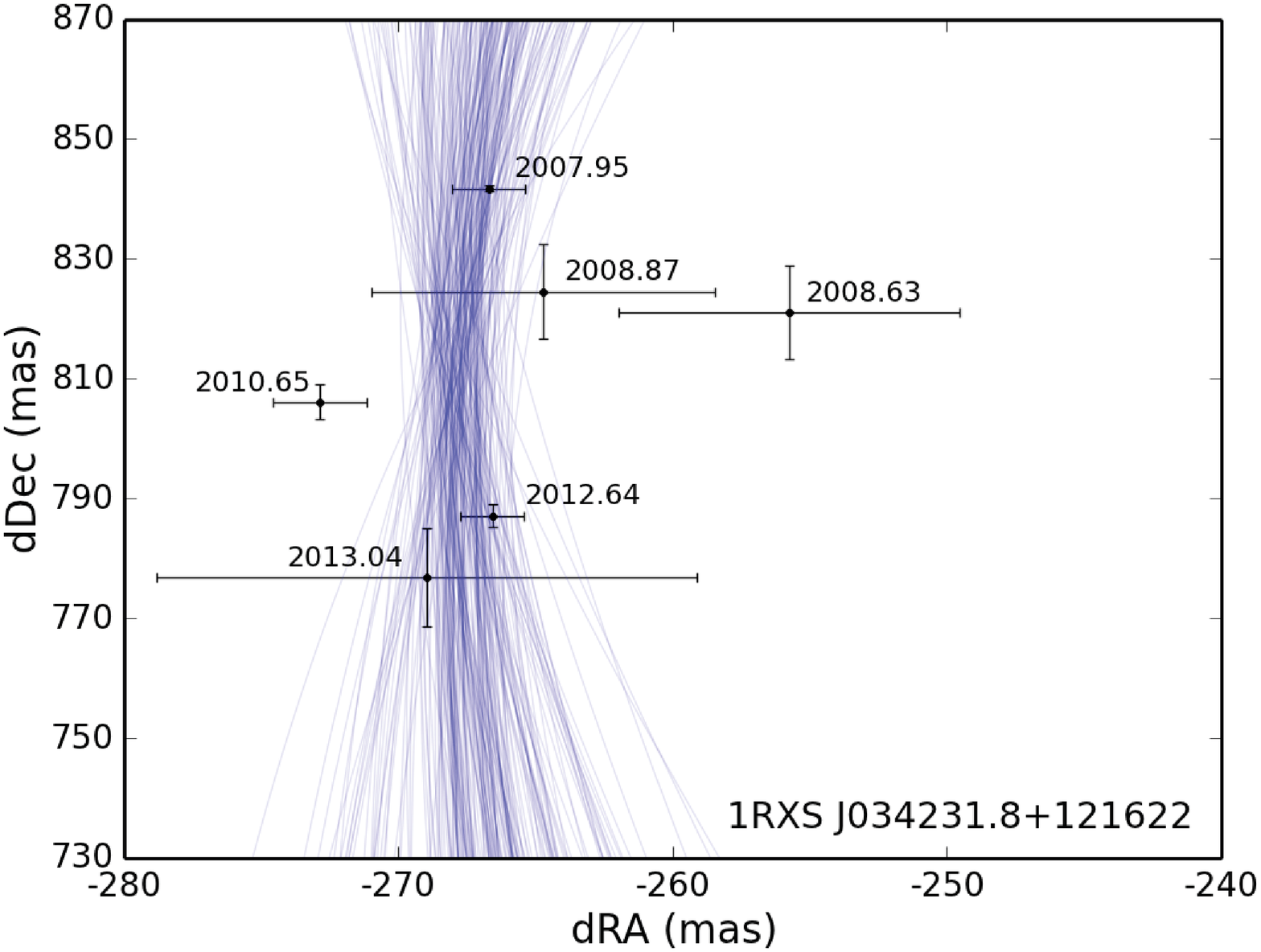}}
  %\plottwo{20141009_RXS0342_zoom.eps}{20141009_RXS0342.eps}
  \caption{Orbit fits for the intermediate-age (60--300 Myr) M4+L0 binary 1RXS~J034231.8+121622 AB.  
  Although the astrometric baseline (2007--2013) spans a small fraction of
  the orbital period, only a subset of Keplerian orbits are consistent with the observations.  Each curve represents a draw 
  from the posterior distribution of allowed orbital elements, conditioned on the astrometric data. The orbits are plotted only from the 
  expected position of the companion in 1980 to the expected position in 2050.  In the upper panel, the black dot denotes
  the position of the primary star.  The lower panel shows 
  a magnified version of the upper panel.    \label{fig:rxs0342orbit} } 
\end{center}
\end{figure}

\clearpage
\newpage

% Figure 9

\begin{figure}
  \vskip -1.in
  \begin{center}
  \resizebox{5in}{!}{\includegraphics{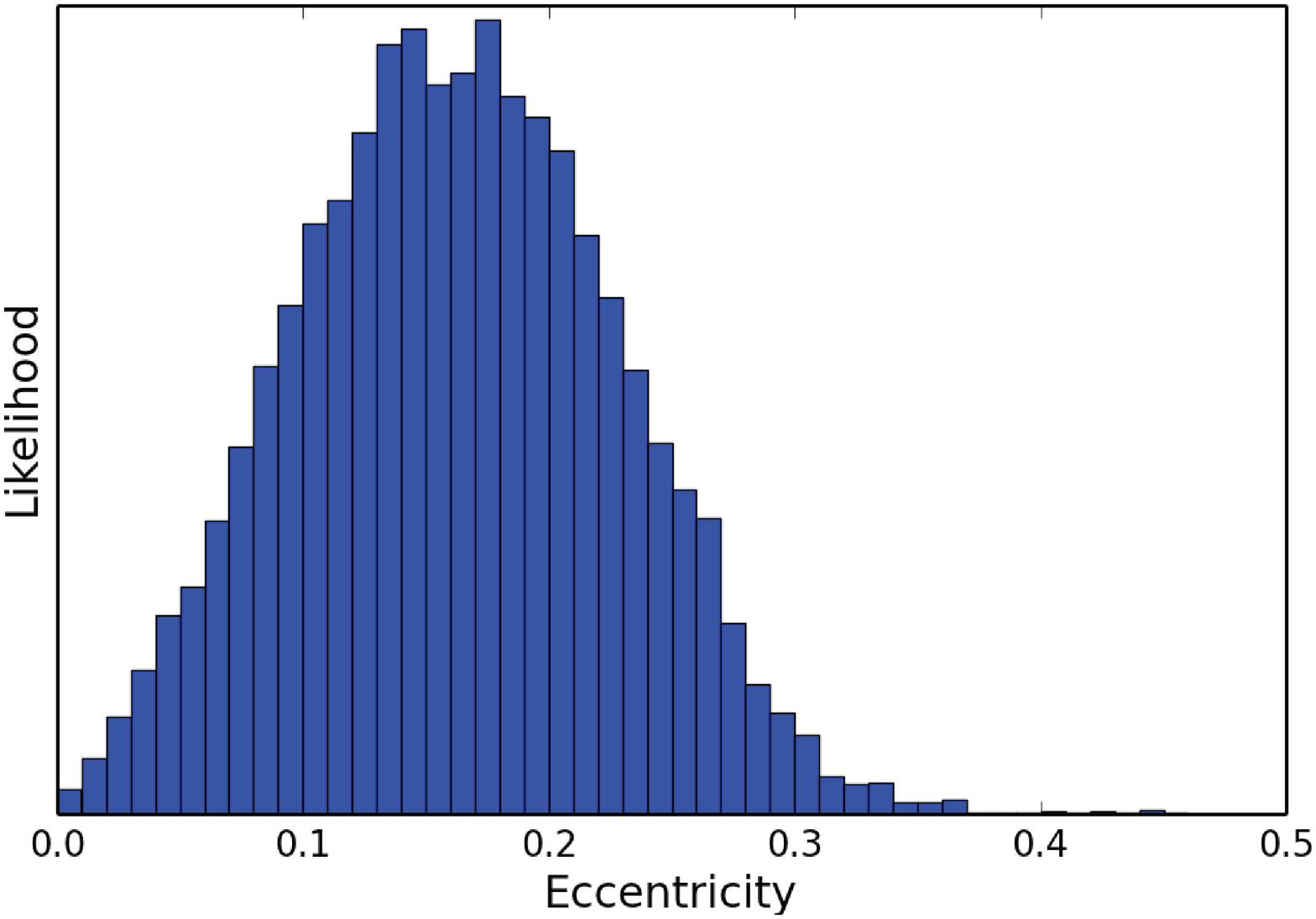}}
  \resizebox{5in}{!}{\includegraphics{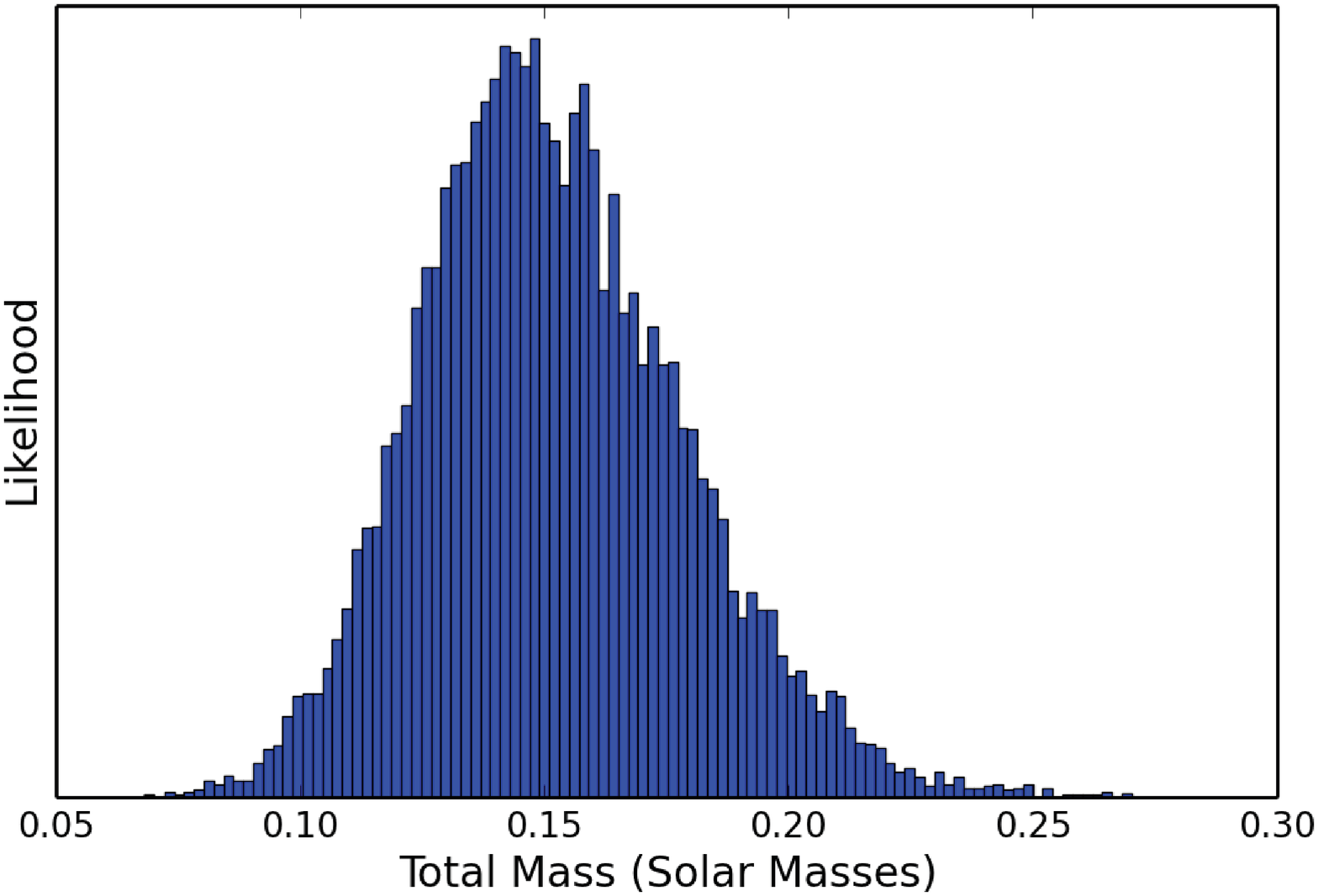}}
 % \plottwo{20141009_RXS0342_ecchist.eps}{20141009_RXS0342_masshist.eps}
  \caption{Posterior probability distributions for the eccentricity and total mass of  1RXS~J034231.8+121622 AB (see Figure~\ref{fig:rxs0342orbit}).
  Initial orbit fits of this system point to a low eccentricity ($\lesssim$0.3) and a total mass of $\approx$0.1--0.2~\Msun.    \label{fig:rxs0342posteriors} } 
\end{center}
\end{figure}

\clearpage
\newpage

% Figure 10

\begin{figure}
  \vskip -2.in
  \hskip -.6in
  \resizebox{7.5in}{!}{\includegraphics{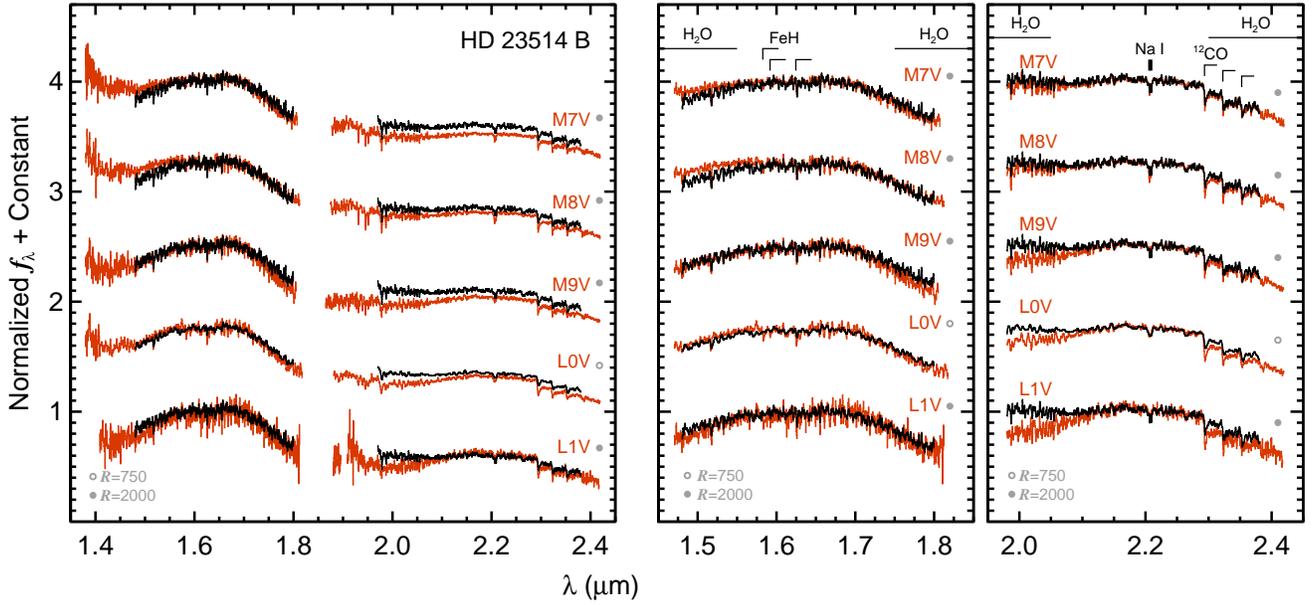}}
  \vskip -1in
  \caption{Keck/OSIRIS 1.4--2.4~$\mu$m spectrum of the substellar companion to the dusty F6 Pleiades star HD~23514.
  HD~23514~B best resembles field M7--L0 templates (red) from the IRTF Spectral Library.  
  Our OSIRIS $H$ and $K$ spectra (black) are flux calibrated to photometry from \citet{Rodriguez:2012ef}.
  All spectra are Gaussian smoothed to a common resolving power, normalized, 
  and offset by a constant.   \label{fig:hd23514b} } 
\end{figure}

\clearpage
\newpage

% Figure 11

\begin{figure}
    \vskip -2.in
  \hskip -.5 in
  \resizebox{7.5in}{!}{\includegraphics{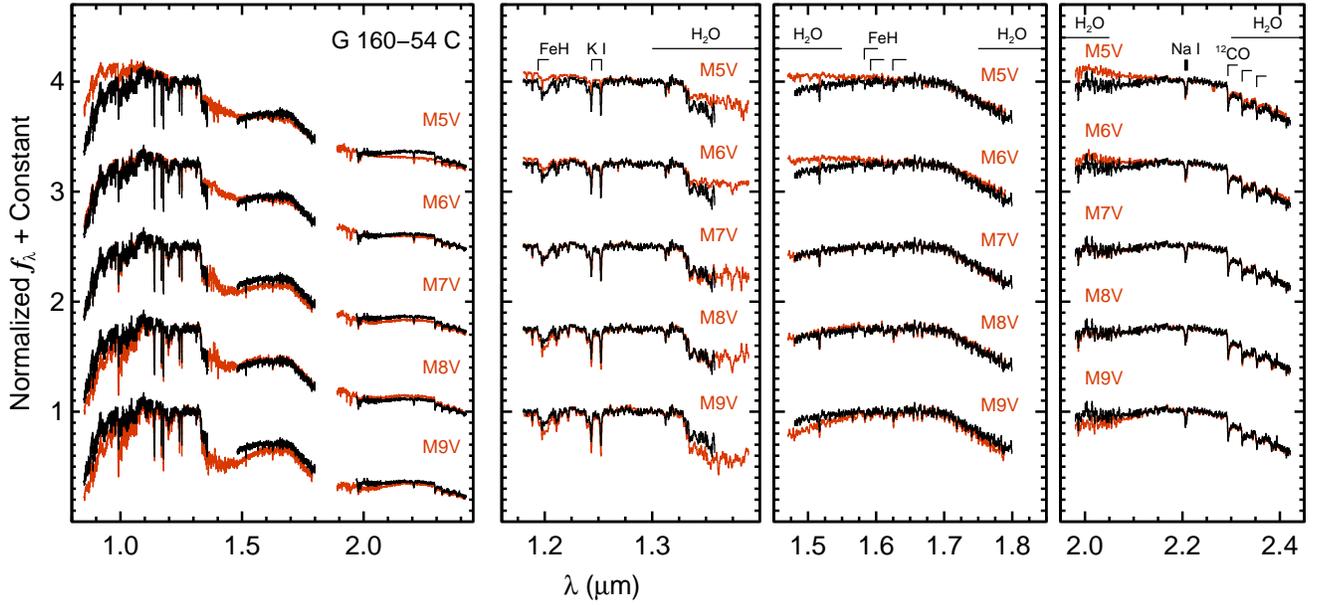}}
    \vskip -1.in
  \caption{IRTF/SpeX 0.8--2.4~$\mu$m spectrum of G~160-54~C (black).  Visual and index-based classification from
   \cite{Allers:2013hk} yields a near-infrared spectral type of M7~$\pm$~0.5 and a field surface gravity (FLD-G).  
   Field templates (red) are from the IRTF Spectral Library and have been Gaussian smoothed to 
   match the resolving power of our SpeX spectrum ($R$$\approx$1200).  
   All spectra are normalized and offset by a constant.    \label{fig:g160sxd} } 
\end{figure}

\clearpage
\newpage

% Figure 12

\begin{figure}
  \vskip -1.5in
  \hskip 0.2 in
  \resizebox{11in}{!}{\includegraphics{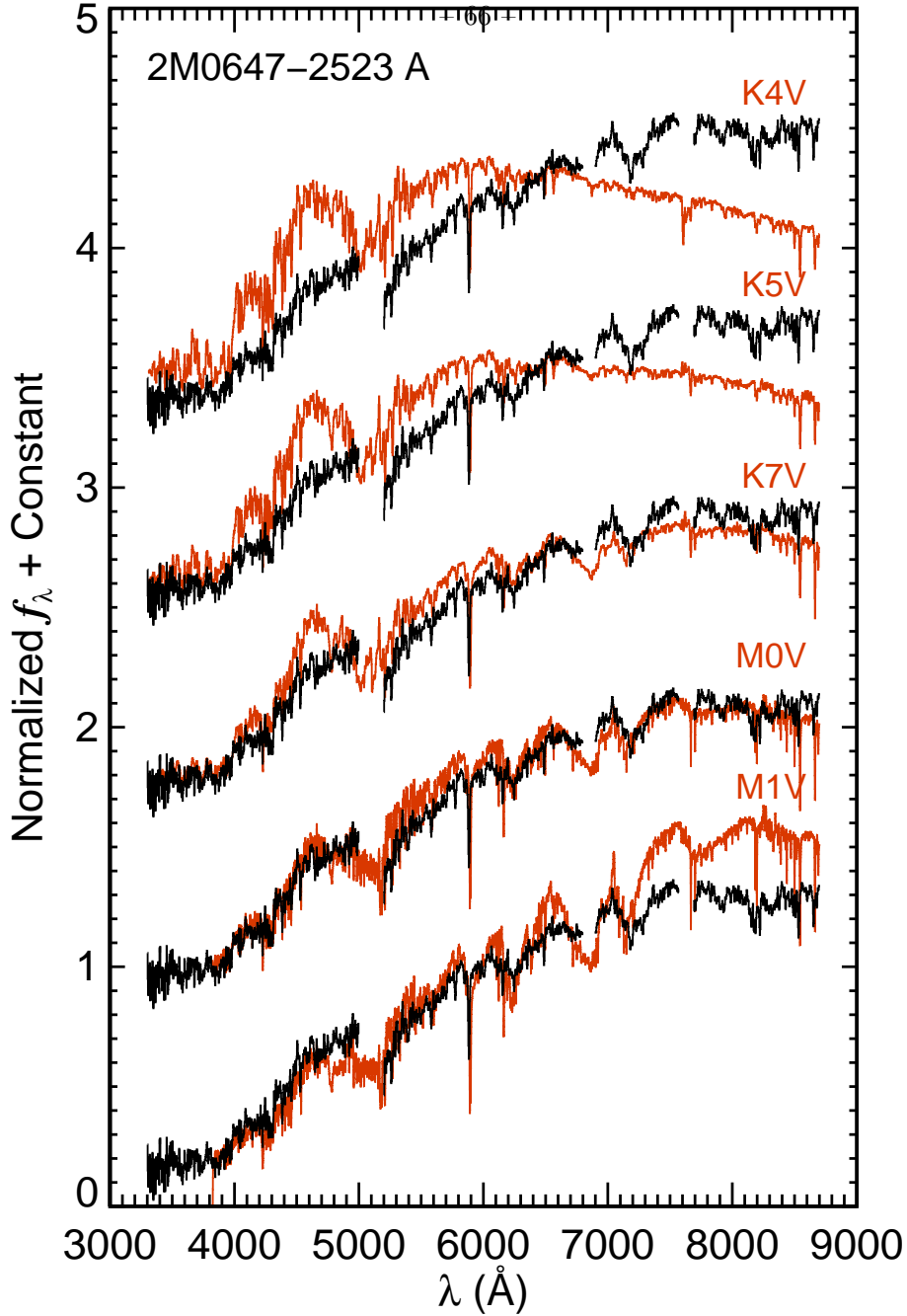}}
  \caption{UH~2.2-m/SNIFS spectrum of 2MASS~J06475229--2523304~A (black) compared to field K- and M-dwarf templates (red) from 
  \citet{Mann:2013fv} and \citet{Bochanski:2007it}, respectively.  2MASS~J06475229--2523304~A broadly 
  resembles the K7V/M0V templates but the detailed molecular 
  band strengths are not a good match.  Instead, index measurements point to a low surface gravity.  
  The H$\alpha$ line is slightly in emission ($EW$=--0.3~\AA).
  Spectral regions with strong telluric absorption have been removed and all spectra are normalized and offset by a constant. 
 \label{fig:twom0647_optspec} } 
\end{figure}

\clearpage
\newpage

% Figure 13

\begin{figure}
  \vskip -2.in
  \hskip -0.4in
  \resizebox{12in}{!}{\includegraphics{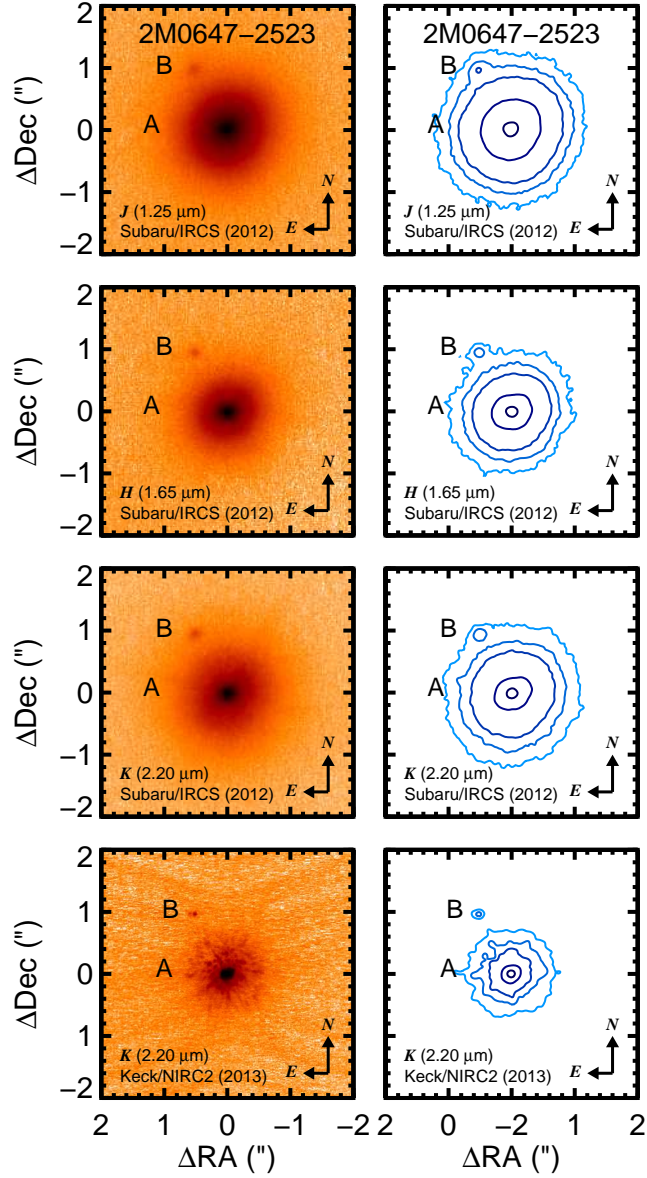}}
  \vskip -1in
  \caption{4$''$$\times$4$''$ Subaru/IRCS and Keck/NIRC2 AO images of 2MASS~J06475229--2523304 AB between 2012 and 2013.
  The companion is easily resolved, with a separation of 1$\farcs$1 at both epochs.  Contours represent 0.02\%, 0.5\%, 1\%, 5\%, and 50\% of the peak flux after convolution with
  a Gaussian kernel with a FWHM equal to that of the image PSF.  North is up and east is left.  \label{fig:twom0647imgs} } 
\end{figure}

\clearpage
\newpage

% Figure 14

\begin{figure}
  \vskip -2.in
  %\hskip -1.3in
  \resizebox{7in}{!}{\includegraphics{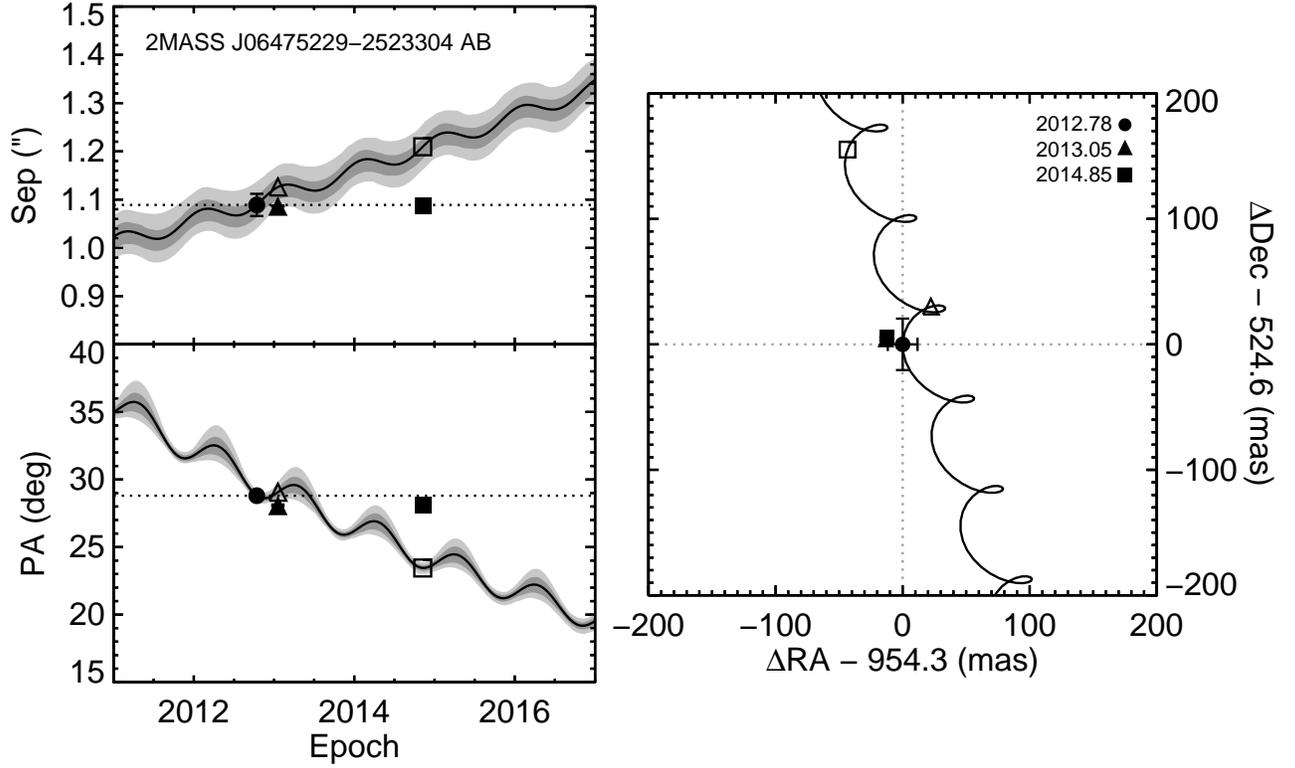}}
  \caption{Relative astrometry of 2MASS~J06475229--2523304 AB between 2012 and 2014 (filled symbols).  The solid curve shows 
  the expected path of a background object and gray shaded regions show 1~$\sigma$ and 2~$\sigma$ confidence intervals.  
  Open symbols represent the predicted astrometry for a stationary object at the epochs of the observations.  
  The pair are clearly comoving and physically bound.  \label{fig:back_twom0647} } 
\end{figure}

\clearpage
\newpage

% Figure 15

\begin{figure}
  \vskip -.7in
  \hskip -1.in
  \resizebox{12in}{!}{\includegraphics{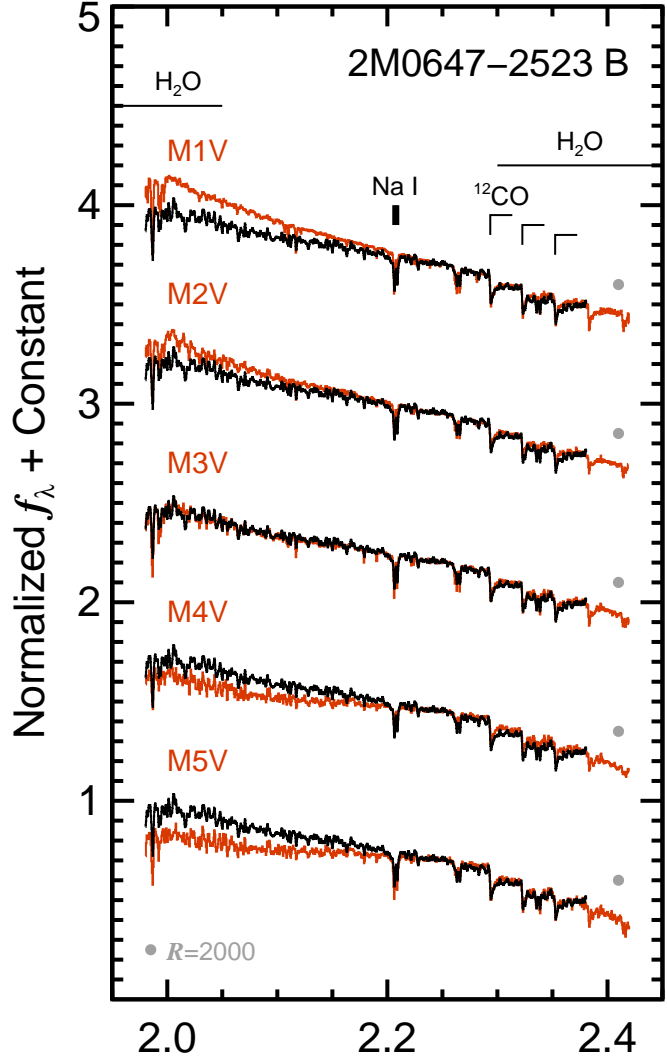}}
  \vskip -1.8 in
  \caption{Keck/OSIRIS $K$-band spectrum of 2MASS~J06475229--2523304 B (black).  Despite being 5.5~mag fainter than
  the K7 host star 2MASS~J06475229--2523304 A in the NIR, this companion has a spectral type of 
  M3V based on comparison to field template (red) from \citet{Rayner:2009ki}.  This suggests one or both components are
  not on the main sequence; the most likely scenario is the system is a distant (240~pc) K7 subgiant with an M3 compaion. 
  Our OSIRIS spectrum has been Gaussian smoothed to match the resolving power of the templates ($R$$\approx$2000),
  and all spectra are normalized and offset by a constant.  
  \label{fig:twom0647b_osiris} } 
\end{figure}

\clearpage
\newpage

% Figure 16

\begin{figure}
  \vskip -.7in
  \hskip -.5in
  \resizebox{7.5in}{!}{\includegraphics{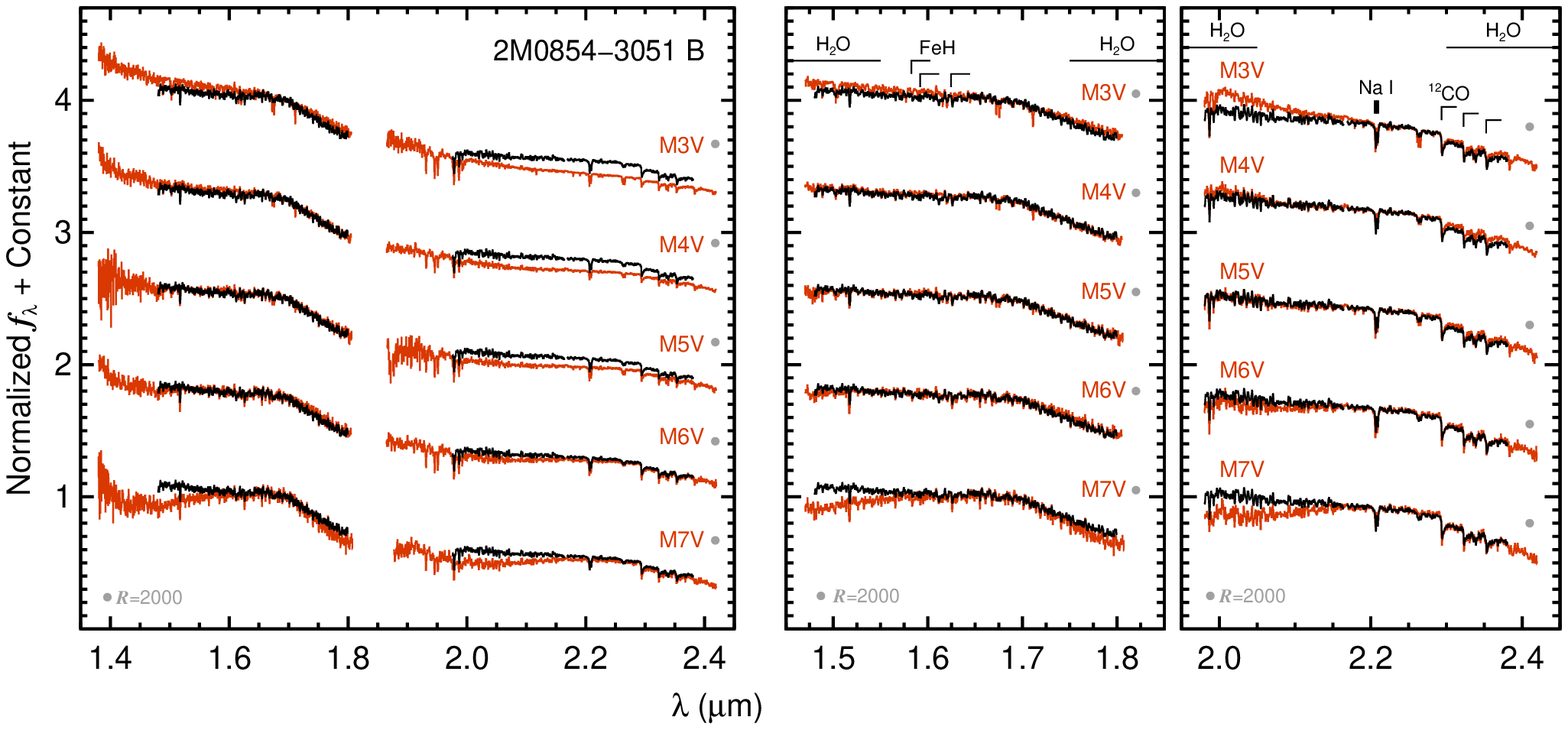}}
  \vskip -1. in
  \caption{Keck/OSIRIS 1.5--2.4~$\mu$m spectrum of 2MASS J08540240--3051366~B (black).  The best match is to M4--M6 field
  templates (red) from the IRTF Spectral Library; we adopt a near-infrared spectral type of M5~$\pm$~0.5.
   Our OSIRIS spectrum has been Gaussian smoothed to match the resolving power of the templates ($R$$\approx$2000), and
   all spectra are normalized and offset by a constant.     \label{fig:twom0854_hgcomp} } 
\end{figure}

\clearpage
\newpage

% Figure 17

\begin{figure}
  \vskip -.7in
  \hskip -.5in
  \resizebox{7.5in}{!}{\includegraphics{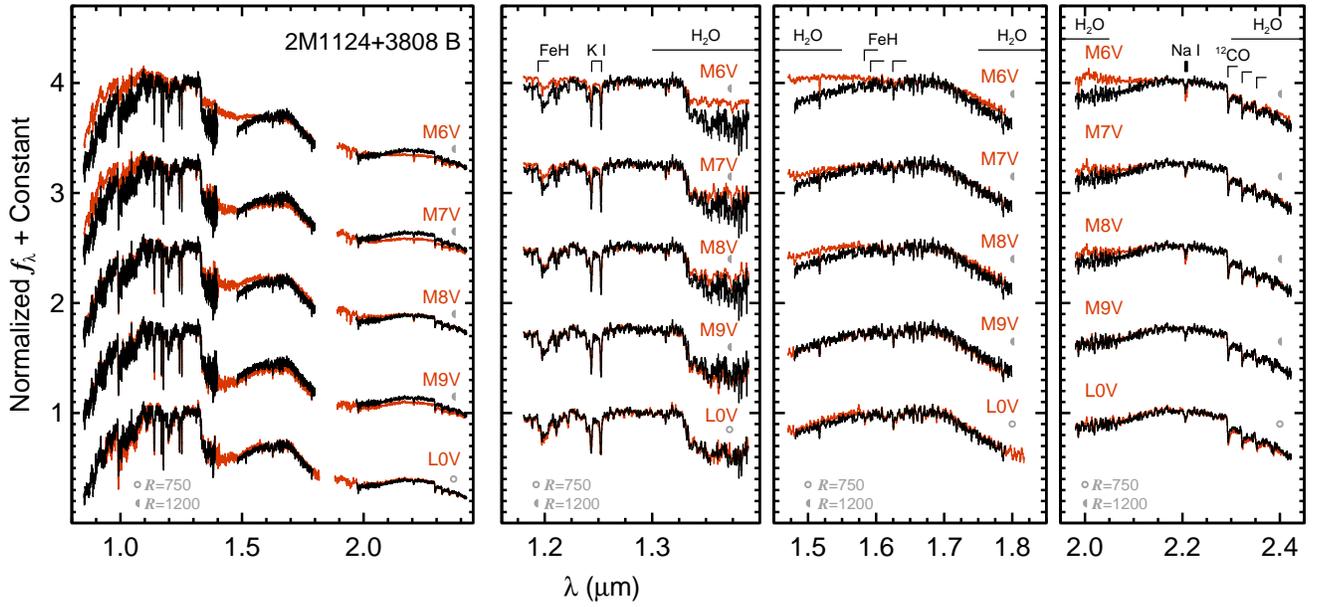}}
  \vskip -1. in
  \caption{IRTF/SpeX 0.8--2.4~$\mu$m spectrum of 2MASS~J11240434+3808108~B (black).  Visual and index-based classification from
   \cite{Allers:2013hk} yields a near-infrared spectral type of M9.5~$\pm$~0.5 and a field surface gravity (FLD-G).  
   Field templates (red) are from the IRTF Spectral Library and the M6V--M9V objects have been Gaussian smoothed to 
   match the resolving power of our SpeX spectrum ($R$$\approx$1200).  The L0 template has $R$$\approx$750 so our
   science spectrum has been smoothed to match that resolution. 
   All spectra are normalized and offset by a constant.     \label{fig:twom1124} } 
\end{figure}

\clearpage
\newpage

% Figure 18

\begin{figure}
  \vskip -2.in
%  \begin{center}
  \resizebox{6in}{!}{\includegraphics{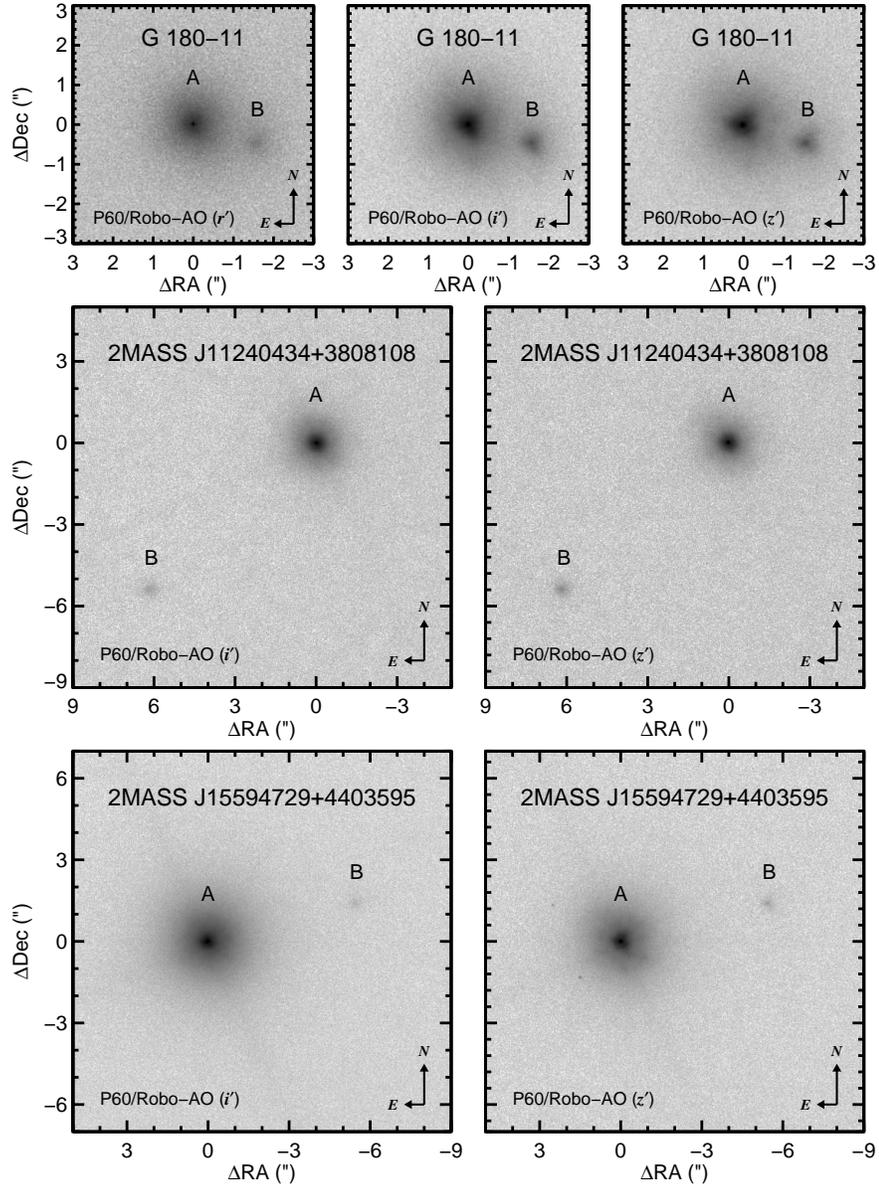}}
  \vskip -0.7in
  \caption{Optical P60/Robo-AO images of G~180-11~AB, 2MASS~J11240434+3808108~AB, and 
  2MASS~J15594729+4403595~AB.  All three late-M companions are easily detected in 60-s frames.
  North is up and east is left.   \label{fig:roboao} } 
%\end{center}
\end{figure}

\clearpage
\newpage

% Figure 19

\begin{figure}
  \vskip -2.in
  \begin{center}
  \resizebox{7in}{!}{\includegraphics{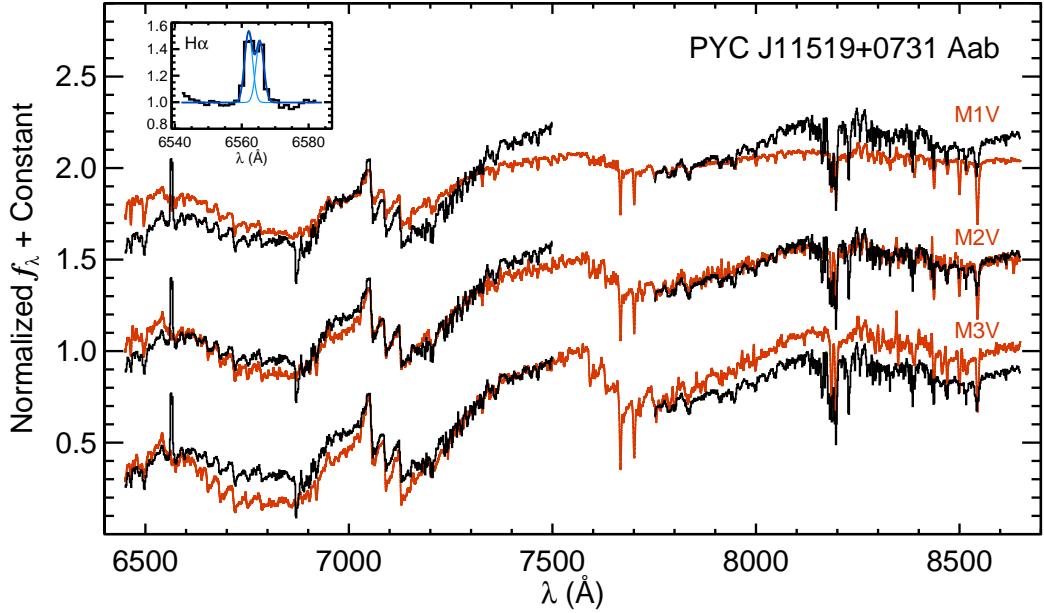}}
  \vskip -1.in
  \caption{Mayall/RC-Spectrograph medium-resolution ($R$$\approx$2500) optical spectrum of PYC~J11519+0731~A (black).  
  Visual classification yields a spectral type of M2, though some peculiarities are evident compared
  to SDSS templates (red) from \citet{Bochanski:2007it}.  The inset shows the H$\alpha$ emission line, which is marginally 
  resolved here but well-separated in our high-resolution spectrum.  A joint fit to both components with the \texttt{AMOEBA} 
  downhill-simplex algorithm yields a differential velocity of $\approx$154 km~s$^{-1}$.  However, for our RV analysis in this work we
  only include the high-resolution spectra.  All spectra are normalized and offset by a constant.  \label{fig:pycoptcomp} } 
\end{center}
\end{figure}

\clearpage
\newpage

% Figure 20

\begin{figure}
  \vskip -1.5in
  \hskip -0.3 in
  \resizebox{7in}{!}{\includegraphics{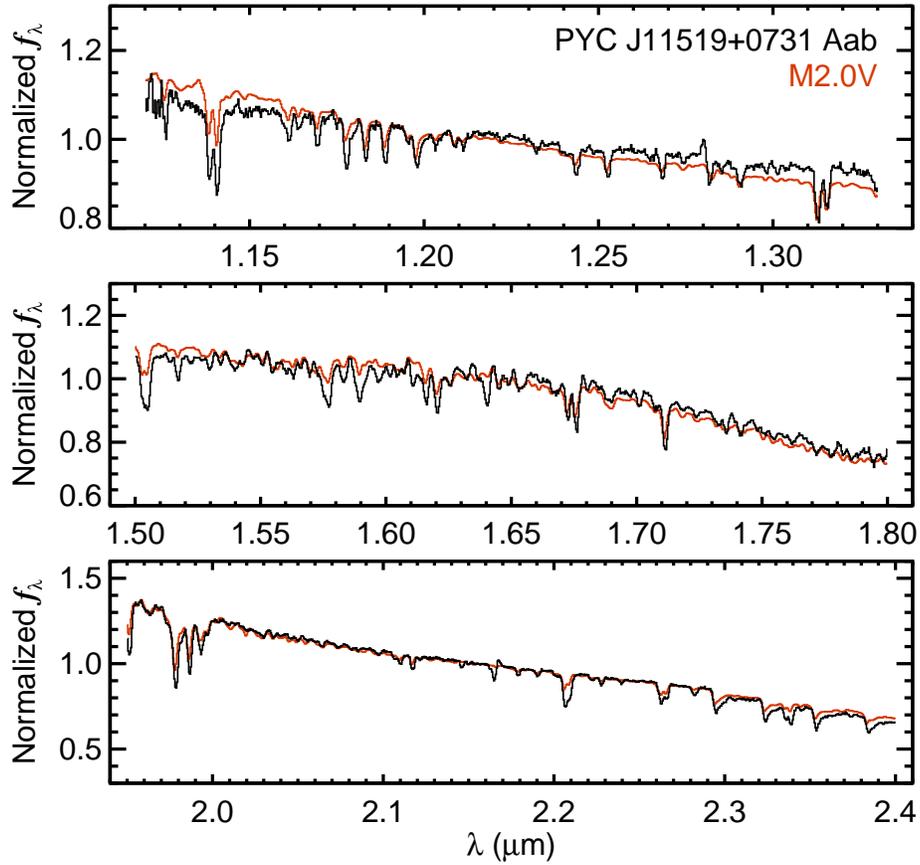}}
  \vskip -3.5 in
  \caption{IRTF/SpeX SXD unresolved spectrum of PYC J11519+0731~Aab (black) compared to the M2.0V template
  Gl~806 (red) from \citet{Rayner:2009ki}.  Several absorption features in our spectrum are unusually deep 
  such as the \ion{Mg}{1} $\lambda$1.503/1.505~$\mu$m and \ion{Si}{1} $\lambda$1.589~$\mu$m lines, possibly reflecting
  a higher metallicity.  The template is smoothed
  to the resolution of our spectrum ($R$$\approx$750) and each individual band is normalized.   \label{fig:pycnircomp} } 
\end{figure}

\clearpage
\newpage

% Figure 21

\begin{figure}
  \vskip -1.in
  \begin{center}
  \resizebox{6.5in}{!}{\includegraphics{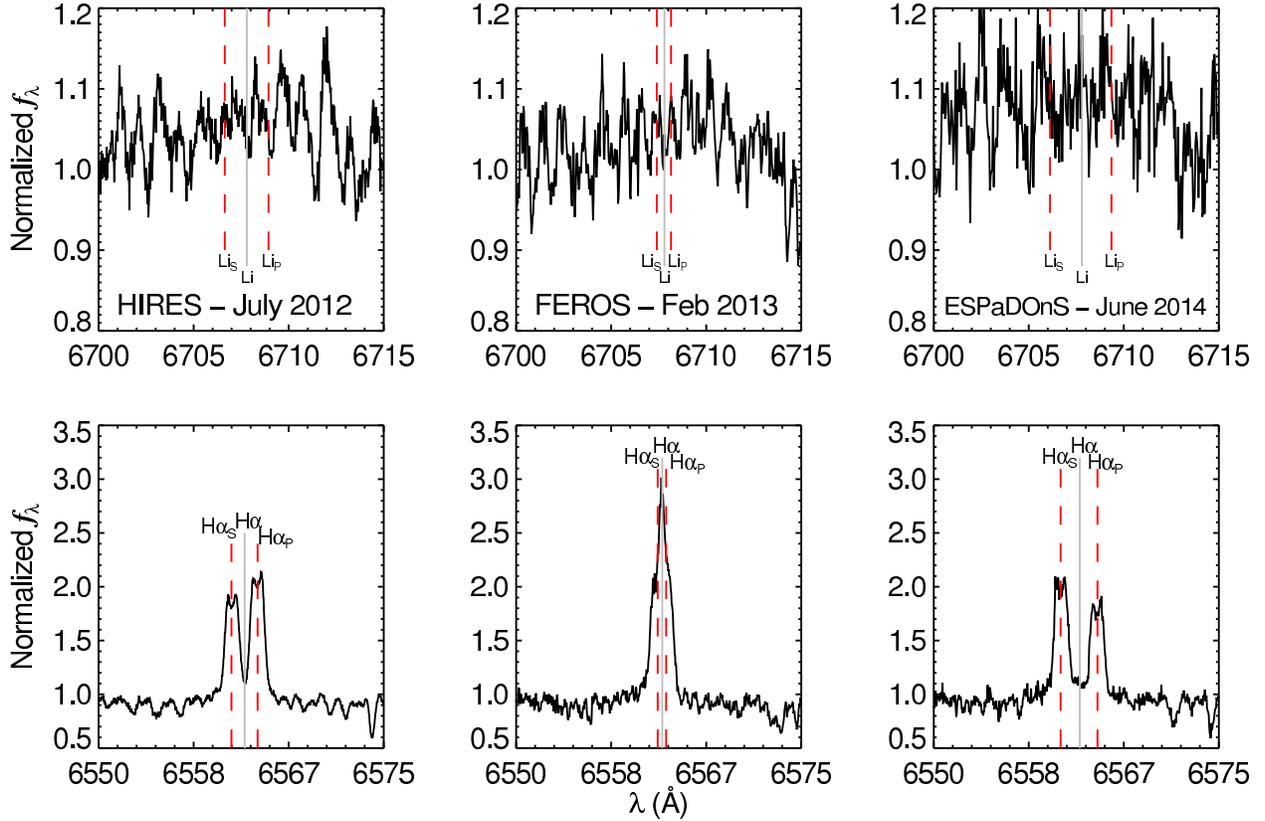}}
  \caption{\emph{Top}: The rest position of the \ion{Li}{1} $\lambda$6708 line in PYC J11519+0731~Aab (solid gray) and predicted positions of the same line in the 
  primary and secondary given their measured RVs (dashed red) at three epochs.  We do not find convincing evidence for lithium absorption in either component.
  \emph{Bottom}: Similar to the top panels but for H$\alpha$ emission lines in the primary and secondary components. The emission lines are easily split
  in our 2012 and 2014 data but have a small relative RV in our 2013 data set.    \label{fig:pychalpha} } 
\end{center}
\end{figure}

\clearpage
\newpage

% Figure 22

\begin{figure}
  \vskip -1.in
  \begin{center}
  \resizebox{4.3in}{!}{\includegraphics{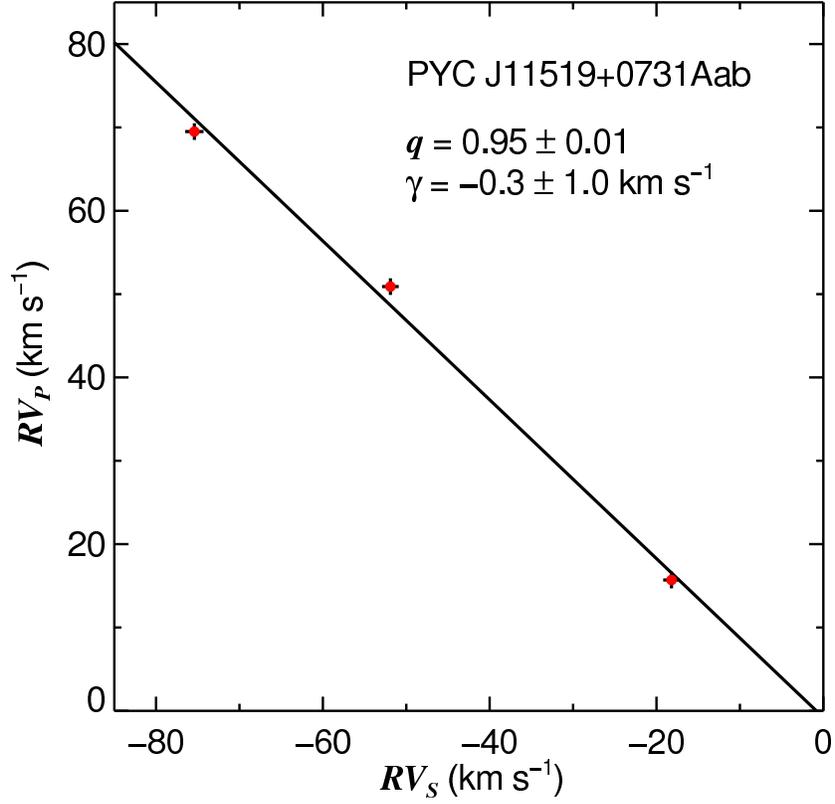}}
  \caption{Radial velocities of PYC J11519+0731~Aa and Ab.  Our three epochs enable a measurement of the 
  mass ratio ($q$) and systemic velocity ($\gamma$) for the system.  The negative slope of the best-fit line (solid line) is equal to the mass
  ratio of the system (0.95~$\pm$~0.01), and we measure a systemic RV of --0.3~$\pm$~1.0 km~s$^{-1}$.    \label{fig:pycsb2} } 
\end{center}
\end{figure}

% Figure 23

\clearpage
\newpage

\begin{figure}
  \vskip -2.in
  \hskip -.3in
  \resizebox{7in}{!}{\includegraphics{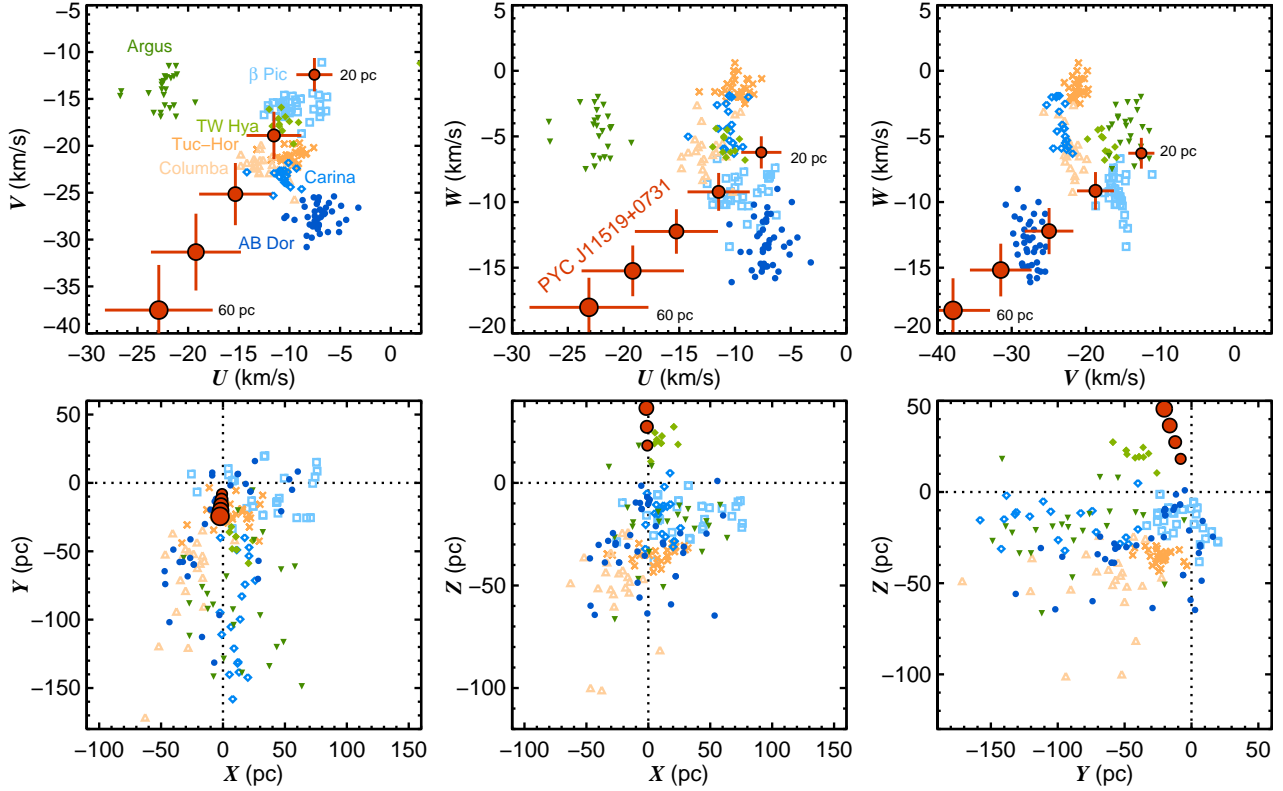}}
  \vskip -0.5 in
  \caption{Partially constrained $UVW$ kinematics and $XYZ$ space positions of PYC J11519+0731~Aab compared to 
  young moving group members from \citet{Torres:2008vq}.  Red circles show the stars' position for distances of 20, 30, 40,
  50, and 60~pc based on the systemic RV.  At close distances of $\approx$20--30 pc, PYC J11519+0731~Aab is
  consistent with $\beta$~Pic moving group, but physically would be tens of parsecs from known members.     \label{fig:pycuvw} } 
\end{figure}

% Figure 24

\clearpage
\newpage

\begin{figure}
  \vskip -1.5in
  \hskip 0.5 in
  \resizebox{7in}{!}{\includegraphics{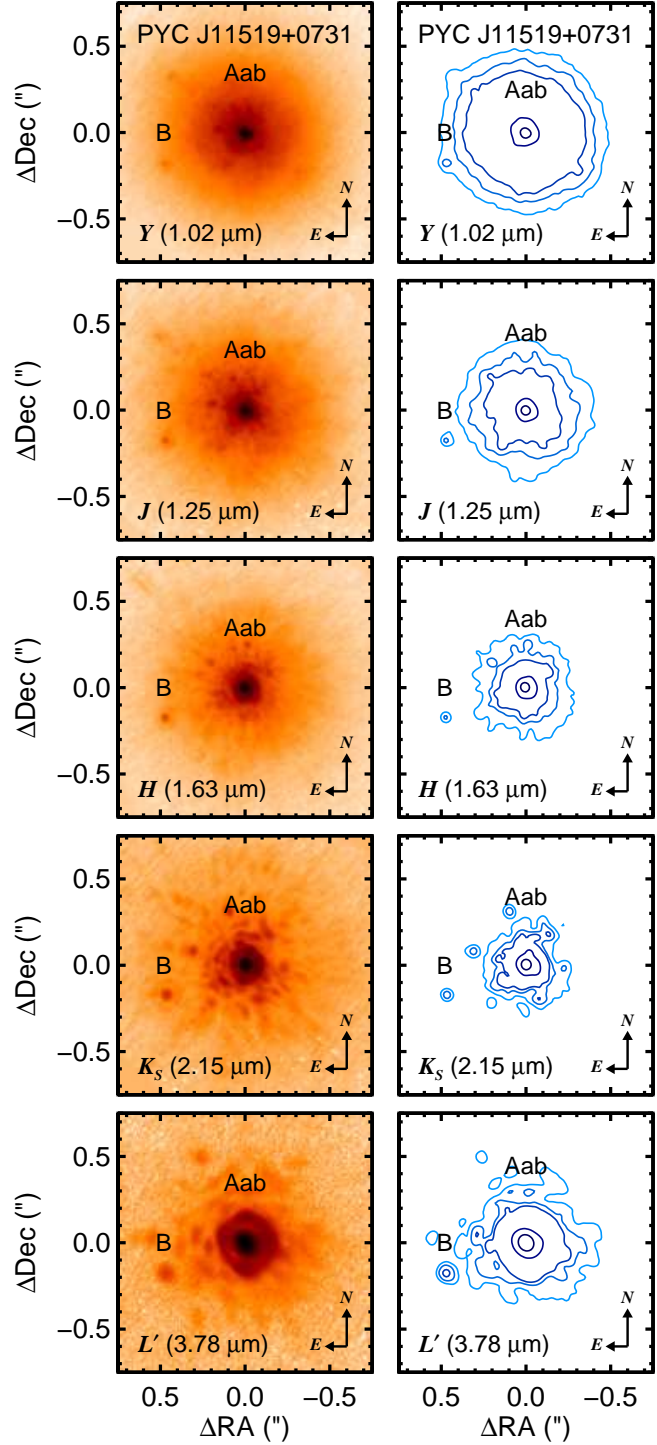}}
  \vskip -.6 in
  \caption{1$\farcs$5$\times$1$\farcs$5 Keck/NIRC2 $Y$-, $J$-, $H$-, $K_S$-, and $L'$-band AO images of PYC J11519+0731.  The 0$\farcs$5 companion 
  PYC J11519+073~B is clearly visible in these unsaturated coadded frames.  Contours represent 0.3\%, 0.6\%, 1\%, 10\%, and 50\% of the peak flux 
  after convolution with a Gaussian kernel with a FWHM equal to that of the image PSF.  North is up and east is left.     \label{fig:pycimgs} } 
\end{figure}

% Figure 25

\clearpage
\newpage

\begin{figure}
  \vskip -3in
  \hskip -1.1 in
  \resizebox{9in}{!}{\includegraphics{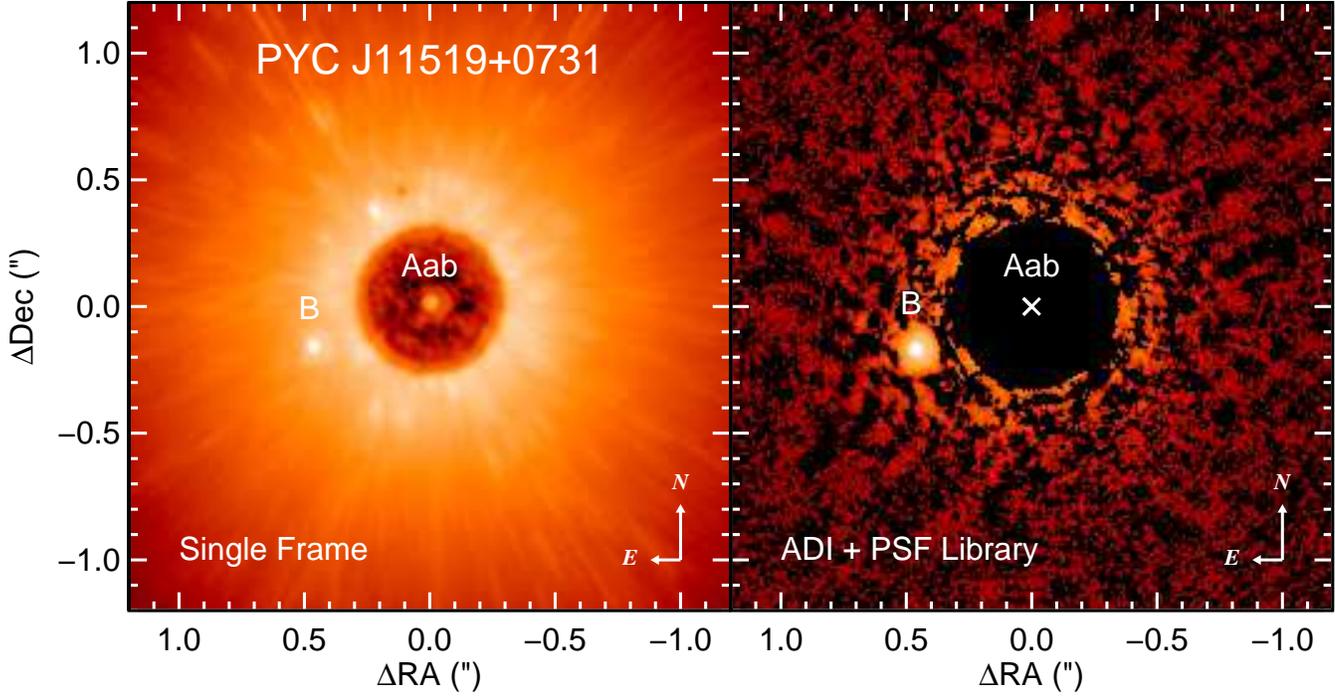}}
  \vskip -.8in
  \caption{NIRC2 $H$-band coronagraphic observations of PYC J11519+0731~AabB.  The Aab component, unresolved here, is centered behind
  the $\approx$0.1\%-throughput 600~mas diameter coronagraph.  The image on the left is a single frame in our ADI sequence; PYC J11519+0731~B
  ($\Delta$$H$=5.2~mag) is visible but resembles speckle noise.  The right panel shows the result after PSF subtraction with a library of NIRC2 PSF reference images.
  North is up and east is left.   \label{fig:pycloci} } 
\end{figure}

% Figure 26

\clearpage
\newpage

\begin{figure}
  \vskip -2.in
  \begin{center}
  \resizebox{7in}{!}{\includegraphics{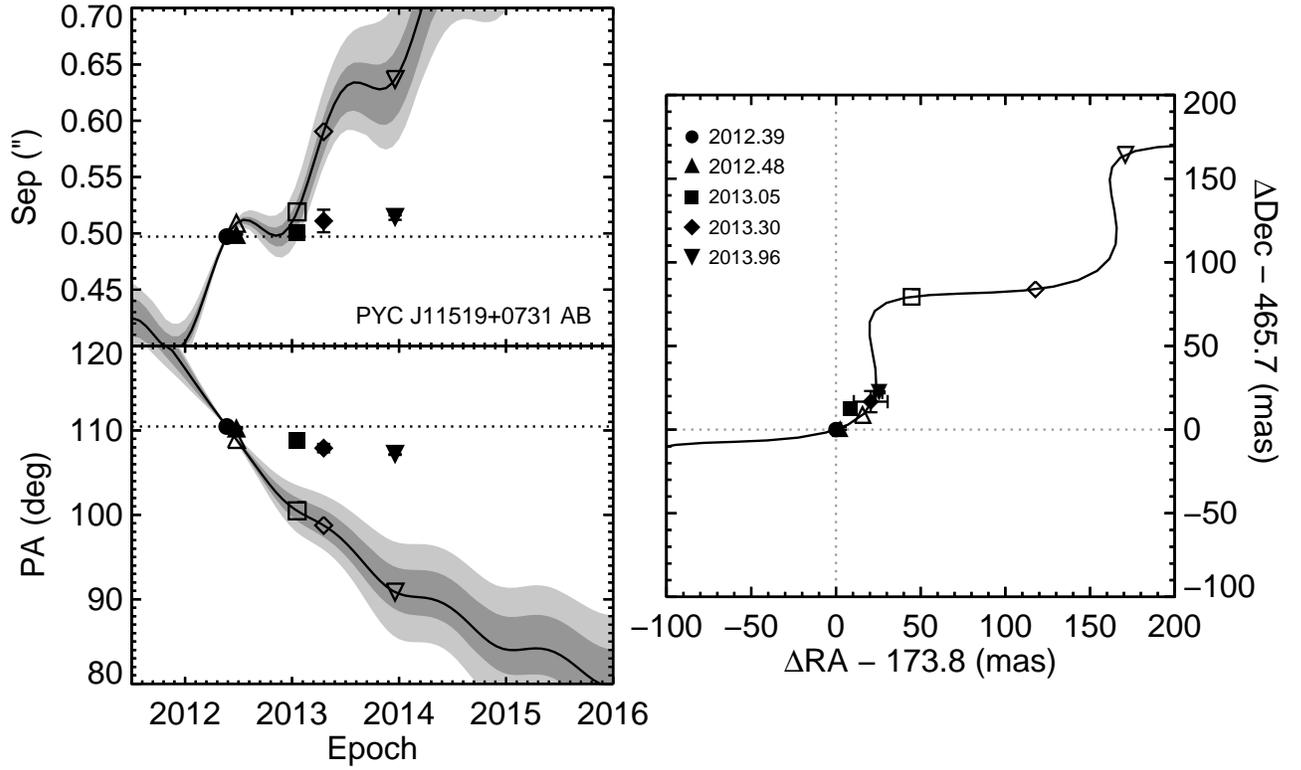}}
  \vskip -.2in
  \caption{Relative astrometry of PYC J11519+0731~Aab and B between 2012 and 2014 (filled symbols).  The solid curve shows 
  the expected path of a background object and gray shaded regions show 1~$\sigma$ and 2~$\sigma$ confidence intervals.  
  Open symbols represent the predicted astrometry for a stationary object at the epochs of the observations.  
  The pair is clearly comoving and exhibits slight but significant orbital motion.    \label{fig:pycback} } 
\end{center}
\end{figure}

% Figure 27

\clearpage
\newpage

\begin{figure}
  \vskip -1.5in
  \hskip -1.6in
  %\begin{center}
  \resizebox{10in}{!}{\includegraphics{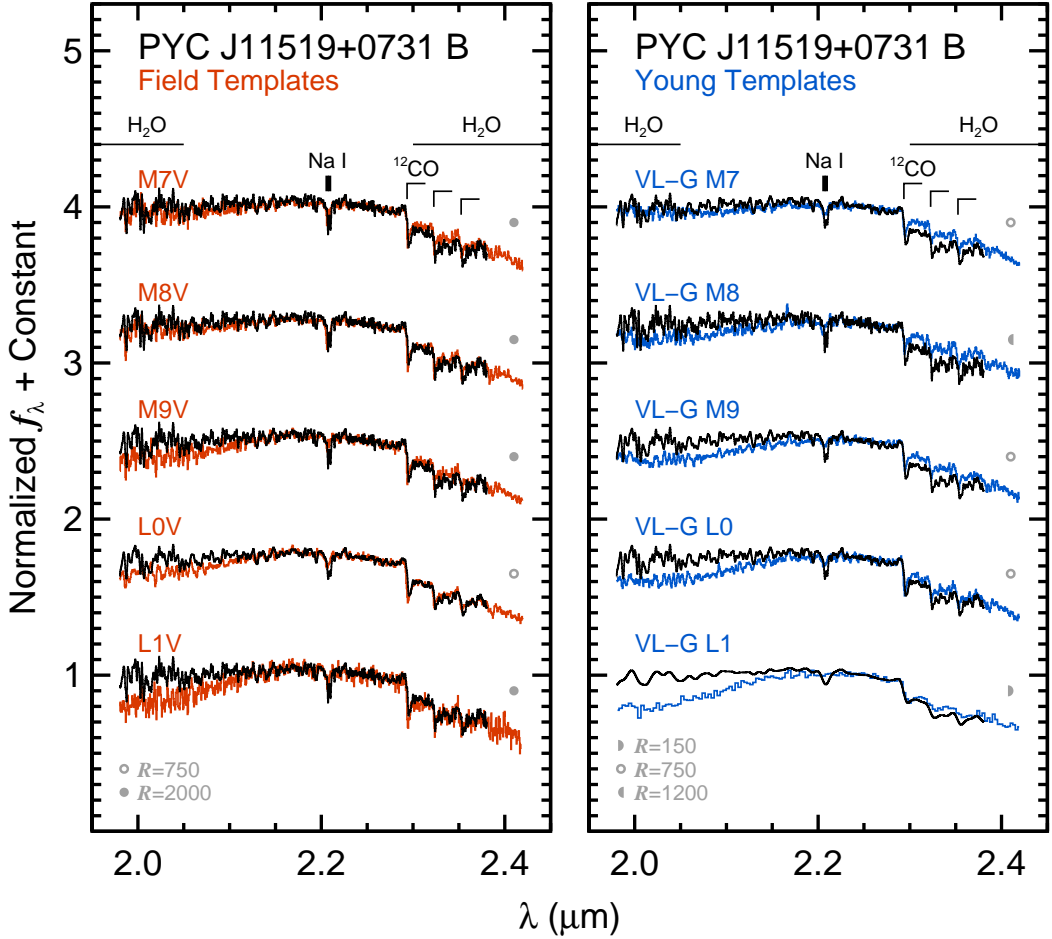}}
  \vskip -1.2 in
  \caption{Keck/OSIRIS $K$-band spectrum of PYC J11519+0731~B (black).  Compared to field templates (red) from the IRTF Spectral Library (left panel),
   PYC J11519+0731~B best resembles the M8 object so we adopt a near-infrared spectral type of M8~$\pm$~1.  The right panel shows a comparison
   to ``very low-gravity'' templates (blue) from  \citet{Allers:2013hk}, none of which match our spectrum as well as the high-gravity counterparts.   
   Our spectrum has been Gaussian smoothed to match the resolving power of the comparison object.  
   All spectra are normalized and offset by a constant.      \label{fig:pycspec} } 
%\end{center}
\end{figure}

% Figure 28

\clearpage
\newpage

\begin{figure}
  \vskip -1.5in
  \hskip -0.8 in
  \resizebox{8in}{!}{\includegraphics{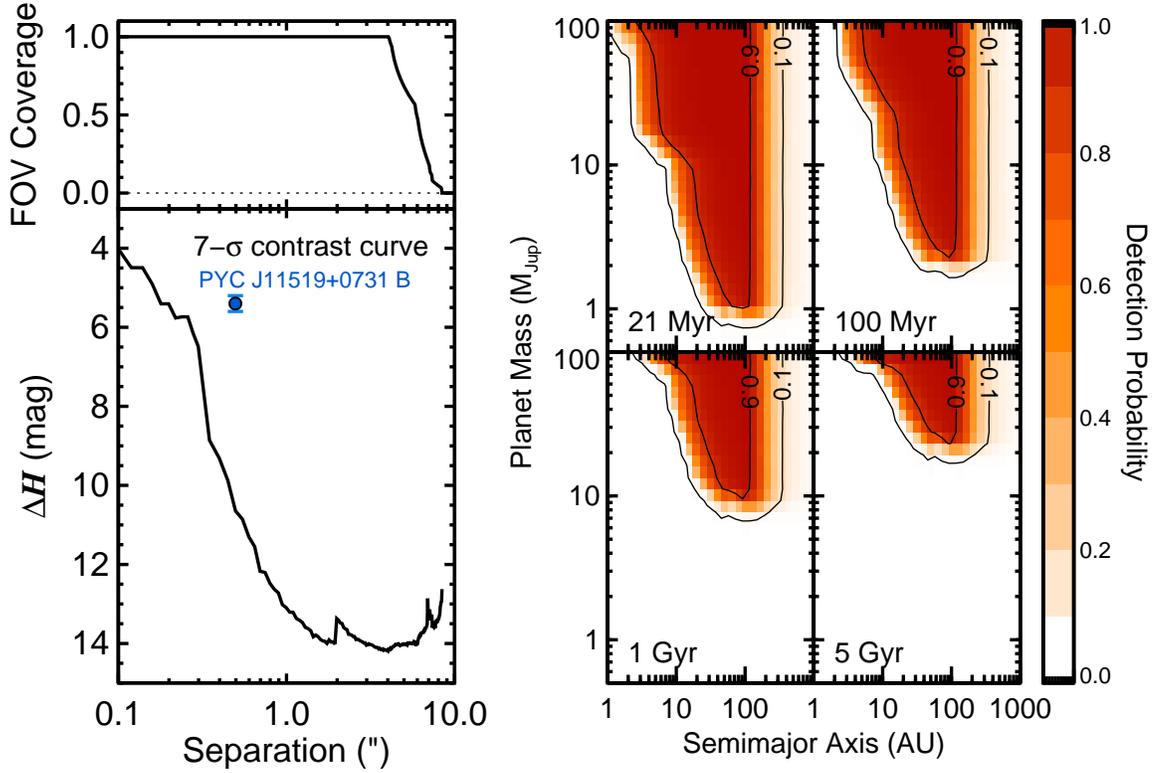}}
  \vskip -.9 in
  \caption{7~$\sigma$ contrast curve (left) and companion sensitivity map (right) for the PYC J11519+0731 system.
  Our limiting contrast of $\Delta$$H$$\approx$13~mag at 1$''$ corresponds to planetary masses ($\lesssim$13~\Mjup) at young ages 
  ($\lesssim$1~Gyr) and masses in the brown dwarf regime for ages beyond $\sim$1~Gyr.  The conversion from contrast to planet
  mass assumes hot start (Cond) evolutionary models from \citet{Baraffe:2003bj} and circular orbits.  See \citet{Bowler:2015ja}
  for details.
     \label{fig:pycsens} } 
\end{figure}

% Figure 29

\clearpage
\newpage

\begin{figure}
  \vskip -2.3in
  \hskip -.7 in
  \resizebox{8in}{!}{\includegraphics{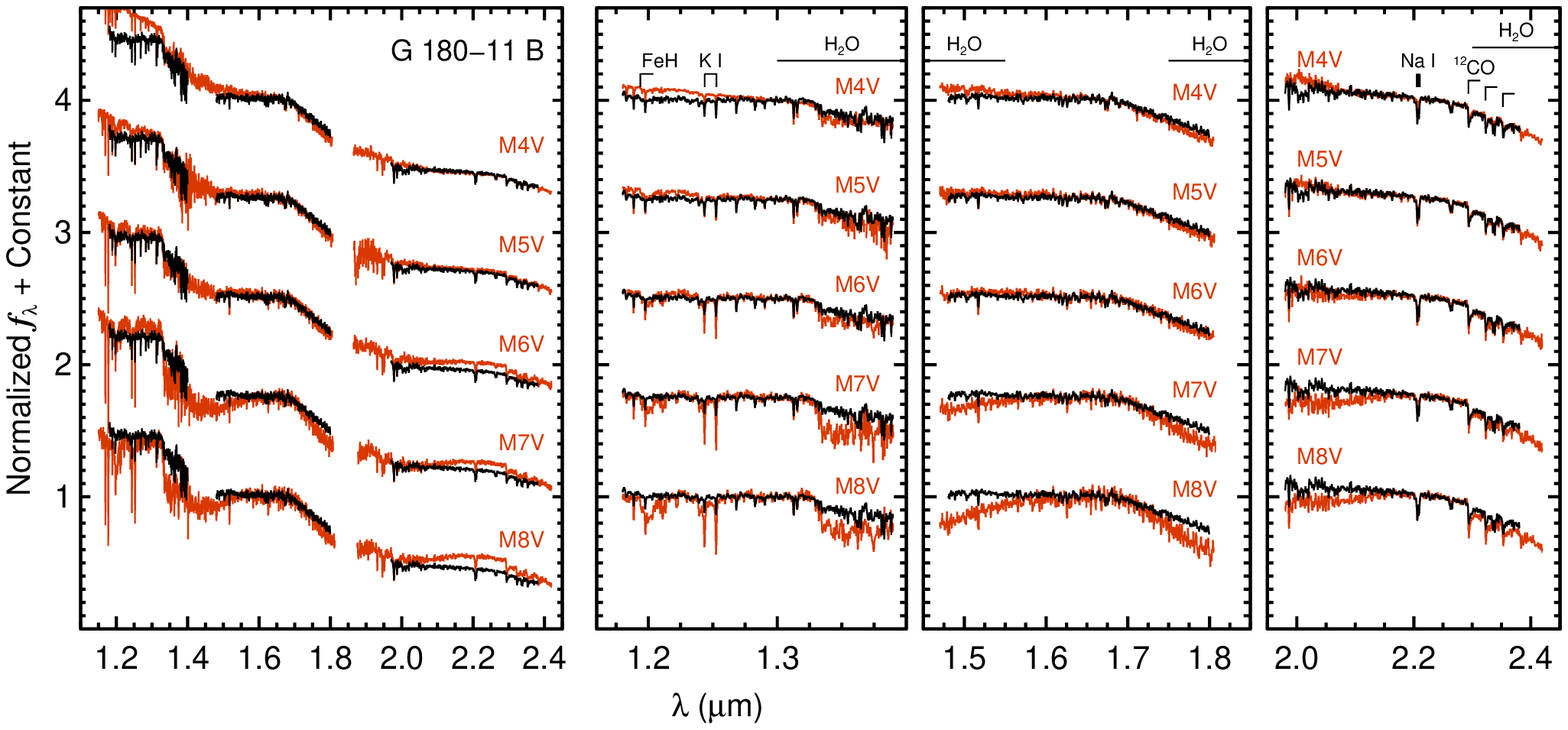}}
  \vskip -1.2 in
  \caption{Keck/OSIRIS 1.2--2.4~$\mu$m spectrum of G~180-11~B (black).  The best match is to M5--M6 field
  templates (red) from the IRTF Spectral Library.  We adopt a near-infrared spectral type of M6~$\pm$~0.5.
   Our OSIRIS spectrum has been Gaussian smoothed to match the resolving power of the templates ($R$$\approx$2000) and
   all spectra are normalized and offset by a constant.    \label{fig:g180hgcomp} } 
\end{figure}

% Figure 30

\begin{figure}
  \vskip -1.5in
  %\begin{center}
  \hskip -1.5 in
  \resizebox{10in}{!}{\includegraphics{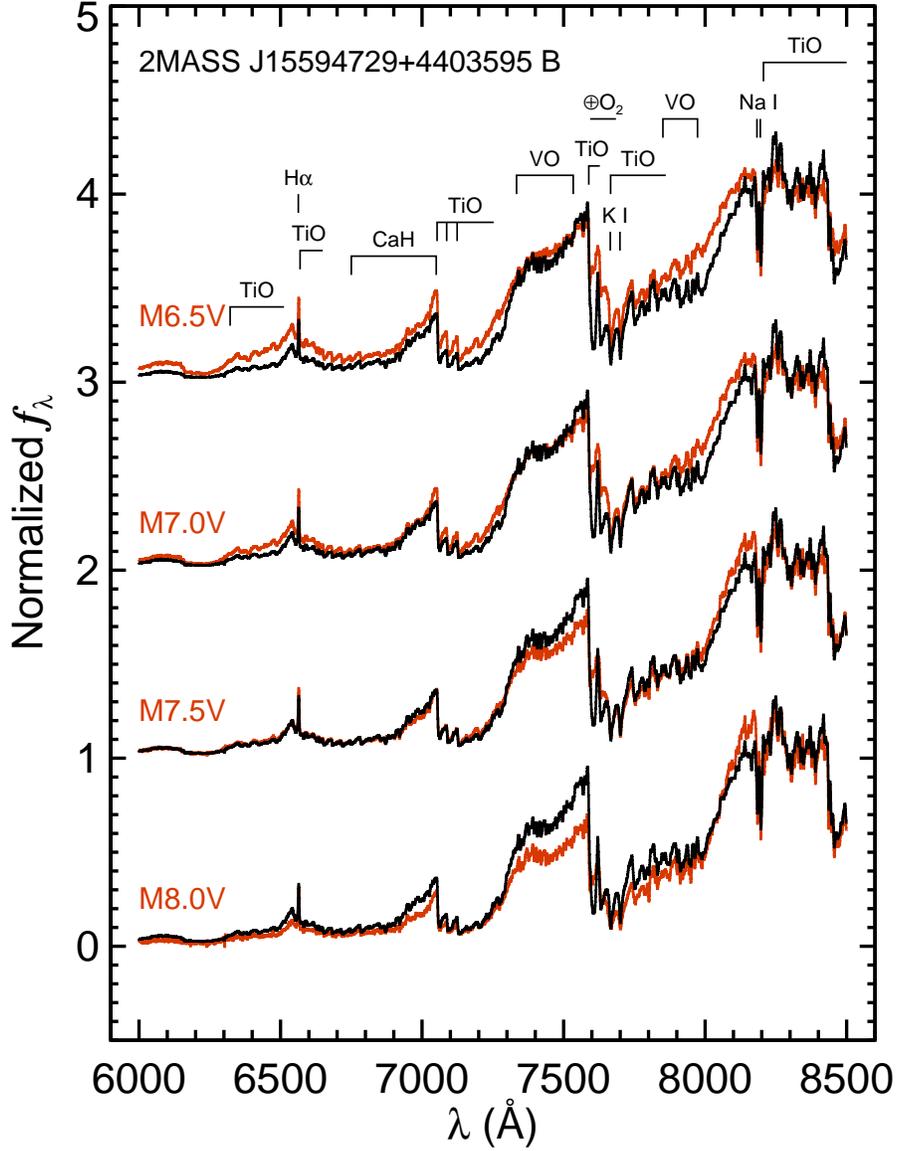}}
  \vskip -.5in
  \caption{Keck/ESI optical spectrum of 2MASS~J15594729+4403595~B (black).  The best matches are M7.0 and M7.5 templates (red) 
  from \citet{Bochanski:2007it}, where the half-types are formed from the average of neighboring integer templates.  We adopt
  an optical spectral type of M7.5~$\pm$~0.5 but note that the $\approx$7400~\AA \ VO feature appears weaker than the field template.  
  \label{fig:twom1559_optspec} } 
%\end{center}
\end{figure}

\clearpage
\newpage

% Figure 31

\begin{figure}
  \vskip -2.in
  \hskip -.4 in
  \resizebox{7.5in}{!}{\includegraphics{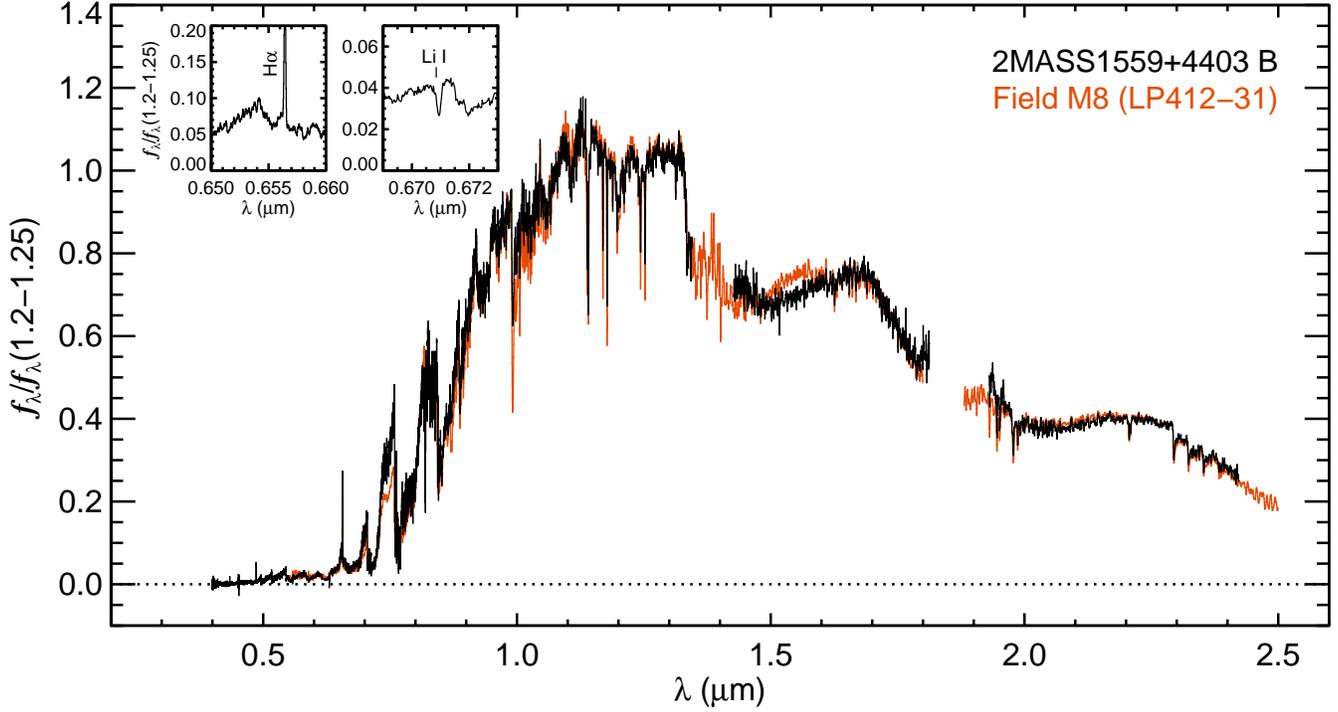}}
  \vskip -1 in
  \caption{Complete 0.5--2.4 $\mu$m spectrum of 2MASS~J15594729+4403595~B (black) compared to the high-gravity M8 field object
  LP~412-31 (red) from the IRTF Spectral Library.  The optical (0.4--0.9 $\mu$m) region is
  our new Keck/ESI spectrum while the near-infrared (0.9--2.4~$\mu$m) region is our IRTF/SpeX spectrum 
  from \citet{Bowler:2015ja}.  The lithium absorption and angular $H$-band shape in 2MASS~J15594729+4403595~B
  point to an age of $\approx$50--200~Myr, making this a nearby young system that does not appear to belong to any known moving groups.
   \label{fig:twom1559_allplot} } 
\end{figure}

\clearpage
\newpage

% Figure 32

\clearpage
\newpage

\begin{figure}
  \vskip 0.in
  \begin{center}
  \resizebox{6in}{!}{\includegraphics{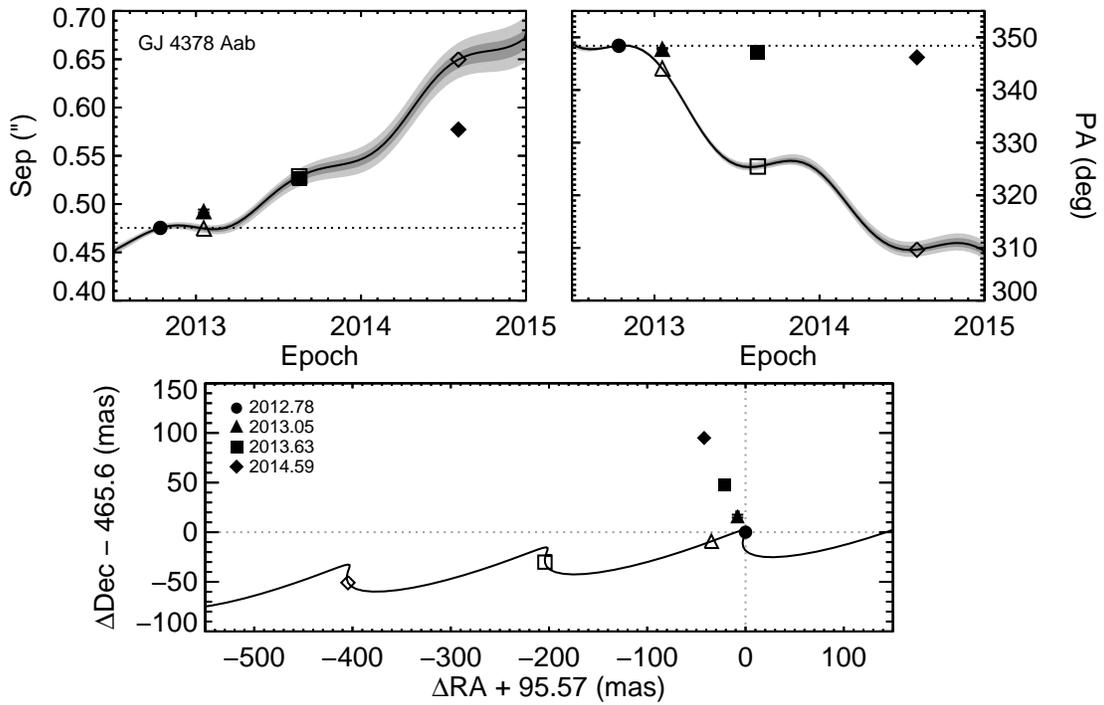}}
  \vskip -.2 in
  \caption{Relative astrometry of GJ~4378~Aa and Ab between 2012 and 2015.  The solid curve shows 
  the expected path of a background object and gray shaded regions show 1~$\sigma$ and 2~$\sigma$ confidence intervals.  
  The pair are clearly comoving and exhibit significant orbital motion.     \label{fig:gj4378backtracks} } 
\end{center}
\end{figure}

\clearpage
\newpage

% Figure 33

\begin{figure}
  \vskip -1.5in
  \hskip 0.5 in
  \resizebox{7in}{!}{\includegraphics{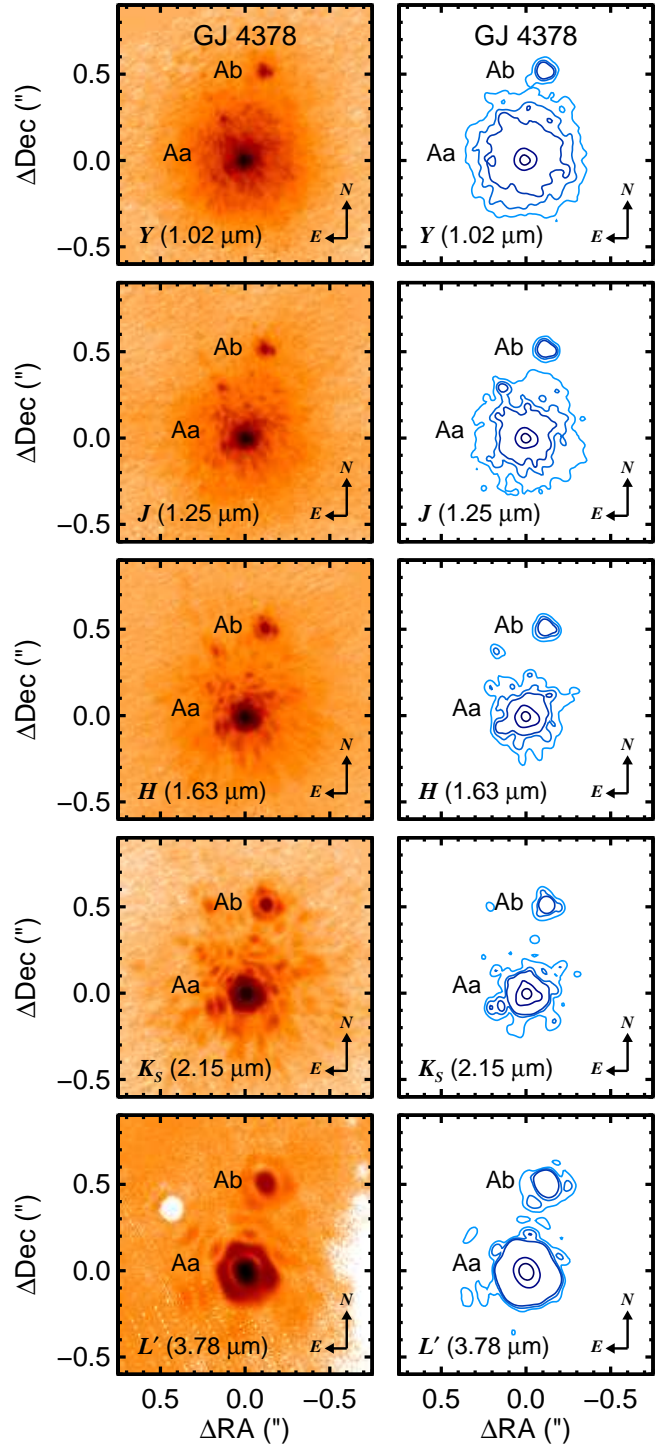}}
  \vskip -.5 in
  \caption{1$\farcs$5$\times$1$\farcs$5 Keck/NIRC2 $Y$-, $J$-, $H$-, $K_S$-, and $L'$-band AO images of GJ~4378~Aab.  Contours 
  represent 0.3\%, 0.6\%, 1\%, 10\%, and 50\% of the peak flux 
  after convolution with a Gaussian kernel with a FWHM equal to that of the image PSF.  North is up and east is left.     \label{fig:gj4378imgs} } 
\end{figure}

\clearpage
\newpage

% Figure 34

\begin{figure}
  \vskip -2.in
  \hskip -.6 in
  \resizebox{7.5in}{!}{\includegraphics{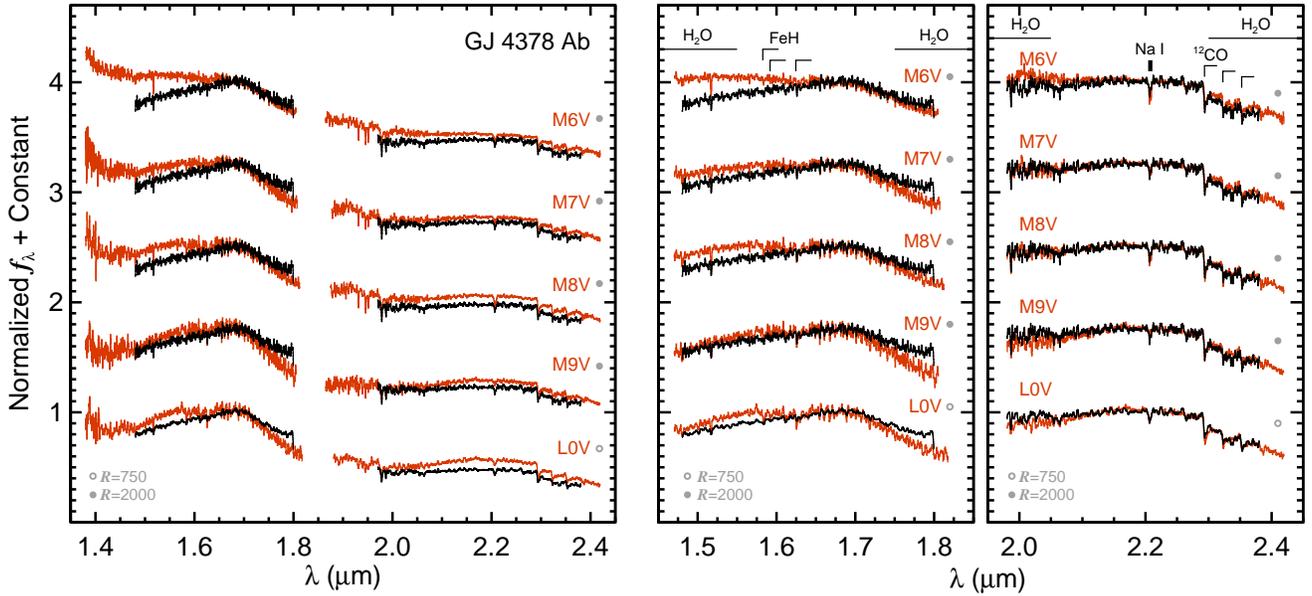}}
  \vskip -1 in
  \caption{Keck/OSIRIS 1.5--2.4~$\mu$m spectrum of GJ~4378~Ab (black).  The best match is to M8--M9 field
  templates (red) from the IRTF Spectral Library.  We adopt a near-infrared spectral type of M8~$\pm$~1.
  Note that the slightly angular $H$-band shape, which points to a young age of $\lesssim$150~Myr, is at odds with
  the lack of H$\alpha$ emission in the wide comoving companion GJ~4379~B, which suggests an old age
  of several Gyr.   Our OSIRIS spectrum has been Gaussian smoothed to match the resolving power of the templates ($R$$\approx$750--2000) and
   all spectra are normalized and offset by a constant.    \label{fig:gj4378spec} } 
\end{figure}

\clearpage
\newpage

% Figure 35

\begin{figure}
  \vskip -2.in
  \hskip 0 in
  \resizebox{6.5in}{!}{\includegraphics{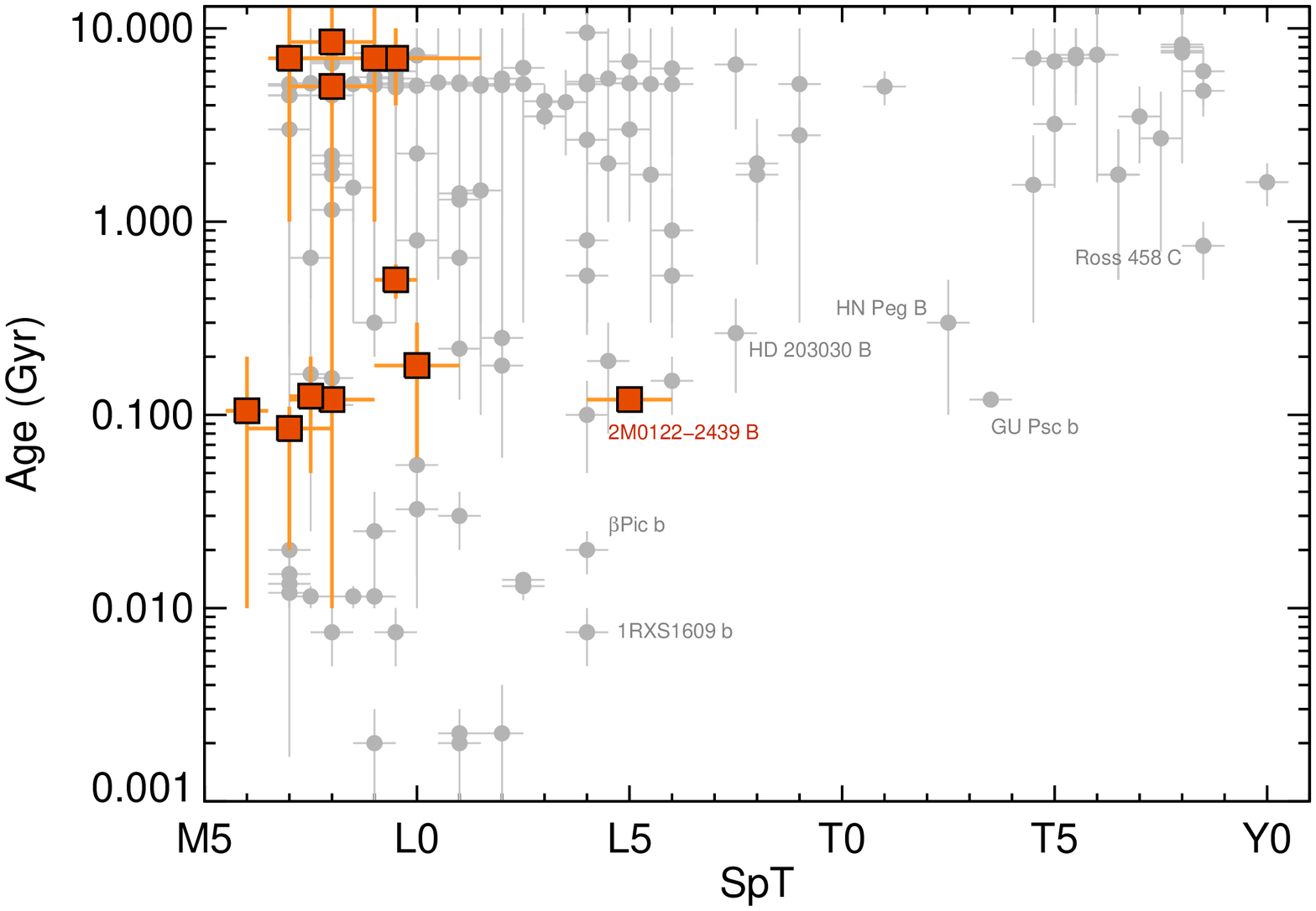}}
  \vskip -.2 in
  \caption{The known ultracool companions to stars.  Systems analyzed in this work are denoted with red
  squares and mostly have late-M spectral types.  The dearth of young ($\lesssim$100~Myr) companions later than $\sim$L5 reflects the paucity of young planets
  discovered via direct imaging.  Note that we have excluded the HR~8799 
  planets (\citealt{Marois:2008ei}; \citealt{Marois:2010gpa}), GJ~504~b (\citealt{Kuzuhara:2013jz}), and HD~95086~b (\citealt{Rameau:2013ds}) because
  their spectral types are either poorly constrained or may defy conventional classification schemes.   Known companions (grey circles) are from \citet{Deacon:2014ey} and are supplemented with our own compilation.   \label{fig:benchmarks} } 
\end{figure}

\end{document}